# Chemical Shifts in Molecular Solids by Machine Learning

Federico M. Paruzzo,[†] Albert Hofstetter,[†] Félix Musil,[‡] Sandip De,[‡] Michele Ceriotti,[*,‡] and Lyndon Emsley[*,†]

[†]Institut des Sciences et Ingénierie Chimiques, Ecole Polytechnique Fédérale de Lausanne (EPFL), 1015 Lausanne, Switzerland
[‡]Institut des Sciences et Génie Matériaux, Ecole Polytechnique Fédérale de Lausanne (EPFL), 1015 Lausanne, Switzerland.

The calculation of chemical shifts in solids has enabled methods to determine crystal structures in powders. The dependence of chemical shifts on local atomic environments sets them among the most powerful tools for structure elucidation of powdered solids or amorphous materials. Unfortunately, this dependency comes with the cost of high accuracy first-principle calculations to qualitatively predict chemical shifts in solids. Machine learning methods have recently emerged as a way to overcome the need for explicit high accuracy first-principle calculations. However, the vast chemical and combinatorial space spanned by molecular solids, together with the strong dependency of chemical shifts of atoms on their environment, poses a huge challenge for any machine learning method. Here we propose a machine learning method based on local environments to accurately predict chemical shifts of different molecular solids and of different polymorphs within DFT accuracy (RMSE of 0.49 ppm ($^{1}$H), 4.3ppm ($^{13}$C), 13.3 ppm ($^{15}$N), and 17.7 ppm ($^{17}$O) with $R^2$ of 0.97 for $^{1}$H, 0.99 for $^{13}$C, 0.99 for $^{15}$N, and 0.99 for $^{17}$O). We also demonstrate that the trained model is able to correctly determine, based on the match between experimentally-measured and ML-predicted shifts, structures of cocaine and the drug 4-[4-(2-adamantylcarbamoyl)-5-tert-butylpyrazol-1-yl]benzoic acid in an chemical shift based NMR crystallography approach.

Solid-state nuclear magnetic resonance (NMR) spectroscopy is among the most powerful methods for determining the atomic-level structure and dynamics of powdered and amorphous solids. Notably, solid-state NMR directly probes the local atomic environments and thus allows for characterization without the need for long-range order. This has led to its broad use today in many fields including for instance materials and pharmaceutical chemistry. In the latter the determination of structure and packing is essential to elaborate structure-property relations for formulations in the drug development process.

A revolution in solid-state NMR has occurred with the introduction of accurate methods to calculate chemical shifts,[1-3] in particular using plane wave DFT methods developed for periodic systems based on the PAW/GIPAW approach.[4-6] This has enabled very rapid development of chemical shift based NMR crystallography, which is now widely used to validate structures of molecular solids and identify known polymorphs,[7-31] or more recently in combination with crystal structure prediction (CSP) protocols, to determine *de novo* crystal structures from powders.[32-37] Recent studies also suggest that the structural accuracy of chemical shift based solid-state NMR crystallography is at least comparable with more traditional methods, such as single crystal X-ray diffraction.[38]

The power of the method arises from the fact that plane wave DFT with the GIPAW method is accurate enough to reproduce the exquisite sensitivity of chemical shifts to changes in local atomic environments. However, this approach also has severe limitations. The scaling of the computational cost with system size prevents the application to larger and more complex crystals, or non-equilibrium structures. If one wanted to use more accurate *ab initio* calculations, the expense is prohibitive.

Machine learning (ML) is emerging as new tool in many areas of chemical and physical science, and potentially provides a method to bridge the gap between the need for high accuracy calculations and limited computational power.[39-43] Notably, prediction of chemical shifts for the specific case of proteins in solution using methods based on large experimental databases, using traditional[44-51] or machine learning approaches,[52-54] have met with considerable success in predicting shifts based on local sequence and structural motifs, and are widely used today. While there are some examples of machine learned experimental and ab-initio chemical shifts of liquid and gas phase molecules,[55-59] to date there is only one example of machine learning being applied to calculations of chemical shifts in solids, which deals with the specific case of silicas.[60] Molecular solids are characterized by the combinatorial complexity and diversity of organic chemistry, the subtle dependence on conformations, and the long and short range effects of crystal packing, which leads to a considerably broader range of chemical environments and possible chemical shieldings than found e.g. in proteins. All of these aspects make this class of systems particularly challenging for machine learning.

Here, we develop a machine learning framework to predict chemical shifts in solids which is based on capturing the local environments of individual atoms, and which make it well suited for the prediction of such local properties. The protocol is schematically illustrated in Figure 1. We train the model on structures taken from the Cambridge Structural Database (CSD),[61] chosen to be as diverse as possible, and then show that the method can predict chemical shifts in a test set with a $R^2$ coefficients between the chemical shifts calculated with DFT and with ML of 0.97 for $^{1}$H, 0.99 for $^{13}$C, 0.99 for $^{15}$N, and 0.99 for $^{17}$O, corresponding to RMSEs of 0.49ppm for $^{1}$H, 4.3ppm for $^{13}$C, 13.3 ppm for $^{15}$N, and 17.7 ppm for $^{17}$O. Predicting the chemical shifts for a polymorph of cocaine, with 86 atoms in the unit-cell, using the ML method takes 46 seconds of CPU time, thus reducing the computational time by a factor of between 5 to 10 thousand, without any significant loss in accuracy as compared to DFT.

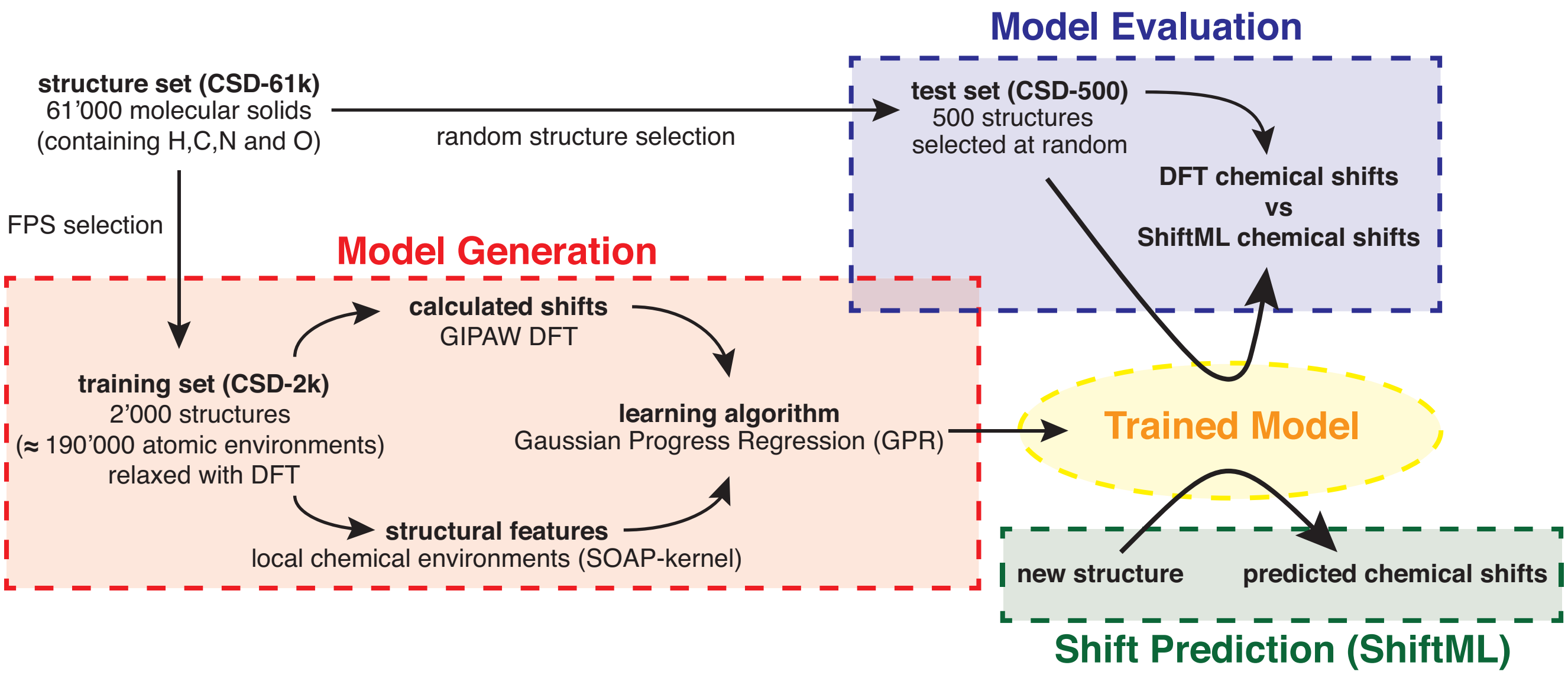

**Figure 1**. Scheme of the machine learning model used for the chemical shift predictions.

Most significantly, we show that the model has sufficient accuracy to be used in a chemical shift driven NMR crystallography protocol to correctly determine, based on the match between experimentally-measured and ML-predicted shifts, the correct structure of cocaine, and the drug 4-[4-(2-adamantylcarbamoyl)-5-tert-butylpyrazol-1-yl]benzoic acid (AZD8329). We also show that it is possible to calculate the NMR spectrum of very large molecular crystals. For this we calculate the chemical shifts of six structures from the CSD with between 768 and 1,584 atoms in the unit-cells.

## Results and discussion

The approach we take to predicting chemical shifts in molecular solids is illustrated in Figure 1. We use the Gaussian Process Regression (GPR) framework[62] to predict the chemical shift of a new atomic configuration based on a statistical model that identifies he correlations between structure and shift for a reference set of training configurations, for which the chemical shifts have been determined by a GIPAW DFT calculation. The predicted chemical shielding for a given atom is given by

$$\sigma(X) = \sum_i \alpha_i k(X, X_i), \quad (1)$$

where $X$ and $X_i$ correspond respectively to a description of the chemical environment of the atom for which we are making a prediction, and that of one of the training configurations. The weights $\alpha_i$ are obtained by requiring that Eq. (1) is consistent with the values computed by DFT for the reference structures. The essential ingredient that differentiates one GPR-based framework from another is the kernel function $k(X, X_i)$ which describes and assesses the similarity between atomic environments, and provides basis functions to approximate the target properties.

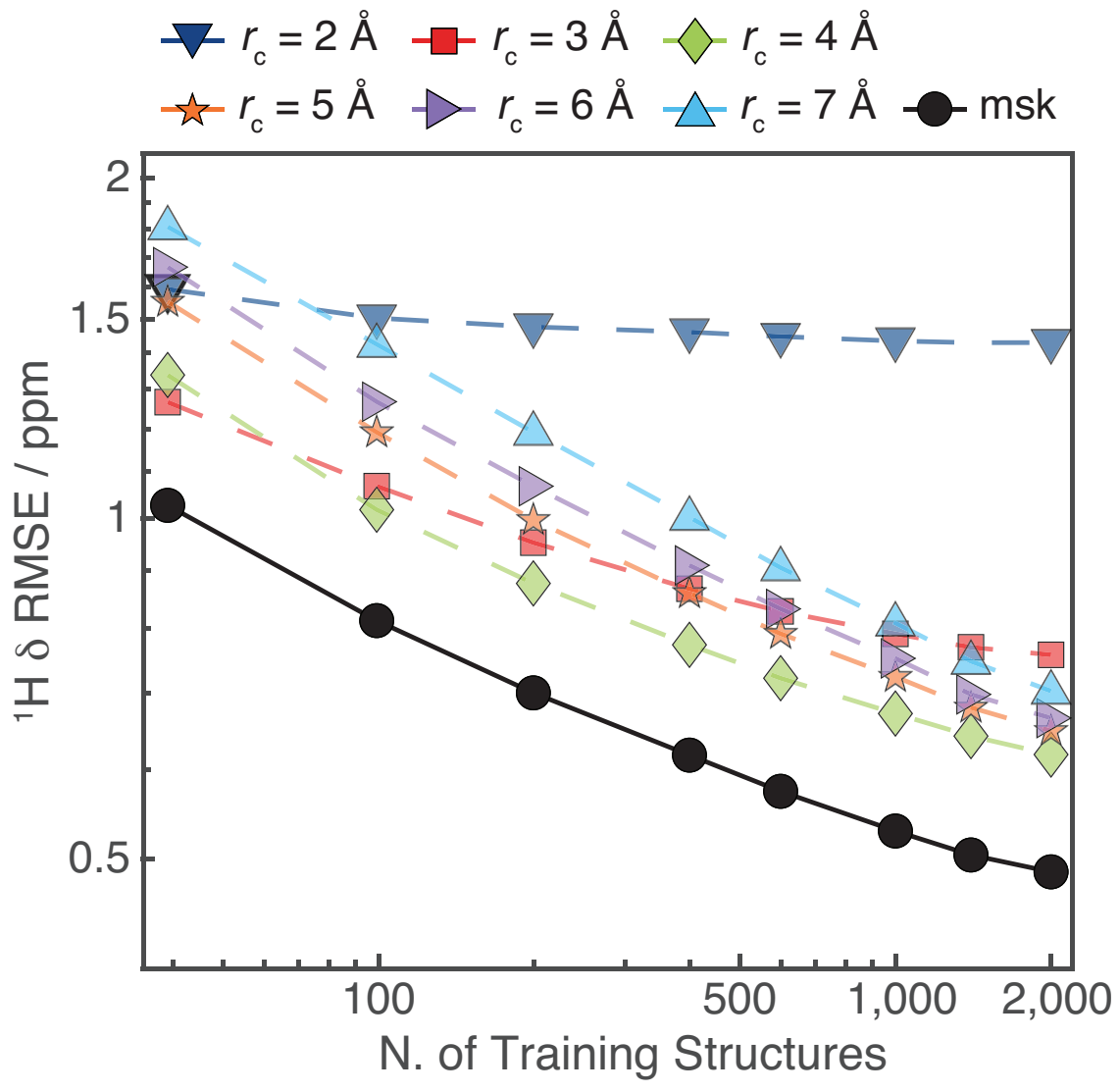

**Figure 2**. $^{1}$H chemical shift prediction error of the trained model on the CSD-500 set. The prediction error as RMSE is shown for the different local environment cut-off radii, and for the multi-kernel (labelled as msk), as a function of the training set size.

Here our model relies on the Smooth Overlap of Atomic Positions (SOAP) kernel,[63,64] in which any atomic environment is represented as a three dimensional neighbourhood density given by a superposition of Gaussians, one centred at each of the atom positions in a spherical neighbourhood within a cut-off radius $r_c$ from the core atom. This framework, combined with GPR, has been used to model the stability and properties of a number of different systems,[40,63,64] and has been extended to the prediction of tensorial properties.[65] We can see that this choice of kernel should be particularly well adapted to predicting chemical shifts, since it describes the local environments around each atom without any simplification, and this is indeed what the chemical shift also probes, as it is determined by the screening of the nucleus from the main magnetic field by the electron density at the nucleus. Note that it should be possible to tune and train other ML methods to accurately predict chemical shifts of molecular crystals. While these possibilities will be explored in future work, the model we present here is already accurate enough to substitute for DFT calculations in chemical shift based NMR crystallography.

As shown in Figure 1, the model is developed by using a reference training set of structures for which chemical shifts are known. To obtain a model which is robust and general, the training set should be as large, as reliable, and as diverse as possible. We first extract from the CSD a large set of about 61,000 structures, corresponding to all the structures in the CSD with fewer than 200 atoms (making DFT chemical shift calculation affordable) and containing C and H and allowing for N and/or O (we call this set CSD-61k). Given that performing a GIPAW calculation for all of these structures would be prohibitively demanding, we then select a random subset of 500 structures (CSD-500) that are representative of the chemical diversity in the CSD, and we use it to test the accuracy of our model. For cross-validation and training, instead, we select 2,000 structures (corresponding to about 70,000 atomic environments) out of the CSD-61k using a farthest point sampling algorithm (FPS)[66,67] (CSD-2k). This step ensures near-uniform sampling of the conformational space, improving the quality of the model when using a relatively small number of reference calculations.

To avoid including spurious environments in the model, e.g. environments which are not well described by DFT, we also automatically detect and discard from the training set all atomic environments with values of the DFT calculated shifts that are anomalous based on a cross validation procedure described in the SI. This pruning as well as the parameter optimization procedure (see below) were done exclusively using cross validation on the CSD-2k set. (Notably the test sets were not subject to any curation.)

In order to reduce the computational cost of the training and testing procedures we then finally remove from the training set all the symmetrically equivalent environments. In case of $^1$H, this reduced the size of the training set from 70,000 to about 35,000 different atomic environments. (Details of the selection method and the members of the different sets used are given in SI).

All the atomic positions of the structures in the training and testing sets were relaxed with DFT, using the Quantum Espresso suite,[68,69] prior to calculation of the chemical shieldings using the GIPAW DFT method.[4,5] Parameters for the DFT calculations are given in the SI. The calculated chemical shieldings $\sigma$ are converted to the corresponding chemical shifts $\delta$ through the relationship $\delta = \sigma_{ref} - \sigma$. Here, we used a $\sigma_{ref}$ of 30.8 ppm (for $^1$H) and 169.5 ppm (for $^{13}$C), found through linear regression between the calculated and experimental chemical shifts for cocaine.

Figure 2 shows the results of the predictions of the chemical shifts of the CSD-500 set, which is representative of the expected accuracy for the entire CSD-61k. The figure shows the overall prediction accuracy for $^1$H chemical shifts as root-mean-square-error (RMSE) in ppm between the shifts calculated with DFT and with ShiftML as a function of the cut-off radius ($r_c$) and as a function of the number of training structures included from CSD-2k. The effect of the different cut-off radii is clearly visible. For example, for $r_c$=2Å the prediction error for a small training set (<10 structures or <100 atomic environments) can be smaller than for the larger radii, but does not improve significantly with increasing size of the training set. On the contrary, for $r_c$=7Å we observe a relatively large prediction error for a small training set, but even with 2,000 structures (35,000 environments), the prediction error is still decreasing. A similar behaviour is observed for the prediction errors of the $^{13}$C, $^{15}$N and $^{17}$O chemical shifts (SI).

The observed differences in the behaviour of the prediction error with respect to $r_c$ clearly indicates the influence of the different extents of the local environment on the chemical shift. Short range interactions are sufficient to explain the rough order of magnitude of the shift, but long range interactions are required to learn about the higher order influences of next-nearest neighbours on shifts. However, for long range interactions, a much larger number of environments is needed in order to determine the correlation between environment and shift.

We exploit these differences to generate a combined SOAP kernel consisting of a linear combination of the single local environment kernels,[40] with weightings of 256 ($r_c$=2Å), 128 ($r_c$=3Å), 32($r_c$=4Å), 8 ($r_c$=5Å and $r_c$=6Å) and 1 ($r_c$=7Å). This weighting was determined by rough optimization around values inspired by previous experience,[40] and by cross-validation on the CSD-2k training set (as described in SI). It is clear that learning with the combined kernel leads consistently to lower prediction errors than any of the single kernels, although the improvement in performance varies between nuclei (SI).

Figure 3a-d shows correlation plots between $^1$H, $^{13}$C, $^{15}$N and $^{17}$O chemical shifts calculated by DFT and by ShiftML for the CSD-500 set trained on the whole CSD-2k combined kernel. Using the combined kernel, we reach a prediction error of 0.49 ppm for $^1$H (4.3 ppm for $^{13}$C, 13.3 ppm for $^{15}$N and 17.7 ppm for $^{17}$O). This is very comparable with reported DFT chemical shift accuracy for $^1$H of 0.33-0.43 ppm,[15,70] while requiring a fraction of the computational time and cost: 46 CPU seconds compared to ~62-150 CPU hours for DFT chemical shift calculation on structures containing 86 atoms (around 350 valence electrons) (see SI). For the other nuclei, the ML accuracy is slightly lower than reported values (1.9-2.2 ppm for $^{13}$C, 5.4 ppm for $^{15}$N and 7.2 ppm for $^{17}$O),[15,70] which is not surprising as there are (currently) significantly less training environments for the heteronuclei than for $^1$H.

The $R^2$ coefficients between the chemical shifts calculated with DFT and with ShiftML are 0.97 for $^1$H, 0.99 for $^{13}$C, 0.99 for $^{15}$N, and 0.99 for $^{17}$O.

We observe that the performance of the model degrades considerably if one does not use the procedure we developed to remove environments for which the DFT shifts appear to be inconsistent. Note that the CSD-500 set used for testing is selected randomly from CSD-61k and not curated. Indeed, we find that many of the atomic environments in the CSD-500 set with a relatively high prediction RMSE possess either unusual cavities inside their crystal structure, possibly indicating an organic cage surrounding non-crystalline solvent or other atoms, or exhibit strongly delocalised π-bonding networks. While there is no theoretical reason preventing the machine learning model from correctly describing such environments, they are rare and not well represented within the training set. CSD-500 thus constitutes a fairly demanding test set.

Having evaluated the power of the trained model to predict the diverse CSD-500 set, we now look at the capacity to predict potentially subtler differences by looking at a set of polymorphs of a given structure. Figures 4a and b show the correlation between the $^1$H shifts calculated by GIPAW DFT and by ShiftML for 30 polymorphs of cocaine and 14 polymorphs of AZD8329, all of which were previously generated with a crystal structure prediction (CSP) procedure.[18,32] The figure clearly shows that ShiftML is able to accurately predict the differences in $^1$H chemical shift for different polymorphs.

We find a chemical shift prediction error (RMSE) for $^1$H for the cocaine polymorphs of 0.37 ppm and for AZD8329 of 0.46 ppm, which is very comparable to the expected GIPAW DFT accuracy.

Note that for these cases the DFT structure optimization and GIPAW chemical shift calculation were done with a different DFT program (CASTEP)[71], which suggests that ShiftML is robust with respect to small deviations from the fully optimized structures. (As shown in the SI, performing the prediction using Quantum Espresso consistently leads to comparable prediction accuracy.)

For the heteronuclei we obtain for cocaine 3.8 ppm for $^{13}$C, 12.1 ppm for $^{15}$N and 15.7 ppm for $^{17}$O. These values for cocaine are again very comparable to the best accuracy so far reported for DFT calculations. For AZD8329 the $^{15}$N and $^{17}$O RMSEs are proportionally larger (17.7 and 54.7 ppm), and we attribute this to the fact that the molecule contains a rather unusual C-O…H-N / C-O…H-O H-bonded dimer structure, for which the learning is thus even sparser than for the heteronuclei in general. To illustrate the unusual nature of this motif, we note that the calculated $^{17}$O shifts using DFT also change by up to 50 ppm for structures relaxed either by the CASTEP protocol used in ref. 30, or the Quantum Espresso protocol used here (the RMSE between ML and DFT for the Quantum Espresso relaxed structures is reduced to 10.9 and 11.5 ppm for $^{15}$N and $^{17}$O!). The RMSE of 4.0 ppm for $^{13}$C for AZD8329 is in line with the other systems.

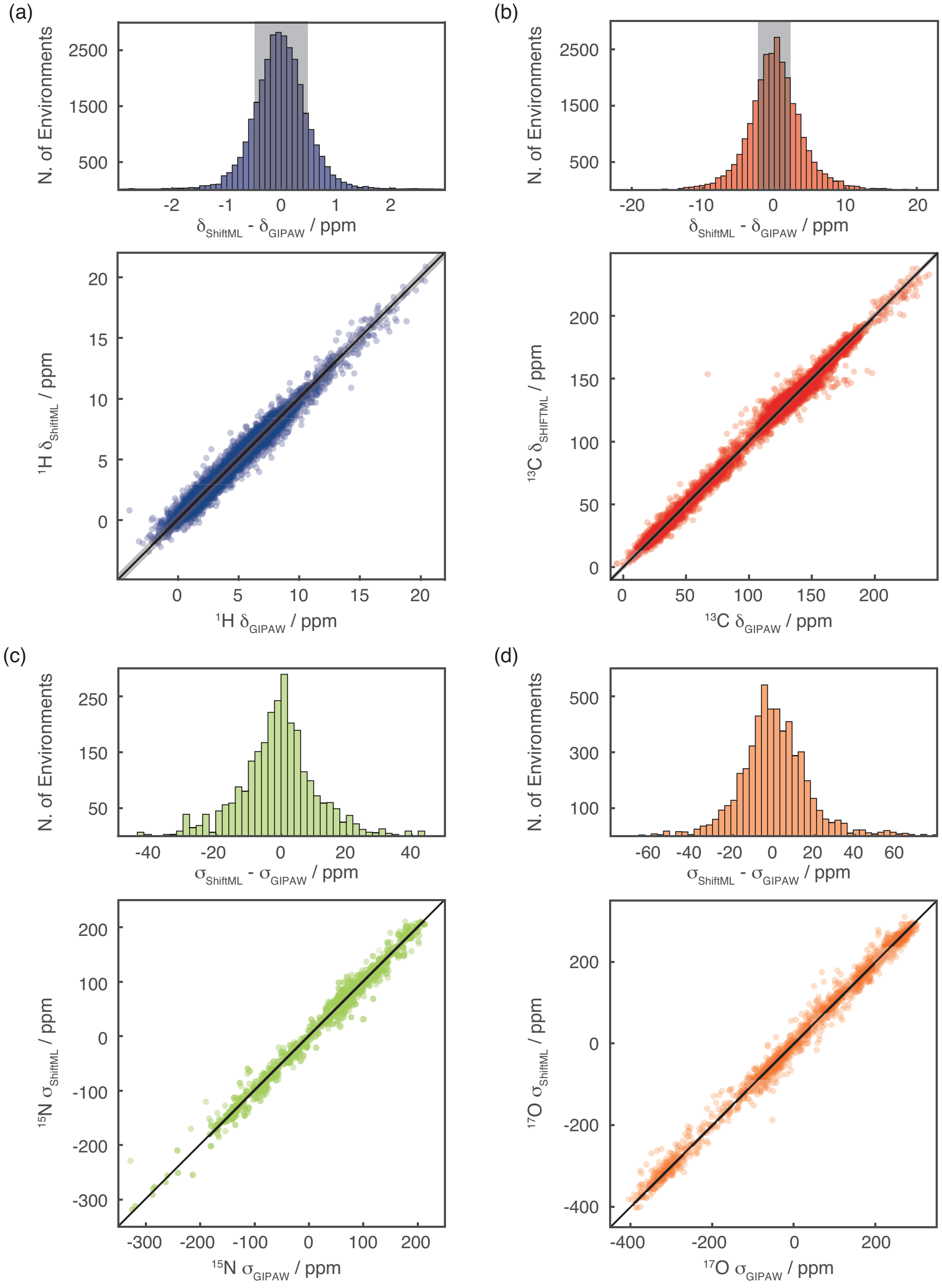

**Figure 3**. Histograms and scatterplots showing the correlation between $^{1}$H (a), $^{13}$C (b), $^{15}$N (c) and $^{17}$O (d) chemical shifts (shieldings) calculated with GIPAW DFT and ShiftML. The black lines indicate a perfect correlation and the grey zones represent the confidence limits between experiment and DFT calculated shifts currently used.[15]

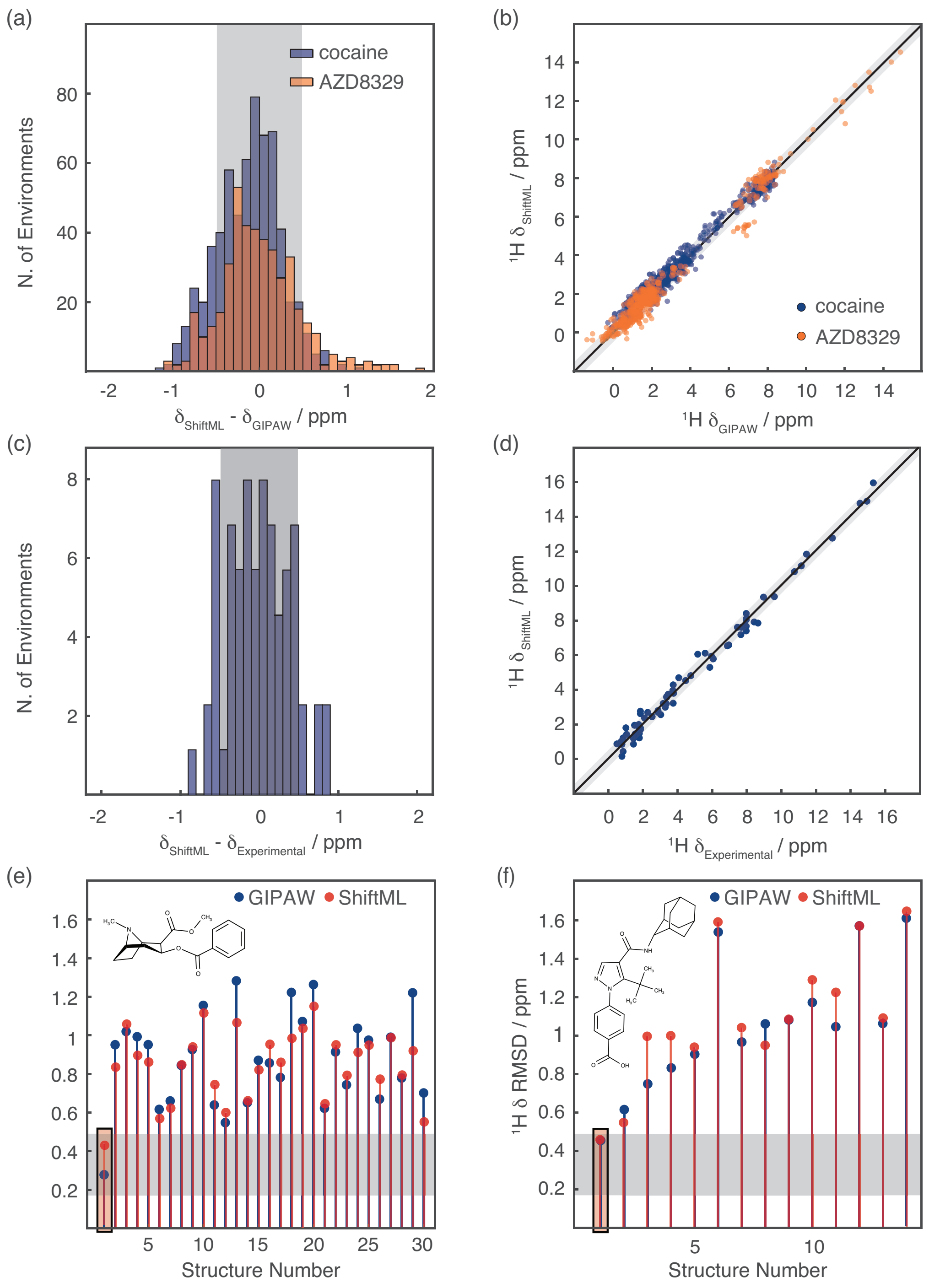

**Figure 4.** Comparison of ShiftML predictions to experimental shifts and to GIPAW predictions for polymorphs of cocaine and AZD8329. (a) Histogram showing the distribution of the differences between $^1H$ chemical shifts calculated with GIPAW and ShiftML. The blue bars were calculated for the polymorphs of cocaine, and the orange ones for the polymorphs of AZD8329. (b) Scatterplot showing the correlation between GIPAW and ShiftML $^1H$ chemical shifts for cocaine (blue) and AZD8329 (orange). (c) Histogram showing the distribution of differences between experimentally measured $^1H$ chemical shifts and $^1H$ chemical shifts calculated with ShiftML for six different crystal structures (see SI for the structures and numerical values of the shifts). (d) Scatterplot showing the correlation between these experimentally measured $^1H$ chemical shifts and shifts calculated with ShiftML. (e-f) Comparison between calculated and experimental $^1H$ chemical shifts for the most stable structures obtained with CSP for cocaine (e) and AZD8329 (f). For each candidate structure an aggregate RMSE is shown between experimentally measured shifts and shifts calculated using either GIPAW (blue) or ShiftML (red). The highlighted bar correspond to the expected errors, and candidates that have RMSEs within this range would be determined as correct crystal structures using a chemical shift driven solid-state NMR crystallography protocol. The black line in (b) and (d) indicates a perfect correlation. In (a-f) the grey zones represent the confidence intervals of the $^1H$ chemical shift RMSD, as described in the text.[15]

Further, the significance of the method is illustrated by comparison to experimentally measured shifts, where we find that the predicted shifts are accurate enough to allow crystal structure determination for both cocaine and AZD8329 from powder samples in a chemical shift driven NMR crystallography approach.

Figures 4c and d show the correlation between experimentally measured $^1$H chemical shifts and the $^1$H chemical shifts calculated by ShiftML for crystal structures of the six molecules shown in scheme 1 (numerical values of the experimental chemical shifts, the crystal structures, and the shifts calculated with ShiftML are given in the SI). The comparison between experimental and calculated $^1$H chemical shifts for all crystal structures (for a total of 68 shifts) gives an error (RMSE) of 0.39 ppm and a $R^2$ coefficient of 0.99.

Figures 4e and f show in blue the RMSE between DFT calculated and experimental $^1$H chemical shifts for the 30 polymorphs predicted by CSP to have the lowest energy for cocaine and the 14 *cis* polymorphs of AZD8329. For both molecules the only structure in agreement with the GIPAW DFT calculations, to below a $^1$H DFT chemical shift confidence interval of 0.49 ppm,[15] is the correct crystal structure. In the same plots we overlay the result where the experimental shifts are now compared to shifts predicted with ShiftML. Note that the RMSE between experiment and the predicted chemical shifts follows the same trends as for the DFT calculated shifts, and that here again the only structures below the confidence interval of 0.49 ppm are the two correct crystal structures. Note, that the cut-off of 0.49 ppm with respect to experiment has been evaluated for GIPAW DFT chemical shifts[15,70] and to rigorously repeat the CSP procedure for the ML method, the accuracy should be re-evaluated using more extensive benchmarking of ShiftML to experiment, which will be the subject of further work.

Finally, we note that the accuracy of the method does not depend on the size of the structure, and that the prediction time is linear in the number of atoms. For the structures we calculate here the prediction time actually appears nearly constant, because it is dominated by the loading time of the reference SOAP vector (see Figure 5a). We have used this method to calculate the NMR spectra (shown in Figure 5b-g) for six structures from the CSD having among the largest numbers of atoms per unit cell (containing only H,C,N,O), with between 768 and 1,584 atoms per unit cell. (See scheme S1 for the chemical formula). The values of the predicted chemical shifts are given as CSD-6 in the SI. Figure 5a shows the comparison between the GIPAW calculation time and the required ML prediction time. We estimate that the whole calculation would require around 16 CPU years by GIPAW. ShiftML requires less than 6 CPU minutes to calculate the shifts for all the compounds.

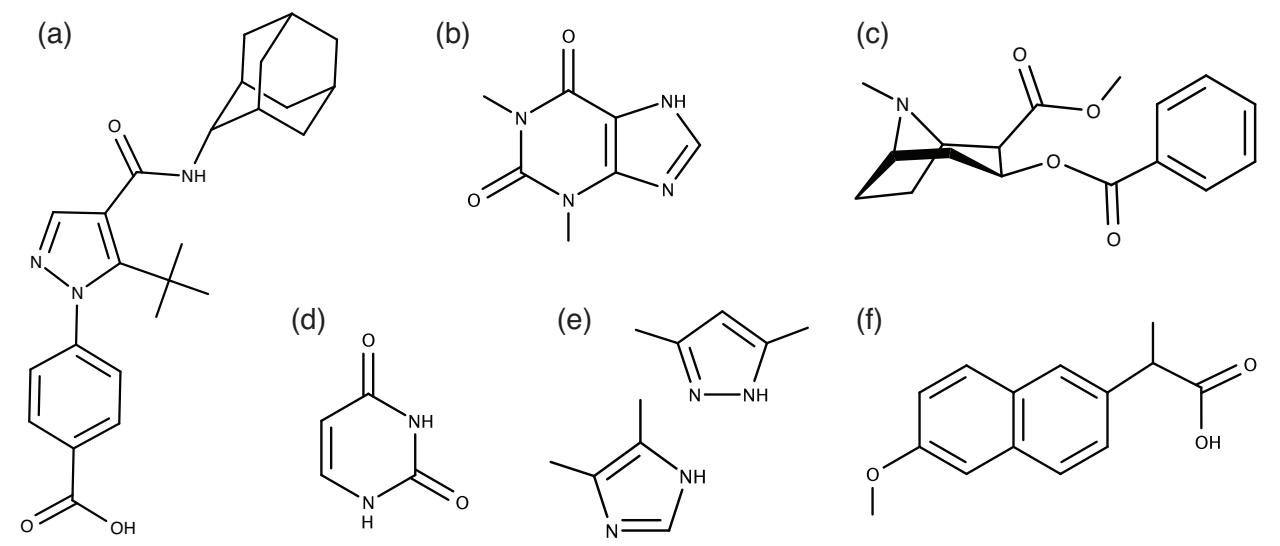

**Scheme 1.** Chemical structures of the six molecules used to evaluate the correlation between experimentally measured $^1$H chemical shifts and the $^1$H chemical shifts calculated by ShiftML. The structures are given as AZD8329 (a), theophylline (b), cocaine (c), uracil (d), 3,5-dimethylimidazole and 4,5-dimethylimidazole (e) and naproxen (f).

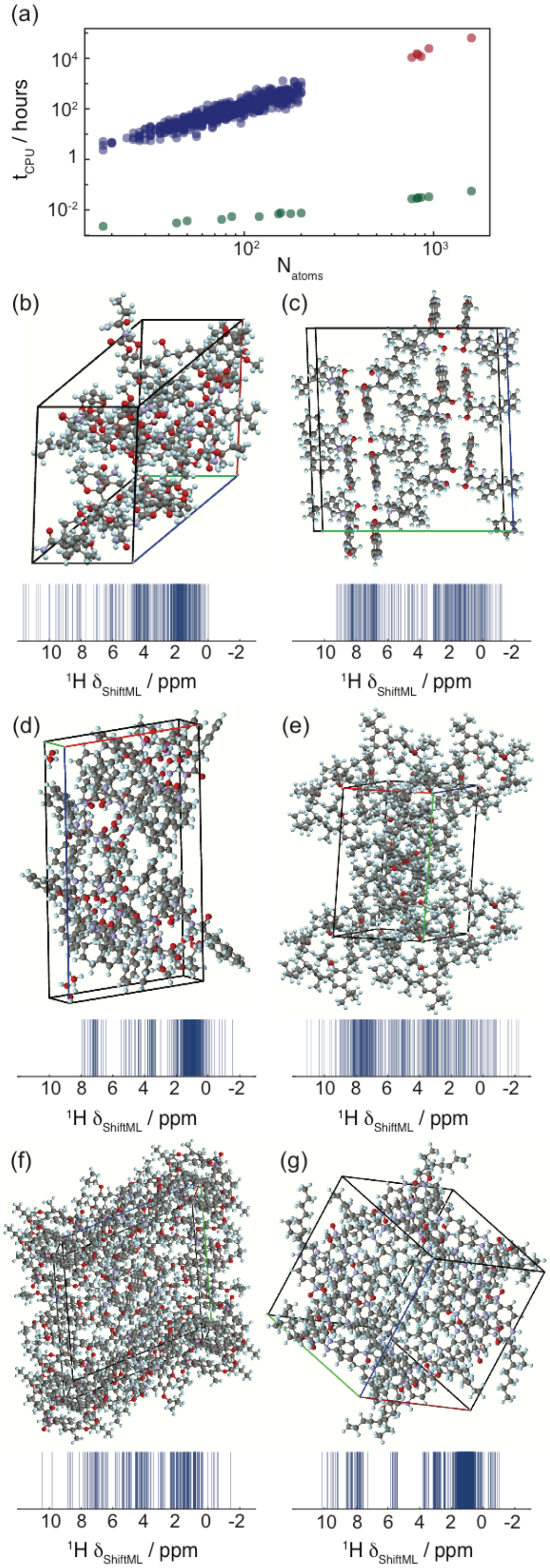

**Figure 5.** (a) DFT GIPAW calculation time (blue) and ShiftML prediction time (green) for different system sizes. The GIPAW DFT calculation time for the six large structures (red) is estimated from a cubic dependence on the number of valence electrons in the structure (see SI). (b-g) 3D-shemes and $^1$H NMR spectra predicted with ShiftML, of the six large molecular crystals with CSD Refcodes: (b) CAJVUH,[72] $N_{atoms}$ = 828, (c) RUKTOI,[73] $N_{atoms}$ = 768, (d) EMEMUE,[74] $N_{atoms}$ = 860, (e) GOKXOV,[75] $N_{atoms}$ = 945, (f) HEJBUW,[76] $N_{atoms}$ = 816, (g) RAYFEF,[77] $N_{atoms}$ = 1,584.

In summary, we have presented a ML model based on local environments to predict chemical shifts of molecular solids containing HCNO to within current DFT accuracy. The $R^2$ coefficients between the chemical shifts calculated with DFT and with ShiftML are 0.97 for $^{1}$H, 0.99 for $^{13}$C, 0.99 for $^{15}$N, and 0.99 for $^{17}$O. The approach allows the calculation of chemical shifts for structures with ~100 atoms in less than 1 minute, reducing the computational cost of chemical shift predictions in solids by a factor of between 5 to 10 thousand compared to current DFT chemical shift calculations, and thereby relieves a major bottleneck in the use of calculated chemical shifts for structure determination in solids.

Far from being just a benchmark of a machine-learning scheme, the method is accurate enough to be used to determine structures by comparison to experimental shifts in chemical shift based NMR crystallography approaches to structure determination, as shown here for cocaine and AZD8329. The ML model only scales linearly with the number of atoms and, for the prediction of individual structures, is dominated by a constant I/O overhead. Here it allows the calculation of chemical shifts for a set of six structures with between 768 and 1584 atoms in their unit cells in less than six minutes (an acceleration of a factor $10^6$ for the largest structure).

The code used for the ML shift calculations is available as an online server at http://shiftml.epfl.ch. The accuracy of the method is likely to increase further with the size of the training set, and subsequently with the future evolution of the accuracy of the method used to calculate the reference shifts used in training (here DFT), or by using experimental shifts if a large enough set were available.

The model used here can easily be extended to organic solids including halides or other nuclei, and to network materials such as oxides, and these will be the subject of further work.

## Methods

For the SOAP kernels,[63,64] each atomic environment is represented as a three dimensional neighbourhood density given by a superposition of Gaussians, one centred at each of the atom positions in a spherical neighbourhood within a cut-off radius $r_c$ from the core atom. The Gaussians have a variance $\varsigma^2$, and a separate density is built for each atomic species. The kernel is then constructed as the symmetrized overlap between the amplitudes representing $X$ and $X'$. This degree of overlap thus measures the similarity between the environments $X$ and $X'$.

The SOAP and GPR parameters are given in the SI. SOAP-based structural kernels contain several adjustable hyper-parameters, which are discussed in refs.[64] However, we have not systematically explored the full parametric space here, instead we chose reasonable values of the parameters without extensive fine-tuning, based on previous experience[40] and with some optimization by cross-validation on the CSD-2k training set (see SI for details).We also combine kernels computed for different cut-off radii to capture the contributions to shifts from different length scales,[40] as is described in detail above. The calculations of the local environment, the similarity kernel and the weighted correlations were done using the GLOSIM2 package.[78]

The program used to calculate shifts using the protocol described here is called ShiftML, and is publically available as an online server at http://shiftml.epfl.ch.

## Supporting Information.

Details on the structure selection, crystal structure prediction, the DFT calculations, the GPR method, the SOAP kernels, the FPS algorithm, the detection procedure of unusual environments, NMR crystallography and the DFT calculation time estimates. Prediction parameters optimization, learning curves and evaluation curves for $^{13}$C, $^{15}$N and $^{17}$O. Predicted and GIPAW chemical shieldings for all cocaine and AZD8329 polymorphs. Chemical formula and predicted chemical shieldings of the CSD-6 set predicted with ShiftML. The Refcodes for CSD-2k and CSD-500.

## Corresponding Author(s)

E-mail: lyndon.emsley@epfl.ch, michele.ceriotti@epfl.ch.

## Acknowledgment

We are grateful for financial support from Swiss National Science Foundation Grant No. 200021_160112. FM and SD were supported by the NCCR MARVEL, funded by the Swiss National Science Foundation. MC acknowledges funding by the European Research Council under the European Union's Horizon 2020 research and innovation program (grant agreement no. 677013-HBMAP). This work was also supported by EPFL through the use of the facilities of its Scientific IT and Application Support Center.

## References

1 Dedios, A. C., Pearson, J. G. & Oldfield, E. Secondary and Tertiary Structural Effects on Protein Nmr Chemical-Shifts - an Abinitio Approach. *Science* **260**, 1491-1496, doi:DOI 10.1126/science.8502992 (1993).
2 Facelli, J. C. & Grant, D. M. Determination of molecular symmetry in crystalline naphthalene using solid-state NMR. *Nature* **365**, 325-327, doi:10.1038/365325a0 (1993).
3 Sebastiani, D. & Parrinello, M. A new ab-initio approach for NMR chemical shifts in periodic systems. *J. Phys. Chem. A* **105**, 1951-1958, doi:10.1021/jp002807j (2001).
4 Pickard, C. J. & Mauri, F. All-electron magnetic response with pseudopotentials: NMR chemical shifts. *Phys. Rev. B* **63**, doi:ARTN 245101 DOIDOI 10.1103/PhysRevB.63.245101 (2001).
5 Yates, J. R., Pickard, C. J. & Mauri, F. Calculation of NMR chemical shifts for extended systems using ultrasoft pseudopotentials. *Phys. Rev. B* **76**, 024401, doi:ARTN 024401 DOI 10.1103/PhysRevB.76.024401 (2007).
6 Blochl, P. E. Projector augmented-wave method. *Phys Rev B Condens Matter* **50**, 17953-17979 (1994).
7 Ochsenfeld, C., Brown, S. P., Schnell, I., Gauss, J. & Spiess, H. W. Structure assignment in the solid state by the coupling of quantum chemical calculations with NMR experiments: a columnar hexabenzocoronene derivative. *J Am Chem Soc* **123**, 2597-2606, doi:10.1021/ja0021823 (2001).
8 Goward, G. R. *et al.* Benzoxazine oligomers: evidence for a helical structure from solid-state NMR spectroscopy and DFT-based dynamics and chemical shift calculations. *J Am Chem Soc* **125**, 5792-5800, doi:10.1021/ja029059r (2003).
9 Harris, R. K. NMR crystallography: the use of chemical shifts. *Solid State Sci.* **6**, 1025-1037, doi:10.1016/j.solidstatesciences.2004.03.040 (2004).
10 Harper, J. K. & Grant, D. M. Enhancing crystal-structure prediction with NMR tensor data. *Cryst. Growth Des.* **6**, 2315-2321, doi:10.1021/cg060244g (2006).
11 Harris, R. K. NMR studies of organic polymorphs and solvates. *Analyst* **131**, 351-373, doi:10.1039/b516057j (2006).
12 Harris, R. K. Applications of solid-state NMR to pharmaceutical polymorphism and related matters. *J Pharm Pharmacol* **59**, 225-239, doi:10.1211/jpp.59.2.0009 (2007).
13 Othman, A., Evans, J. S., Evans, I. R., Harris, R. K. & Hodgkinson, P. Structural study of polymorphs and solvates of finasteride. *J Pharm Sci* **96**, 1380-1397, doi:10.1002/jps.20940 (2007).

14 Salager, E., Stein, R. S., Pickard, C. J., Elena, B. & Emsley, L. Powder NMR crystallography of thymol. *Phys Chem Chem Phys* **11**, 2610-2621, doi:10.1039/b821018g (2009).
15 Salager, E. *et al.* Powder crystallography by combined crystal structure prediction and high-resolution 1H solid-state NMR spectroscopy. *J Am Chem Soc* **132**, 2564-2566, doi:10.1021/ja909449k (2010).
16 Webber, A. L., Emsley, L., Claramunt, R. M. & Brown, S. P. NMR crystallography of campho[2,3-c]pyrazole (Z' = 6): combining high-resolution 1H-13C solid-state MAS NMR spectroscopy and GIPAW chemical-shift calculations. *J Phys Chem A* **114**, 10435-10442, doi:10.1021/jp104901j (2010).
17 Dudenko, D. *et al.* A strategy for revealing the packing in semicrystalline pi-conjugated polymers: crystal structure of bulk poly-3-hexyl-thiophene (P3HT). *Angew. Chem., Int. Ed. Engl.* **51**, 11068-11072, doi:10.1002/anie.201205075 (2012).
18 Baias, M. *et al.* Powder crystallography of pharmaceutical materials by combined crystal structure prediction and solid-state 1H NMR spectroscopy. *Phys Chem Chem Phys* **15**, 8069-8080, doi:10.1039/c3cp41095a (2013).
19 Pawlak, T., Jaworska, M. & Potrzebowski, M. J. NMR crystallography of alpha-poly(L-lactide). *Phys Chem Chem Phys* **15**, 3137-3145, doi:10.1039/c2cp43174b (2013).
20 Santos, S. M., Rocha, J. & Mafra, L. NMR Crystallography: Toward Chemical Shift-Driven Crystal Structure Determination of the beta-Lactam Antibiotic Amoxicillin Trihydrate. *Cryst. Growth Des.* **13**, 2390-2395, doi:10.1021/cg4002785 (2013).
21 Koike, R. *et al.* Structural Determination of a Novel Polymorph of Sulfathiazole-Oxalic Acid Complex in Powder Form by Solid-State NMR Spectroscopy on the Basis of Crystallographic Structure of Another Polymorph. *Cryst. Growth Des.* **14**, 4510-4518, doi:10.1021/cg5005903 (2014).
22 Ludeker, D. & Brunklaus, G. NMR crystallography of ezetimibe co-crystals. *Solid State Nucl Magn Reson* **65**, 29-40, doi:10.1016/j.ssnmr.2014.11.002 (2015).
23 Paluch, P., Pawlak, T., Oszajca, M., Lasocha, W. & Potrzebowski, M. J. Fine refinement of solid state structure of racemic form of phospho-tyrosine employing NMR Crystallography approach. *Solid State Nucl Magn Reson* **65**, 2-11, doi:10.1016/j.ssnmr.2014.08.002 (2015).
24 Pinon, A. C., Rossini, A. J., Widdifield, C. M., Gajan, D. & Emsley, L. Polymorphs of Theophylline Characterized by DNP Enhanced Solid-State NMR. *Mol Pharm* **12**, 4146-4153, doi:10.1021/acs.molpharmaceut.5b00610 (2015).
25 Sardo, M. *et al.* Diazole-based powdered cocrystal featuring a helical hydrogen-bonded network: structure determination from PXRD, solid-state NMR and computer modeling. *Solid State Nucl Magn Reson* **65**, 49-63, doi:10.1016/j.ssnmr.2014.12.005 (2015).
26 Watts, A. E., Maruyoshi, K., Hughes, C. E., Brown, S. P. & Harris, K. D. M. Combining the Advantages of Powder X-ray Diffraction and NMR Crystallography in Structure Determination of the Pharmaceutical Material Cimetidine Hydrochloride. *Cryst. Growth Des.* **16**, 1798-1804, doi:10.1021/acs.cgd.6b00016 (2016).
27 Widdifield, C. M., Robson, H. & Hodgkinson, P. Furosemide's one little hydrogen atom: NMR crystallography structure verification of powdered molecular organics. *Chem Commun (Camb)* **52**, 6685-6688, doi:10.1039/c6cc02171a (2016).
28 Mali, G. Ab initio crystal structure prediction of magnesium (poly)sulfides and calculation of their NMR parameters. *Acta Crystallogr C Struct Chem* **73**, 229-233, doi:10.1107/S2053229617000687 (2017).
29 Harris, R. K., Joyce, S. A., Pickard, C. J., Cadars, S. & Emsley, L. Assigning carbon-13 NMR spectra to crystal structures by the INADEQUATE pulse sequence and first principles computation: a case study of two forms of testosterone. *Phys Chem Chem Phys* **8**, 137-143, doi:10.1039/b513392k (2006).
30 Mifsud, N., Elena, B., Pickard, C. J., Lesage, A. & Emsley, L. Assigning powders to crystal structures by high-resolution (1)H-(1)H double quantum and (1)H-(13)C J-INEPT solid-state NMR spectroscopy and first principles computation. A case study of penicillin G. *Phys Chem Chem Phys* **8**, 3418-3422, doi:10.1039/b605227d (2006).
31 Heider, E. M., Harper, J. K. & Grant, D. M. Structural characterization of an anhydrous polymorph of paclitaxel by solid-state NMR. *Phys Chem Chem Phys* **9**, 6083-6097, doi:10.1039/b711027h (2007).
32 Baias, M. *et al.* De novo determination of the crystal structure of a large drug molecule by crystal structure prediction-based powder NMR crystallography. *J Am Chem Soc* **135**, 17501-17507, doi:10.1021/ja4088874 (2013).
33 Fernandes, J. A., Sardo, M., Mafra, L., Choquesillo-Lazarte, D. & Masciocchi, N. X-ray and NMR Crystallography Studies of Novel Theophylline Cocrystals Prepared by Liquid Assisted Grinding. *Cryst. Growth Des.* **15**, 3674-3683, doi:10.1021/acs.cgd.5b00279 (2015).
34 Leclaire, J. *et al.* Structure elucidation of a complex CO2-based organic framework material by NMR crystallography. *Chem. Sci.* **7**, 4379-4390, doi:10.1039/c5sc03810c (2016).
35 Selent, M. *et al.* Clathrate Structure Determination by Combining Crystal Structure Prediction with Computational and Experimental (129) Xe NMR Spectroscopy. *Chemistry* **23**, 5258-5269, doi:10.1002/chem.201604797 (2017).
36 Widdifield, C. M. *et al.* Does Z' equal 1 or 2? Enhanced powder NMR crystallography verification of a disordered room temperature crystal structure of a p38 inhibitor for chronic obstructive pulmonary disease. *Phys Chem Chem Phys* **19**, 16650-16661, doi:10.1039/c7cp02349a (2017).
37 Nilsson Lill, S. O. *et al.* Elucidating an Amorphous Form Stabilization Mechanism for Tenapanor Hydrochloride: Crystal Structure Analysis Using X-ray Diffraction, NMR Crystallography, and Molecular Modeling. *Mol Pharm* **15**, 1476-1487, doi:10.1021/acs.molpharmaceut.7b01047 (2018).
38 Hofstetter, A. & Emsley, L. Positional Variance in NMR Crystallography. *J Am Chem Soc* **139**, 2573-2576, doi:10.1021/jacs.6b12705 (2017).
39 Curtarolo, S. *et al.* The high-throughput highway to computational materials design. *Nat Mater* **12**, 191-201, doi:10.1038/nmat3568 (2013).
40 Bartok, A. P. *et al.* Machine learning unifies the modeling of materials and molecules. *Sci Adv* **3**, e1701816, doi:10.1126/sciadv.1701816 (2017).
41 Xue, D. *et al.* Accelerated search for materials with targeted properties by adaptive design. *Nat Commun* **7**, 11241, doi:10.1038/ncomms11241 (2016).
42 Ward, L., Agrawal, A., Choudhary, A. & Wolverton, C. A general-purpose machine learning framework for predicting properties of inorganic materials. *npj Comput. Mater.* **2**, doi:ARTN 16028 DOI 10.1038/npjcompumats.2016.28 (2016).
43 Rupp, M., Tkatchenko, A., Muller, K. R. & von Lilienfeld, O. A. Fast and accurate modeling of molecular atomization energies with machine learning. *Phys Rev Lett* **108**, 058301, doi:10.1103/PhysRevLett.108.058301 (2012).
44 Shen, Y. & Bax, A. Protein backbone chemical shifts predicted from searching a database for torsion angle and sequence homology. *J Biomol NMR* **38**, 289-302, doi:10.1007/s10858-007-9166-6 (2007).
45 Neal, S., Nip, A. M., Zhang, H. Y. & Wishart, D. S. Rapid and accurate calculation of protein H-1, C-13 and N-15 chemical shifts. *J. Biomol. NMR* **26**, 215-240, doi:Doi 10.1023/A:1023812930288 (2003).
46 Wishart, D. S., Watson, M. S., Boyko, R. F. & Sykes, B. D. Automated H-1 and C-13 chemical shift prediction using the BioMagResBank. *J. Biomol. NMR* **10**, 329-336, doi:Doi 10.1023/A:1018373822088 (1997).
47 Iwadate, M., Asakura, T. & Williamson, M. P. C alpha and C beta carbon-13 chemical shifts in proteins from an empirical database. *J Biomol NMR* **13**, 199-211, doi:Doi 10.1023/A:1008376710086 (1999).
48 Xu, X. P. & Case, D. A. Automated prediction of (15)N, (13)C(alpha), (13)C(beta) and (13)C ' chemical shifts in proteins using a density functional database. *J. Biomol. NMR* **21**, 321-333, doi:Doi 10.1023/A:1013324104681 (2001).
49 Moon, S. & Case, D. A. A new model for chemical shifts of amide hydrogens in proteins. *J Biomol NMR* **38**, 139-150, doi:10.1007/s10858-007-9156-8 (2007).
50 Vila, J. A., Arnautova, Y. A., Martin, O. A. & Scheraga, H. A. Quantum-mechanics-derived 13Calpha chemical shift server

(CheShift) for protein structure validation. *Proc Natl Acad Sci U S A* **106**, 16972-16977, doi:10.1073/pnas.0908833106 (2009).
51 Kohlhoff, K. J., Robustelli, P., Cavalli, A., Salvatella, X. & Vendruscolo, M. Fast and accurate predictions of protein NMR chemical shifts from interatomic distances. *J Am Chem Soc* **131**, 13894-13895, doi:10.1021/ja903772t (2009).
52 Meiler, J. PROSHIFT: Protein chemical shift prediction using artificial neural networks. *J. Biomol. NMR* **26**, 25-37, doi:Doi 10.1023/A:1023060720156 (2003).
53 Han, B., Liu, Y., Ginzinger, S. W. & Wishart, D. S. SHIFTX2: significantly improved protein chemical shift prediction. *J Biomol NMR* **50**, 43-57, doi:10.1007/s10858-011-9478-4 (2011).
54 Shen, Y. & Bax, A. SPARTA+: a modest improvement in empirical NMR chemical shift prediction by means of an artificial neural network. *J Biomol NMR* **48**, 13-22, doi:10.1007/s10858-010-9433-9 (2010).
55 Rupp, M., Ramakrishnan, R. & von Lilienfeld, O. A. Machine Learning for Quantum Mechanical Properties of Atoms in Molecules. *J. Phys. Chem. Lett* **6**, 3309-3313, doi:10.1021/acs.jpclett.5b01456 (2015).
56 Blinov, K. *et al.* Performance validation of neural network based 13C NMR prediction using a publicly available data source. *Journal of chemical information and modeling* **48**, 550-555 (2008).
57 Smurnyy, Y. D., Blinov, K. A., Churanova, T. S., Elyashberg, M. E. & Williams, A. J. Toward more reliable 13C and 1H chemical shift prediction: a systematic comparison of neural-network and least-squares regression based approaches. *J Chem Inf Model* **48**, 128-134, doi:10.1021/ci700256n (2008).
58 Aires-de-Sousa, J., Hemmer, M. C. & Gasteiger, J. Prediction of 1H NMR chemical shifts using neural networks. *Anal Chem* **74**, 80-90 (2002).
59 Kuhn, S., Egert, B., Neumann, S. & Steinbeck, C. Building blocks for automated elucidation of metabolites: machine learning methods for NMR prediction. *BMC Bioinformatics* **9**, 400, doi:10.1186/1471-2105-9-400 (2008).
60 Cuny, J., Xie, Y., Pickard, C. J. & Hassanali, A. A. Ab Initio Quality NMR Parameters in Solid-State Materials Using a High-Dimensional Neural-Network Representation. *J Chem Theory Comput* **12**, 765-773, doi:10.1021/acs.jctc.5b01006 (2016).
61 Groom, C. R., Bruno, I. J., Lightfoot, M. P. & Ward, S. C. The Cambridge Structural Database. *Acta Crystallogr B* **72**, 171-179, doi:10.1107/S2052520616003954 (2016).
62 Rasmussen, C. E. & Williams, C. K. *Gaussian processes for machine learning*. Vol. 1 (MIT press Cambridge, 2006).
63 Bartók, A. P., Kondor, R. & Csányi, G. On representing chemical environments. *Phys. Rev. B* **87**, 1-16, doi:10.1103/PhysRevB.87.184115 (2013).
64 De, S., Bartok, A. P., Csanyi, G. & Ceriotti, M. Comparing molecules and solids across structural and alchemical space. *Phys Chem Chem Phys* **18**, 13754-13769, doi:10.1039/c6cp00415f (2016).
65 Grisafi, A., Wilkins, D. M., Csanyi, G. & Ceriotti, M. Symmetry-Adapted Machine Learning for Tensorial Properties of Atomistic Systems. *Phys Rev Lett* **120**, 036002, doi:10.1103/PhysRevLett.120.036002 (2018).
66 Ceriotti, M., Tribello, G. A. & Parrinello, M. Demonstrating the Transferability and the Descriptive Power of Sketch-Map. *J Chem Theory Comput* **9**, 1521-1532, doi:10.1021/ct3010563 (2013).
67 Campello, R. J. G. B., Moulavi, D., Zimek, A. & Sander, J. Hierarchical Density Estimates for Data Clustering, Visualization, and Outlier Detection. *Acm Transactions on Knowledge Discovery from Data* **10**, 5, doi:Artn 5 DOI 10.1145/2733381 (2015).
68 Giannozzi, P. *et al.* Advanced capabilities for materials modelling with Quantum ESPRESSO. *J Phys Condens Matter* **29**, 465901, doi:10.1088/1361-648X/aa8f79 (2017).
69 Giannozzi, P. *et al.* QUANTUM ESPRESSO: a modular and open-source software project for quantum simulations of materials. *J Phys Condens Matter* **21**, 395502, doi:10.1088/0953-8984/21/39/395502 (2009).
70 Hartman, J. D., Kudla, R. A., Day, G. M., Mueller, L. J. & Beran, G. J. Benchmark fragment-based (1)H, (13)C, (15)N and (17)O chemical shift predictions in molecular crystals. *Phys Chem Chem Phys* **18**, 21686-21709, doi:10.1039/c6cp01831a (2016).
71 Clark, S. J. *et al.* First principles methods using CASTEP. *Zeitschrift Fur Kristallographie* **220**, 567-570, doi:DOI 10.1524/zkri.220.5.567.65075 (2005).
72 Arico-Muendel, C. C. *et al.* Orally active fumagillin analogues: transformations of a reactive warhead in the gastric environment. *ACS Med Chem Lett* **4**, 381-386, doi:10.1021/ml3003633 (2013).
73 Dao, H. T., Li, C., Michaudel, Q., Maxwell, B. D. & Baran, P. S. Hydromethylation of Unactivated Olefins. *J Am Chem Soc* **137**, 8046-8049, doi:10.1021/jacs.5b05144 (2015).
74 Garozzo, D. *et al.* Inclusion networks of a calix[5]arene-based exoditopic receptor and long-chain alkyldiammonium ions. *Org Lett* **5**, 4025-4028, doi:10.1021/ol035310b (2003).
75 Bats, J. W. *CSD Communication* (2010).
76 Huang, G. B. *et al.* Selective recognition of aromatic hydrocarbons by endo-functionalized molecular tubes via C/N-H center dot center dot center dot pi interactions. *Chin. Chem. Lett.* **29**, 91-94, doi:10.1016/j.cclet.2017.07.005 (2018).
77 Plater, M. J., Harrison, W. A., Machado de los Toyos, L. & Hendry, L. The consistent hexameric paddle-wheel crystallisation motif of a family of 2,4-bis(n-alkylamino)nitrobenzenes: alkyl = pentyl, hexyl, heptyl and octyl. *J. Chem. Res.* **41**, 235-238, doi:10.3184/174751917x14902201357356 (2017).
78 F. Musil, S. De & M. Cerrioti. *Glosim2 package*, <https://github.com/cosmo-epfl/glosim2> (2017).

# Supporting Information

## Chemical Shifts in Molecular Solids by Machine Learning

Federico M. Paruzzo,[†] Albert Hofstetter,[†] Félix Musil,[‡] Sandip De,[‡]
Michele Ceriotti,*[,‡] and Lyndon Emsley*[,†]

[†]Institut des Sciences et Ingénierie Chimiques, Ecole Polytechnique Fédérale de Lausanne (EPFL), 1015 Lausanne, Switzerland

[‡]Institut des Sciences et Génie Matériaux, Ecole Polytechnique Fédérale de Lausanne (EPFL), 1015 Lausanne, Switzerland.

Table of Contents

## I. Methods and Theoretical Background

### Crystal Structures.

All the crystal structures of CSD-61k and CSD-500 were obtained from the Cambridge Structural Database (CSD).[1] A total of 88,648 structures was downloaded from the CSD, using two different selection criteria: the maximum number and the type of atoms contained in the unit-cell. We selected only structures with a maximum of 200 atoms, containing either (i) only H and C or (ii) H, C and one heteroatom between N and O or both. From this set we extracted a subset of 61,012 (CSD-61k) structures by removing (i) structures with ambiguous atom positions, and (ii) structures where the distance of at least one pair of atoms was smaller than the sum of their covalent radii minus 0.3 Å. The remaining structures were then used to create both the training (CSD-2k) and the testing set (CSD-500) for the $^{1}$H, $^{13}$C, $^{15}$N and $^{17}$O chemical shift prediction as described in the main text. The test set (CSD-500) was created by randomly picking 500 structures from the CSD-61k excluding the structures already selected for the training set.

### Crystal Structure Prediction.

Here we use a set of polymorphs predicted by CSP for cocaine and the drug 4-[4-(2-adamantylcarbamoyl)-5-tert - butylpyrazol-1-yl]-benzoic acid (also referred as AZD8329). General details on the CSP protocol can be found in ref. 2. In chemical shift based NMR crystallography, the CSP polymorphs are tested against experimental parameters ($^{1}$H chemical shifts) to determine the experimental crystal structure.

In this work we used 30 polymorph structures of cocaine and 14 structures of AZD8329 generated with CSP. The 30 structures of cocaine were obtained from the Electronic Supporting Information (ESI) of ref. 3, and correspond to the most stable polymorphs obtained with CSP. Crystal structures of AZD8329 were obtained from the ESI of ref. 4, and correspond to the 14 most stable predicted polymorphs with the *cis* conformation of the amide bond. From the same sources we obtained chemical shifts for each structure calculated with GIPAW[5,6] using the DFT program CASTEP[7] and the experimental chemical shifts. Labels for the different polymorphs of each structure are based on their energy, with 1 being the most stable polymorph of a given molecule.

### DFT Calculations.

All the DFT calculations were carried out using the DFT program Quantum ESPRESSO.[8,9] For all structures in the CSD-2k and CSD-500 databases we first carried out geometry optimization using plane wave DFT. We used ultrasoft pseudopotentials with GIPAW[5,6] reconstruction, H.pbe-kjpaw_psl.0.1.UPF, C.pbe-n-kjpaw_psl.0.1.UPF, N.pbe-n-kjpaw_psl.0.1.UPF and O.pbe-n-kjpaw_psl.0.1.UPF from the USSP pseudopotendial database [http://materialscloud.org/sssp].[10] The optimizations were done with the generalized-gradient-approximation (GGA) density functional PBE,[11] using a wave-function energy cut-off of 60 Ry, a charge density energy cut-off of 240 Ry and without k-points. The Grimme van der Waals dispersion correction[12] was included in order to account for van der Waals interactions. The geometry optimization was done relaxing all atomic positions while keeping the lattice parameters fixed.

A single point energy (scf) was then computed for the relaxed geometry, using higher wave-function and charge density energy cut-offs which were set to 100 Ry and 400 Ry respectively. For this calculation we also used a Monkhorst-Pack grid of *k*-points[13] corresponding to a maximum spacing of 0.06 Å$^{-1}$ in the reciprocal space. The *k*-points and energy cut-off values were optimized to ensure scf convergence. Finally, we calculated the chemical shielding $\sigma_{\mathrm{DFT}}$ using the GIPAW method, with the same parameters as used in the scf calculation.

### Machine Learning.

We model the isotropic chemical shielding as a function of the local environment $A$ using a Gaussian Process Regression framework, that assumes that chemical shift values predicted by the model can be written as

$$\sigma(A) = f(A) + \varepsilon,$$

where the function $f$ is a Gaussian Process[14] and $\varepsilon$ represents the error of the prediction, which is modeled as independent identically distributed Gaussian variates, with variance $\sigma_n^2$. Following the Gaussian Process Regression framework, the isotropic chemical shielding function becomes:

$$\sigma(A) = \sum_{i=1}^{N} \alpha_i k(A, X_i)^{\zeta},$$

where $\{X_i\}_{i=1}^{N}$ is a training set of *N* reference local environments for which the isotropic chemical shieldings are known, $k$ is a kernel function measuring the covariance between local environments and $\zeta$ is a hyperparameter controlling the sensitivity of the kernel.

The weights can be computed by inverting the kernel matrix $K_{ij} = k(X_i, X_j)$ computed between the reference configurations, including a regularization that depends on an estimate of the intrinsic uncertainty in the fit, due to errors in the training set, the limitations of the model or the reduced number of training configurations

$$\alpha_i = \sum_j [K^{\zeta} + \sigma_n^2 1]^{-1}{}_{ij}\, \sigma(X_j)$$

To assess the correlation between local atomic environment A and B, we use the SOAP kernel[15] defined by the rotationally invariant overlap between smooth representations of their atomic density:

$$k(A,B) = \int_{SO(3)} \left| \int_{R^3} \rho_A(\vec{r}) \rho_B(\hat{R}\vec{r}) d\vec{r} \right|^2 d\hat{R},$$

where the density is built as a superimposition of Gaussians having width ς, centered on the atoms within a cutoff distance of the central atom in the environment

$$\rho_A(\vec{r}) = \sum_{i \in A} exp\,[\|\vec{r} - (\vec{r}_i - \vec{r}_A)\|^2/2\varsigma^2] f_c(|\vec{r}_i - \vec{r}_A|)$$

The details of the construction, and the extension to the case with many atomic species, are given in refs. 16 and 17.

### Farthest Point Sampling Algorithm.

Given that a GPR model is essentially an interpolation procedure between the reference configurations, it is crucial that training points are chosen to cover as uniformly as possible the space of structures for which one wants to perform predictions. To achieve this uniform sampling we use a farthest point selection algorithm[18,19] to sort the CSD-61k in descending order of "diversity". Essentially, we select a first structure at random, and then pick the others in the sequence such that

$$k = \underset{k \in CSD-61k}{argmax} \; \underset{j \in selection}{min} |X_k - X_j|$$

where the distance is the kernel-induced distance associated with an average SOAP kernel for the entire structure.[16] The CSD-2k set corresponds to the first 2,000 configurations identified with this procedure.

### Detection of Unusual Environments.

The quality of the training set is essential to ensure the optimal performance of a machine learning algorithm. However, the individual curation of the 2,000 molecular crystals of the CSD-2k dataset would be very time consuming and cumbersome. Note, that the 2,000 molecular crystals correspond to around 35,000 symmetrically non-equivalent atomic environments for $^1$H alone and the following detection procedure is applied directly to the individual atomic environments instead of the whole molecular crystals.

We automate this detection procedure by assessing the 'instability' of the prediction of the shielding of a given local environment using the difference between the predictions of several GPR models and the reference DFT-shielding.

We define this indicator as:

$$\varepsilon(X) = \frac{1}{M} \sum_{i=1}^{M} (y_i(X) - y(X)),$$

where each of the M models is made using a 2-fold split of the shuffled training set that does not include the structure X. In total we generate M=40 models, where each is generated using a different random shuffling of the data. Environments with a large value of $|\varepsilon(X)|$ are not well-described by the rest of the training set within the SOAP-GPR framework. Note, that the error would cancel out in the case of random noise within the prediction, while a large value of $|\varepsilon(X)|$ corresponds to a systematic error in the predicted chemical shielding, that could be associated to the limitations listed below. We define local environments to be unusual when $|\varepsilon(X)|$ is larger than three times the standard deviation of $|\varepsilon(X)|$ over the whole training set, and we then do not use them for training.

We perform this elimination procedure on the CSD-2k dataset using a single kernel for each element ($r_c$ = 4.5 Å for $^1$H, 4 Å for $^{13}$C, 4 Å for $^{15}$N and 3 Å for $^{17}$O). The hyperparameters of the single kernels used in the elimination procedure were determined using a grid search and 3-fold cross validation on the uncleaned CSD-2k training set. The $^1$H environments excluded with this approach are shown in Figure S1, while further details for $^1$H and the other nuclei are listed in section VII.

It is interesting to see that in several cases we can trace the unusual behavior of the environment to subtle errors in the DFT calculations, or to physical phenomena that are ill described within our DFT model (metallic systems, zwitterions,…).

Most of the environments detected as "unusual" are part of zwitterionic structures or charged structures (such as VIWYEH, ZACSOO or EKUJIF). Others are metallic structures ($E_{LUMO} - E_{HOMO} = 0$), such as HAZQUV, QUICNA02, DMEBQU01 or AYUKIP, or have a partially empty unit cell (QAHVUQ). An intrinsic limit of this procedure is the fact that it might detect structures with uncommon functional groups as "anomalies" (e.g. TIMCHX, which is an aziridine – a three membered heterocycle with one amine group, or FIGMAJ which has a cubane group), due to the fact that these structures are not well represented by the used training set. However, with increasing training size, we expect these structures to be better represented and they will not be detected as anomalies anymore.

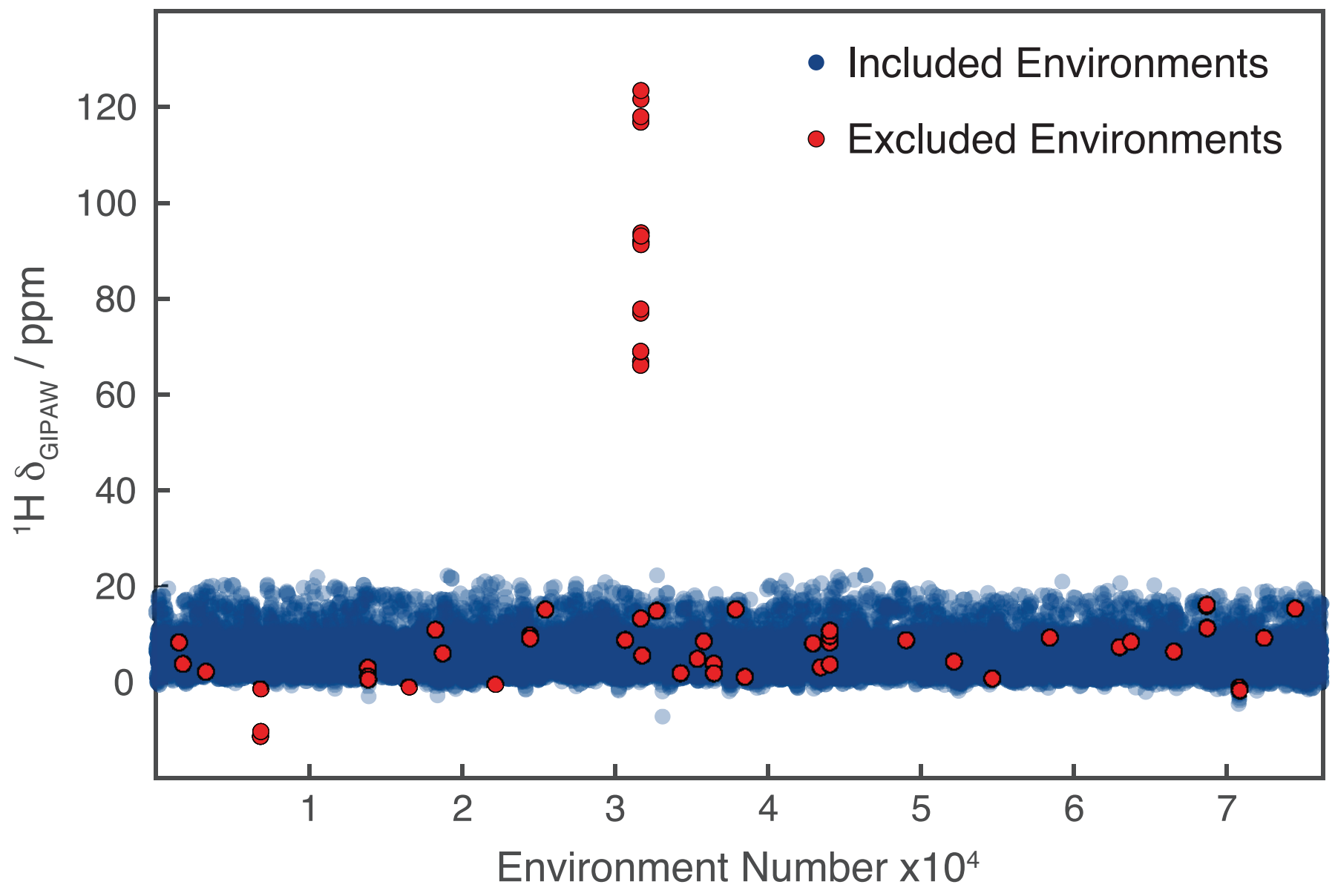

**Figure S1.** $^1$H chemical shifts of the 76,214 environments in the CSD-2k set. The environments excluded using the unusual structures detection procedure described in the text are shown in red.

**NMR Crystallography.**

To validate the accuracy of the chemical shifts calculated with ShiftML, we replicated the last step of the protocol for the *ab initio* crystal structure determination of powdered solids[3,4,20] using predicted shifts. This step consists in the comparison between experimental and predicted $^1$H chemical shifts for the candidate crystal structures selected from a crystal structure prediction method. We perform this analysis for cocaine and form 4 of AZD8329.[3,4] The value $\sigma_{ref}$ for the conversion between chemical shieldings to chemical shifts is calculated for each structure with a linear regression between calculated and experimental shifts, imposing a slope equal to 1. This procedure is done independently for the $^1$H chemical shieldings calculated with DFT and ShiftML. The geometry of the structures predicted with CSP, as well as their relative chemical shift values calculated with GIPAW and the experimental chemical shifts of the observed polymorphs were obtained from refs. 3 and 4. Remarkably, the high accuracy shown in Figure 4 was obtained using crystal structures with only $^1$H positions relaxed and DFT chemical shift calculations carried out using a different program (CASTEP) to the one we used to build our training set (Quantum Espresso). Figure S2 shows the results obtained for cocaine and AZD8329 after all-atom optimization and calculation of GIPAW chemical shifts with Quantum Espresso. Here we show fewer structures compared to Figure 4, due to the fact that we limit ourselves to calculate DFT chemical shifts of structures with less than 250 atoms. This selection removes structures 15 for cocaine and structures 2, 11 and 14 for AZD8329. The accuracy is consistent with that reported in Figure 4, although the all-atom optimization leads to some significant structural differences compared to the only $^1$H relaxed structures, especially for AZD8329. We find a chemical shift prediction error (RMSE) for $^1$H for cocaine of 0.40 ppm and for AZD8329 of 0.51ppm, which is very comparable to the expected GIPAW DFT accuracy. For the heteronuclei we obtain, for cocaine and AZD 8329 respectively, 3.5 and 3.4 ppm for $^{13}$C, 9.3 ppm and 11.0 ppm for $^{15}$N and 12.2 ppm and 11.5 ppm for $^{17}$O.

Experimental chemical shifts were referenced to the $^1$H resonance observed for adamantane at 1.87 ppm with respect to TMS. We used assigned chemical shifts values and we account for rotational dynamics of the methyl groups by averaging the chemical shift values of the three $^1$H positions to a single value for each methyl group. For AZD8329 the chemical shifts of the $CH_2$ groups were also averaged. The RMSE calculation was carried out in MATLAB using a home-written script. The chemical structures of cocaine and AZD8329, together with the assignment of the experimental chemical shifts are shown in Figure S3 and Table S1. $^1$H chemical shieldings calculated with GIPAW and ShiftML are given as separate .cs files, named according to the corresponding structure and method used for DFT calculations. Each file contains the $^1$H chemical shift calculated with GIPAW (first column) and predicted with ShiftML (second column) ordered according to Table S1.

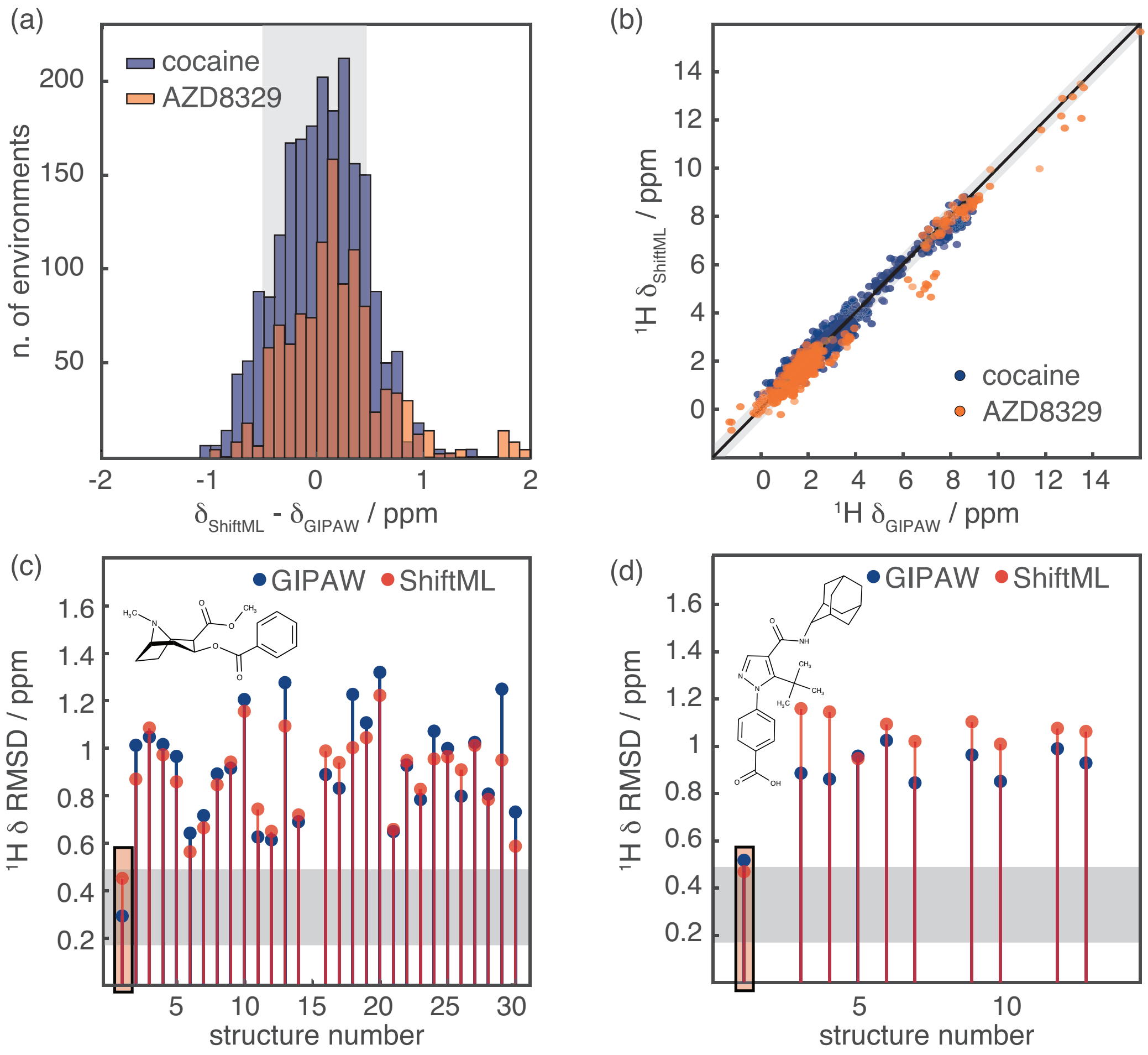

**Figure S2**. (a) Histogram showing the distribution of the differences between chemical shifts calculated with GIPAW and ShiftML. The blue bars were calculated for the polymorphs of cocaine, and the orange ones for the polymorphs of AZD8329. (b) Scatterplot showing the correlation between GIPAW and ShiftML chemical shifts for cocaine (blue) and AZD8329 (orange). The black line indicates a perfect correlation. (c-d) Comparison between calculated and experimental $^{1}H$ chemical shifts for the most stable structures obtained with CSP for cocaine (c) and form 4 of AZD8329 (d). Chemical shifts were calculated using GIPAW (blue) and ShiftML (red). The highlighted bars correspond to the candidates that would be selected as correct crystal structures using the chemical shift based solid-state NMR crystallography protocol. In (a-d) the grey zones represent the confidence intervals of the $^{1}H$ chemical shift RMSD, as described in the text.[20]

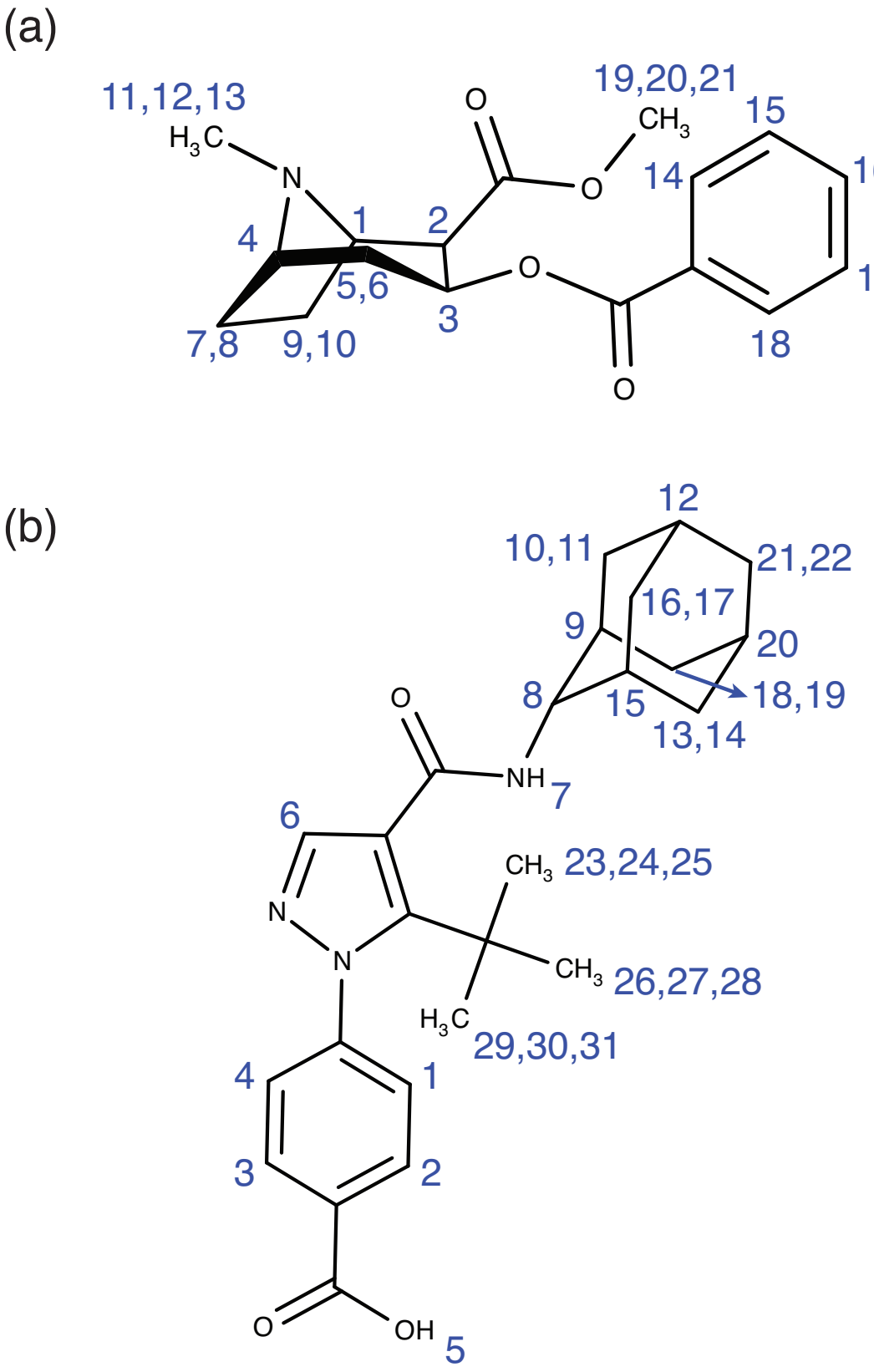

**Figure S3**. Chemical structure of cocaine (a) and AZD8329 (b) and the labelling scheme used here.

| **Cocaine** | | **AZD8329** | |
|---|---|---|---|
| Atom Label | $^1$H δ (ppm) | Atom Label | $^1$H δ (ppm) |
| 1 | 3.76 | 1 | 6.92 |
| 2 | 3.78 | 2 | 8.69 |
| 3 | 5.63 | 3 | 9.01 |
| 4 | 3.32 | 4 | 8.47 |
| 5 | 3.49 | 5 | 15.37 |
| 6 | 3.06 | 6 | 7.73 |
| 7 | 2.91 | 7 | 9.64 |
| 8 | 3.38 | 8 | 2.90 |
| 9 | 2.56 | 9 | 1.78 |
| 10 | 2.12 | 10,11 | 1.88 |
| 11,12,13 | 1.04 | 12 | 1.8 |
| 14 | 8.01 | 13 | 1.6 |
| 15 | 8.01 | 14 | 0.44 |
| 15 | 8.01 | 15 | 1.54 |
| 17 | 8.01 | 16,17 | 1.88 |
| 18 | 8.01 | 18,19 | 0.8 |
| 19,20,21 | 3.78 | 20 | 1 |
| | | 21,22 | 1.74 |
| | | 23,24,25, 26,27,28, 29,30,31 | 0.73 |

**Table S1**. Experimental chemical shifts of cocaine and the form 4 of AZD8329. Labelling scheme is given in Figure S3. When more than one atom corresponds to a single chemical shift value, their values were average

## II. DFT Calculation Times

Figure S4 shows the CPU time needed for part of the GIPAW DFT calculations done for this work. The calculations shown in Figure S4a were done on polymorph 1 of the cocaine dataset, which contains 86 atoms per unit-cell, while the one in Figure S4b were done on 500 structures of the CSD-2k set. In Figure S4a the calculation time is plotted as a function of the number of Monkhorst-pack k-points per axis for three different energy-cut-off ($E_{\mathrm{cutoff}}$) values: 40 Ry (blue), 70 Ry (red), 100 Ry (yellow). When increased, these two parameters improve the accuracy of the calculation, but at the same time they drastically increase the computational time needed to carry out the calculation. Figure S4b shows the CPU time for the GIPAW chemical shift calculations (blue dots) and for the DFT structure optimizations (green squares) as a function of the number of valence electrons ($N_e$) per unit-cell. For the GIPAW chemical shift calculations the energy-cut-off was 100 Ry, using a Monkhorst-pack grid with a k-point spacing of 0.06 Å$^{-1}$. For the DFT structure optimizations the energy-cut-off was 60 Ry and no k-points were used. The red line shows the best fit between the number of valence electrons and the required CPU time for the GIPAW chemical shift calculations as $t_{CPU} = aN_e^2 + bN_e^3$, where the $N_e^3$ scaling accounts for the general DFT scaling and the $N_e^2$ describes the scaling of the matrix inversion, which dominates for small system sizes. The best fit parameters are given as 8.83e-04 (a) and 1.02e-06 (b).

Currently the machine learning model has only been rigorously tested and applied for structures optimized with DFT. Also slight structural changes away from the equilibrium geometry of a molecular crystal have been shown to result in significant changes in the chemical shifts.[21] For this reason, the predictive accuracy of ShiftML for non-equilibrium structures has not yet been quantified. This will be the subject of further work. However, Figure S4b clearly shows that the computational cost for the structure optimization is negligible compared to the computational cost of the GIPAW chemical shift calculations. For structures with $N_e \approx 100$ the GIPAW shift calculations require around 10x more CPU time as the DFT structure optimization, and for $N_e \approx 1{,}000$, 80x more CPU time is required.

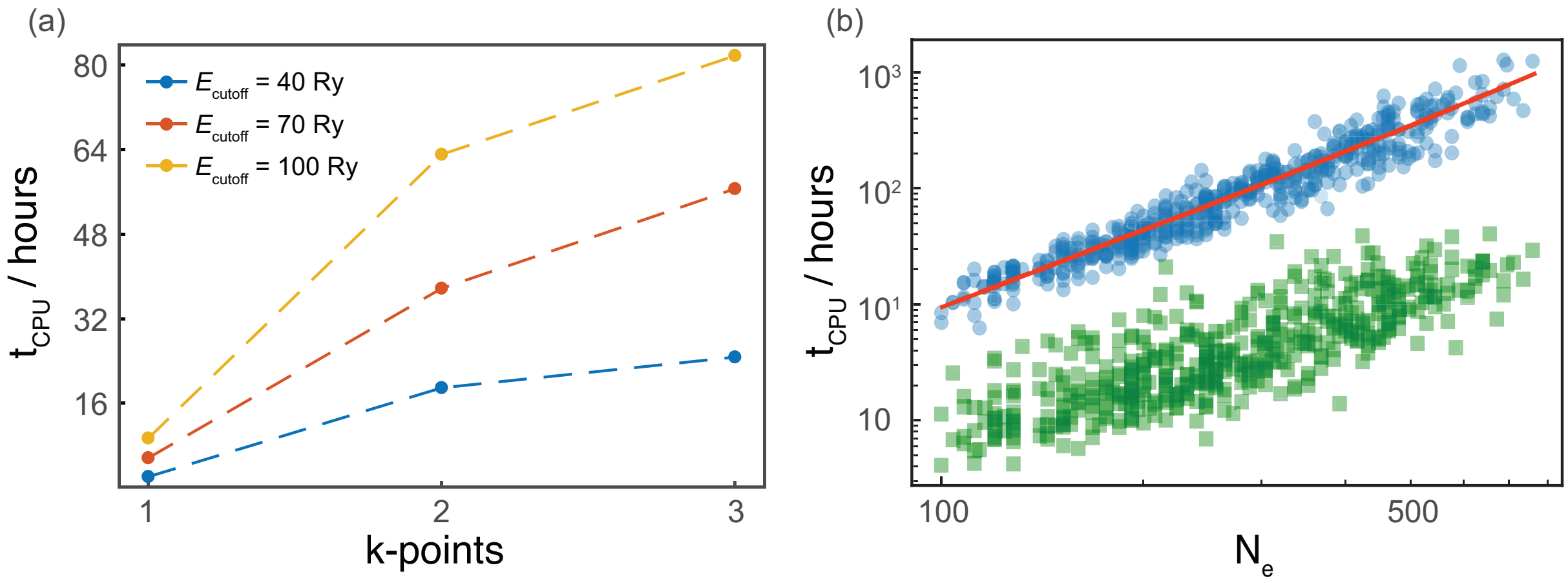

**Figure S4.** CPU time for NMR chemical shift calculations using the GIPAW method. (a) The CPU time is shown as function of the DFT accuracy, determined by the plane-wave cutoff energy $E_{cutoff}$ and the number of k-points in each dimension for polymorph 1 of cocaine. The charge density energy cut-offs were set to $E_\rho = 4E_{cutoff}$. (b) The CPU time is shown as function of increasing system size in CSD-2k. The green squares and blue dots show individual geometry optimization and GIPAW chemical shift DFT calculations, respectively. The red line shows the best fit between the number of valence electrons and the required CPU time as $t_{CPU} = aN_e^2 + bN_e^3$.

## III. ShiftML Prediction Times

The ShiftML prediction times scales linearly with the number of atoms per unit cell. However, for all the structures investigated here (from 20 to 1,500 atoms per unit-cell) the required prediction time is dominated by a constant prefactor associated with the used training set.

Prior to the prediction step, the SOAP reference vector between the test and the training structures is created. This step should be linear in the size of the test-structure, but is currently dominated by the size of the training set. As a result this takes around one CPU minute for any of the investigated structures here.

The actual subsequent chemical shift prediction, which is linear in the number of atoms within the test-structure, requires at most 10-20 CPU seconds for the large investigated structures.

Note that prior to the chemical shift predictions, the single kernels for all the atomic species must be loaded into virtual memory and the multiscale kernel created. On one CPU this currently takes around 45 minutes. Note, that this has to be done only once, independently of the number and size of the test-structures that are subsequently calculated.

## IV. Prediction Parameters and Evaluation Curves of $^{13}C$, $^{15}N$ and $^{17}O$

Table S2 and S3 shows the parameters used for the single and the multi-scale kernel predictions respectively. Using these parameters, we obtained the curves shown in Figure 2 in the main text and the ones shown in Figures S5, S6, S7 and S8. Figures S5 and S6 show the RMSE and MAE learning curves for $^{1}H$, $^{13}C$, $^{15}N$ and $^{17}O$ for the different local environment cut-off radii, and for the multi-kernel. The training was done on up to 1500 randomly selected frames, while testing on 400 structures selected randomly from the CSD-2k set excluding the structures already selected for the training set. For each point, the random sampling was repeated N times (where N is equal to 300, 255, 215, 170, 130, 85, 45, 5 respectively for training set sizes of 40, 100, 200, 400, 600, 1000, 1400 and 1500 structures)

Figure S7 and S8 show the results of the predictions of the chemical shifts of the CSD-500 set as a function of the cut-off value and the size of the training set. The parameters for the multi-scale kernel prediction were optimized using 3-fold cross validation on the CSD-2k set. The results for $^{1}H$ are listed in table S4.

| Atom | Cut-off ($r_c$) | Gaussian width ($\varsigma$) | $l_{max}$ | $n_{max}$ | $\sigma_n$ | $\zeta$ |
|---|---|---|---|---|---|---|
| $^{1}H$ | 2 | 0.3 | 9 | 9 | 0.1 | 2 |
| | 3 | 0.3 | 9 | 9 | 0.1 | 2 |
| | 4 | 0.4 | 9 | 9 | 0.1 | 2 |
| | 5 | 0.4 | 9 | 9 | 0.1 | 2 |
| | 6 | 0.5 | 9 | 12 | 0.1 | 2 |
| | 7 | 0.5 | 9 | 12 | 0.1 | 2 |
| $^{13}C$ | 2 | 0.3 | 9 | 9 | 0.01 | 2 |
| | 3 | 0.3 | 9 | 9 | 3.0 | 2 |
| | 4 | 0.4 | 9 | 9 | 5.0 | 2 |
| | 5 | 0.4 | 9 | 9 | 3.0 | 2 |
| | 6 | 0.5 | 9 | 12 | 1.0 | 2 |
| | 7 | 0.5 | 9 | 12 | 1.0 | 1 |
| $^{15}N$ | 2 | 0.3 | 9 | 9 | 0.5 | 2 |
| | 3 | 0.3 | 9 | 9 | 1.0 | 2 |
| | 4 | 0.4 | 9 | 9 | 0.1 | 2 |
| | 5 | 0.4 | 9 | 9 | 0.1 | 2 |
| | 6 | 0.5 | 9 | 12 | 0.1 | 2 |
| | 7 | 0.5 | 9 | 12 | 0.05 | 2 |
| $^{17}O$ | 2 | 0.3 | 9 | 9 | 0.5 | 2 |
| | 3 | 0.3 | 9 | 9 | 5.0 | 2 |
| | 4 | 0.4 | 9 | 9 | 5.0 | 2 |
| | 5 | 0.4 | 9 | 9 | 5.0 | 2 |
| | 6 | 0.5 | 9 | 12 | 1.0 | 2 |
| | 7 | 0.5 | 9 | 12 | 7.0 | 2 |

**Table S2.** Kernel and GPR parameters. The GPR parameters ($\sigma_n$ and $\zeta$) are the ones used in single kernel predictions.

| Atom | Multi-Scale Kernel Weights | | | | | | $\sigma_n$ | $\zeta$ |
|---|---|---|---|---|---|---|---|---|
| | $r_c$ = 2 Å | $r_c$ = 3 Å | $r_c$ = 4 Å | $r_c$ =5 Å | $r_c$ = 6 Å | $r_c$ = 7 Å | | |
| $^{1}H$ | 256 | 128 | 32 | 8 | 8 | 1 | 0.1 | 2 |
| $^{13}C$ | 256 | 512 | 64 | 8 | 8 | 1 | 2.0 | 2 |
| $^{15}N$ | 256 | 128 | 32 | 8 | 8 | 1 | 0.1 | 2 |
| $^{17}O$ | 256 | 128 | 32 | 8 | 8 | 1 | 5.0 | 2 |

**Table S3.** Kernel weights and GPR parameters used for multi-scale kernel prediction.

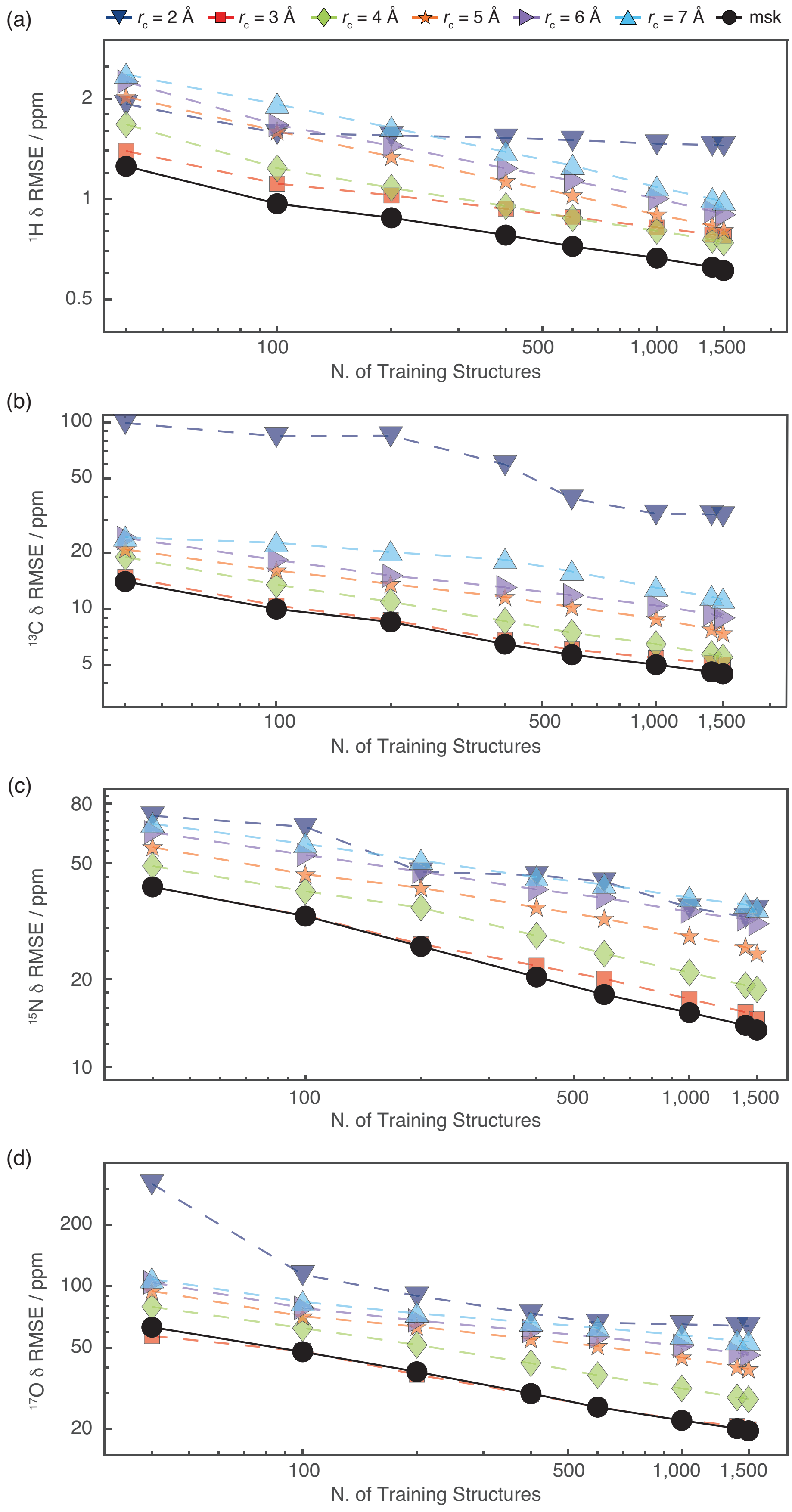

**Figure S5.** RMSE learning curves for $^{13}$C (a), $^{15}$N (b) and $^{17}$O (c). The multi-kernel learning-curve is labelled as msk.

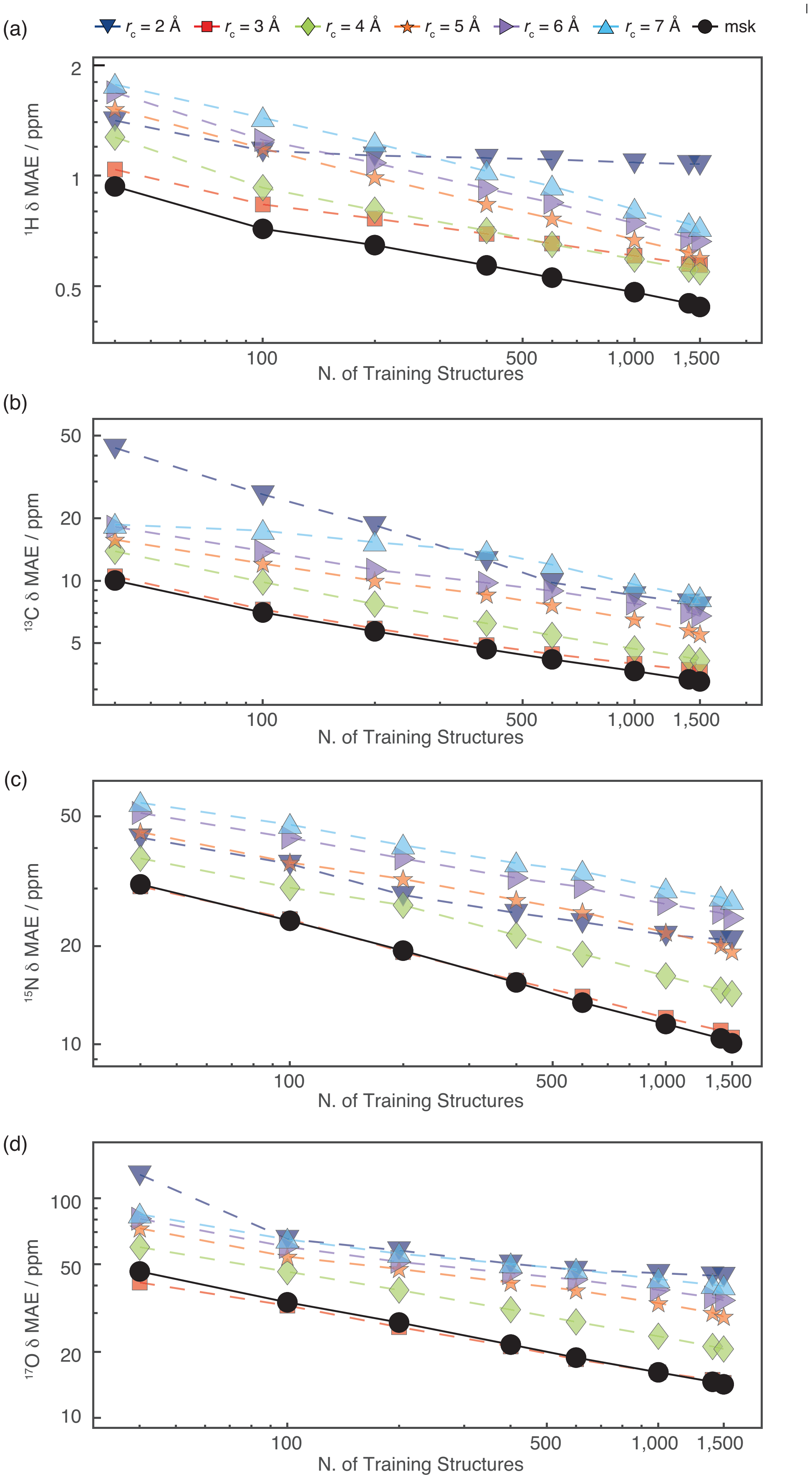

**Figure S6.** MAE learning curves of $^{13}C$ (a), $^{15}N$ (b) and $^{17}O$ (c). The multi-kernel learning-curve is labelled as msk.

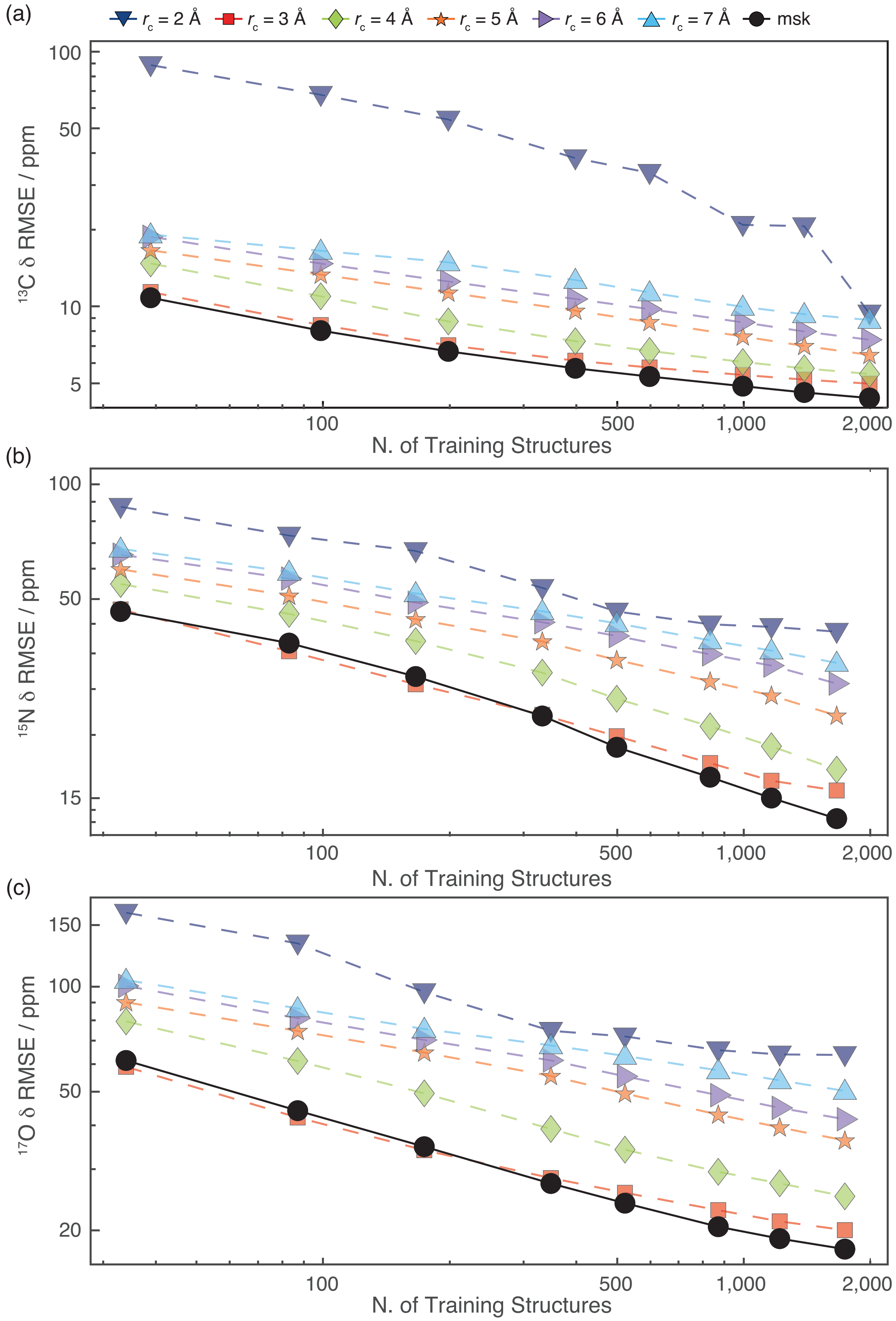

**Figure S7.** RMSE evaluation curves of $^{13}C$ (a), $^{15}N$ (b) and $^{17}O$ (c) for different training set sizes, evaluated on the CSD-500 test set. The RMSE evaluation curves were acquired as described in the paper. The multi-kernel learning-curve is labelled as msk.

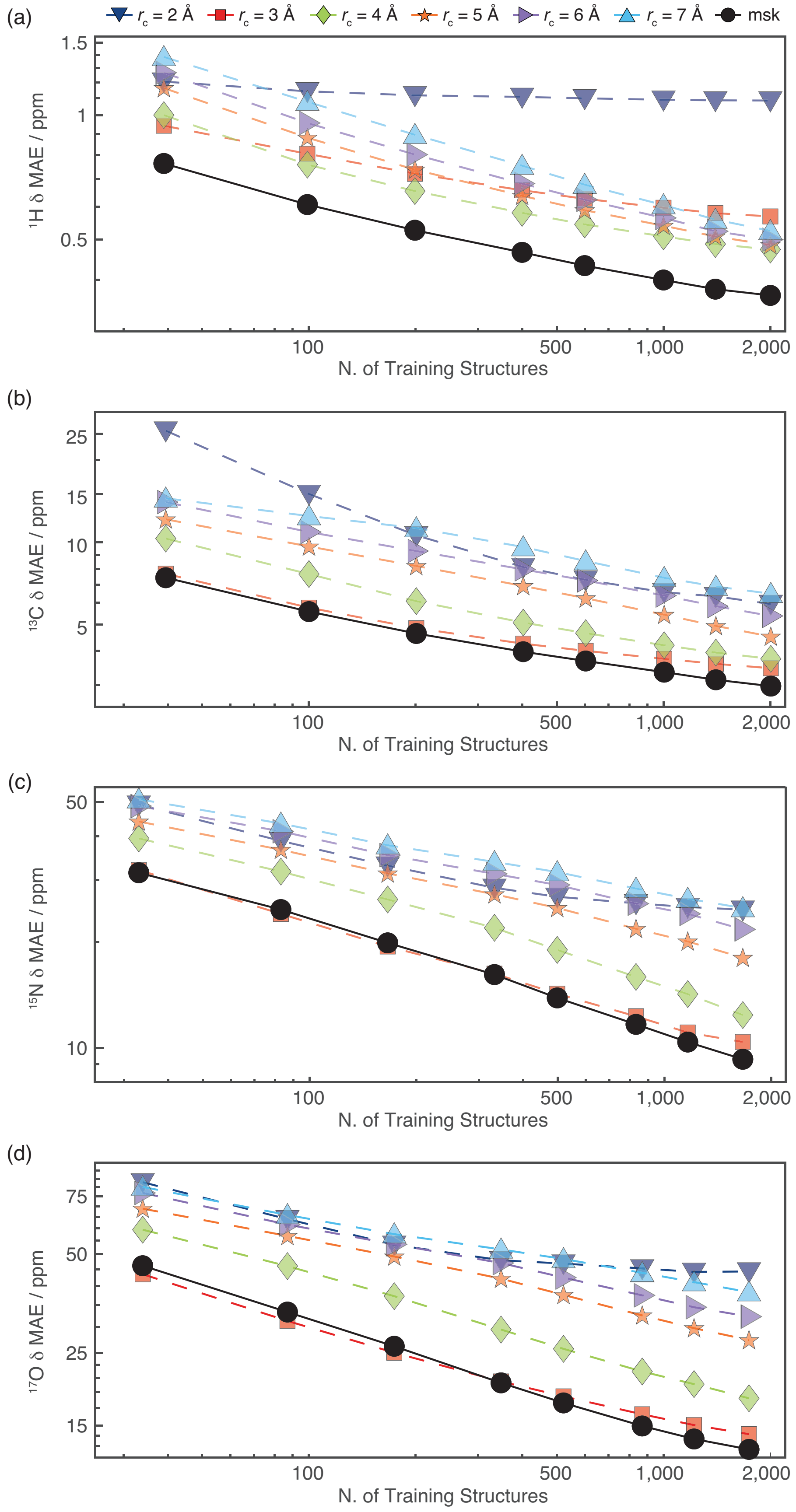

**Figure S8.** MAE evaluation curves of $^{13}C$ (a), $^{15}N$ (b) and $^{17}O$ (c) for different training set sizes, evaluated on the CSD-500 test set. The MAE evaluation curves were acquired as described in the paper. The multi-kernel learning-curve is labelled as msk.

| Multi-Scale Kernel Weights | | | | | | $\zeta$ | $\sigma_n$ | MAE (ppm) | RMSE (ppm) | $R^2$ | SUP (ppm) |
|---|---|---|---|---|---|---|---|---|---|---|---|
| $r_c$ =2Å | $r_c$ =3Å | $r_c$ =4Å | $r_c$ =5Å | $r_c$ =6Å | $r_c$ =7Å | | | | | | |
| **256** | **128** | **32** | **8** | **8** | **1** | **2** | **0.1** | **0.453** | **0.632** | **0.969** | **6.543** |
| 256 | 256 | 32 | 8 | 8 | 1 | 2 | 0.1 | 0.453 | 0.632 | 0.969 | 6.654 |
| 256 | 64 | 32 | 8 | 8 | 1 | 2 | 0.1 | 0.454 | 0.634 | 0.969 | 6.443 |
| 256 | 128 | 32 | 16 | 8 | 1 | 2 | 0.1 | 0.455 | 0.634 | 0.969 | 6.477 |
| 256 | 256 | 32 | 8 | 16 | 1 | 2 | 0.1 | 0.455 | 0.634 | 0.969 | 6.616 |
| 256 | 256 | 32 | 16 | 8 | 1 | 2 | 0.1 | 0.455 | 0.634 | 0.969 | 6.598 |
| 256 | 128 | 32 | 8 | 16 | 1 | 2 | 0.1 | 0.455 | 0.635 | 0.969 | 6.505 |
| 256 | 256 | 64 | 8 | 8 | 1 | 2 | 0.1 | 0.456 | 0.635 | 0.969 | 6.659 |
| 256 | 128 | 64 | 8 | 8 | 1 | 2 | 0.1 | 0.456 | 0.636 | 0.969 | 6.574 |
| 256 | 256 | 32 | 16 | 16 | 1 | 2 | 0.1 | 0.457 | 0.636 | 0.969 | 6.559 |
| 256 | 256 | 64 | 8 | 16 | 1 | 2 | 0.1 | 0.457 | 0.636 | 0.969 | 6.628 |
| 256 | 256 | 64 | 16 | 8 | 1 | 2 | 0.1 | 0.457 | 0.637 | 0.969 | 6.613 |
| 256 | 128 | 32 | 16 | 16 | 1 | 2 | 0.1 | 0.457 | 0.637 | 0.969 | 6.438 |
| 256 | 64 | 32 | 16 | 8 | 1 | 2 | 0.1 | 0.457 | 0.637 | 0.969 | 6.374 |
| 256 | 64 | 32 | 8 | 16 | 1 | 2 | 0.1 | 0.457 | 0.638 | 0.969 | 6.411 |
| 256 | 128 | 64 | 8 | 16 | 1 | 2 | 0.1 | 0.458 | 0.638 | 0.969 | 6.542 |
| 256 | 256 | 64 | 16 | 16 | 1 | 2 | 0.1 | 0.459 | 0.638 | 0.969 | 6.579 |
| 256 | 128 | 64 | 16 | 8 | 1 | 2 | 0.1 | 0.458 | 0.638 | 0.969 | 6.520 |
| 256 | 64 | 64 | 8 | 8 | 1 | 2 | 0.1 | 0.459 | 0.639 | 0.968 | 6.507 |
| 256 | 128 | 64 | 16 | 16 | 1 | 2 | 0.1 | 0.460 | 0.640 | 0.968 | 6.485 |
| 256 | 64 | 32 | 16 | 16 | 1 | 2 | 0.1 | 0.460 | 0.640 | 0.968 | 6.340 |
| 256 | 64 | 64 | 8 | 16 | 1 | 2 | 0.1 | 0.461 | 0.642 | 0.968 | 6.479 |
| 256 | 64 | 64 | 16 | 8 | 1 | 2 | 0.1 | 0.461 | 0.642 | 0.968 | 6.449 |
| 256 | 64 | 64 | 16 | 16 | 1 | 2 | 0.1 | 0.463 | 0.644 | 0.968 | 6.417 |
| 256 | 128 | 32 | 8 | 8 | 1 | 4 | 0.1 | 0.467 | 0.649 | 0.967 | 6.337 |
| 256 | 64 | 32 | 8 | 8 | 1 | 4 | 0.1 | 0.467 | 0.650 | 0.967 | 6.307 |
| 256 | 128 | 32 | 8 | 16 | 1 | 4 | 0.1 | 0.469 | 0.652 | 0.967 | 6.275 |
| 256 | 128 | 32 | 16 | 8 | 1 | 4 | 0.1 | 0.469 | 0.652 | 0.967 | 6.245 |
| 256 | 256 | 32 | 8 | 8 | 1 | 4 | 0.1 | 0.470 | 0.652 | 0.967 | 6.355 |
| 256 | 256 | 32 | 8 | 16 | 1 | 4 | 0.1 | 0.471 | 0.653 | 0.967 | 6.298 |
| 256 | 128 | 64 | 8 | 8 | 1 | 4 | 0.1 | 0.471 | 0.653 | 0.967 | 6.215 |
| 256 | 256 | 32 | 16 | 8 | 1 | 4 | 0.1 | 0.471 | 0.654 | 0.967 | 6.282 |
| 256 | 64 | 32 | 16 | 8 | 1 | 4 | 0.1 | 0.470 | 0.654 | 0.967 | 6.209 |
| 256 | 128 | 32 | 16 | 16 | 1 | 4 | 0.1 | 0.470 | 0.654 | 0.967 | 6.186 |
| 256 | 256 | 64 | 8 | 8 | 1 | 4 | 0.1 | 0.472 | 0.654 | 0.967 | 6.217 |
| 256 | 64 | 32 | 8 | 16 | 1 | 4 | 0.1 | 0.470 | 0.654 | 0.967 | 6.254 |
| 256 | 256 | 32 | 16 | 16 | 1 | 4 | 0.1 | 0.472 | 0.654 | 0.967 | 6.226 |
| 256 | 256 | 64 | 8 | 16 | 1 | 4 | 0.1 | 0.472 | 0.655 | 0.967 | 6.184 |
| 256 | 128 | 64 | 8 | 16 | 1 | 4 | 0.1 | 0.472 | 0.655 | 0.967 | 6.178 |
| 256 | 256 | 64 | 16 | 8 | 1 | 4 | 0.1 | 0.473 | 0.655 | 0.967 | 6.166 |
| 256 | 128 | 64 | 16 | 8 | 1 | 4 | 0.1 | 0.472 | 0.655 | 0.967 | 6.149 |
| 256 | 256 | 64 | 16 | 16 | 1 | 4 | 0.1 | 0.473 | 0.656 | 0.967 | 6.130 |
| 256 | 64 | 64 | 8 | 8 | 1 | 4 | 0.1 | 0.472 | 0.656 | 0.967 | 6.217 |
| 256 | 128 | 64 | 16 | 16 | 1 | 4 | 0.1 | 0.473 | 0.657 | 0.967 | 6.111 |
| 256 | 64 | 32 | 16 | 16 | 1 | 4 | 0.1 | 0.472 | 0.657 | 0.967 | 6.157 |
| 256 | 64 | 64 | 8 | 16 | 1 | 4 | 0.1 | 0.474 | 0.659 | 0.966 | 6.183 |
| 256 | 64 | 64 | 16 | 8 | 1 | 4 | 0.1 | 0.474 | 0.659 | 0.966 | 6.145 |
| 256 | 64 | 64 | 16 | 16 | 1 | 4 | 0.1 | 0.476 | 0.661 | 0.966 | 6.132 |
| 256 | 256 | 32 | 16 | 8 | 1 | 4 | 1 | 0.474 | 0.662 | 0.965 | 6.071 |
| 256 | 256 | 32 | 16 | 16 | 1 | 4 | 1 | 0.474 | 0.662 | 0.965 | 6.080 |
| 256 | 256 | 64 | 16 | 8 | 1 | 4 | 1 | 0.474 | 0.662 | 0.965 | 6.103 |
| 256 | 256 | 32 | 8 | 8 | 1 | 1 | 0.1 | 0.473 | 0.662 | 0.966 | 6.622 |
| 256 | 256 | 32 | 8 | 16 | 1 | 4 | 1 | 0.474 | 0.662 | 0.965 | 6.086 |
| 256 | 256 | 32 | 8 | 8 | 1 | 4 | 1 | 0.475 | 0.662 | 0.965 | 6.073 |
| 256 | 256 | 64 | 16 | 16 | 1 | 4 | 1 | 0.474 | 0.662 | 0.965 | 6.115 |
| 256 | 256 | 64 | 8 | 16 | 1 | 4 | 1 | 0.474 | 0.662 | 0.965 | 6.139 |
| 256 | 256 | 64 | 8 | 8 | 1 | 4 | 1 | 0.474 | 0.662 | 0.965 | 6.126 |
| 256 | 256 | 64 | 8 | 8 | 1 | 1 | 0.1 | 0.473 | 0.663 | 0.966 | 6.720 |
| 256 | 256 | 32 | 16 | 8 | 1 | 1 | 0.1 | 0.475 | 0.664 | 0.966 | 6.573 |
| 256 | 256 | 64 | 16 | 8 | 1 | 1 | 0.1 | 0.474 | 0.665 | 0.966 | 6.662 |
| 256 | 128 | 32 | 8 | 8 | 1 | 1 | 0.1 | 0.476 | 0.665 | 0.966 | 6.627 |
| 256 | 256 | 32 | 8 | 16 | 1 | 1 | 0.1 | 0.476 | 0.665 | 0.966 | 6.628 |
| 256 | 128 | 64 | 16 | 16 | 1 | 4 | 1 | 0.476 | 0.666 | 0.965 | 6.233 |
| 256 | 256 | 64 | 8 | 16 | 1 | 1 | 0.1 | 0.475 | 0.666 | 0.966 | 6.724 |
| 256 | 128 | 64 | 8 | 8 | 1 | 1 | 0.1 | 0.476 | 0.666 | 0.966 | 6.741 |
| 256 | 256 | 32 | 16 | 16 | 1 | 1 | 0.1 | 0.477 | 0.667 | 0.966 | 6.579 |
| 256 | 128 | 32 | 16 | 8 | 1 | 1 | 0.1 | 0.478 | 0.667 | 0.966 | 6.565 |
| 256 | 256 | 64 | 16 | 16 | 1 | 1 | 0.1 | 0.477 | 0.668 | 0.966 | 6.666 |
| 256 | 128 | 64 | 16 | 8 | 1 | 1 | 0.1 | 0.478 | 0.668 | 0.965 | 6.672 |
| 256 | 64 | 32 | 8 | 8 | 1 | 1 | 0.1 | 0.479 | 0.669 | 0.965 | 6.597 |
| 256 | 128 | 32 | 8 | 16 | 1 | 1 | 0.1 | 0.479 | 0.669 | 0.965 | 6.630 |
| 256 | 128 | 64 | 8 | 16 | 1 | 1 | 0.1 | 0.479 | 0.669 | 0.965 | 6.741 |
| 256 | 64 | 64 | 8 | 8 | 1 | 1 | 0.1 | 0.479 | 0.670 | 0.965 | 6.740 |
| 256 | 128 | 32 | 16 | 16 | 1 | 1 | 0.1 | 0.481 | 0.671 | 0.965 | 6.569 |
| 256 | 64 | 32 | 16 | 8 | 1 | 1 | 0.1 | 0.481 | 0.671 | 0.965 | 6.524 |
| 256 | 128 | 64 | 16 | 16 | 1 | 1 | 0.1 | 0.481 | 0.672 | 0.965 | 6.671 |
| 256 | 256 | 64 | 16 | 16 | 1 | 4 | 0.01 | 0.488 | 0.672 | 0.965 | 6.022 |
| 256 | 64 | 64 | 16 | 8 | 1 | 1 | 0.1 | 0.481 | 0.672 | 0.965 | 6.659 |
| 256 | 256 | 64 | 8 | 16 | 1 | 4 | 0.01 | 0.489 | 0.672 | 0.965 | 6.098 |
| 256 | 128 | 64 | 16 | 8 | 1 | 4 | 0.01 | 0.488 | 0.673 | 0.965 | 6.077 |
| 256 | 64 | 32 | 8 | 16 | 1 | 1 | 0.1 | 0.482 | 0.673 | 0.965 | 6.596 |
| 256 | 128 | 64 | 8 | 8 | 1 | 4 | 0.01 | 0.488 | 0.673 | 0.965 | 6.174 |
| 256 | 128 | 64 | 8 | 16 | 1 | 4 | 0.01 | 0.488 | 0.673 | 0.965 | 6.101 |
| 256 | 128 | 64 | 16 | 16 | 1 | 4 | 0.01 | 0.488 | 0.673 | 0.965 | 6.059 |
| 256 | 256 | 64 | 16 | 8 | 1 | 4 | 0.01 | 0.489 | 0.673 | 0.965 | 6.089 |
| 256 | 256 | 64 | 8 | 8 | 1 | 4 | 0.01 | 0.490 | 0.673 | 0.965 | 6.171 |
| 256 | 64 | 64 | 8 | 16 | 1 | 1 | 0.1 | 0.482 | 0.674 | 0.965 | 6.735 |
| 256 | 128 | 32 | 16 | 8 | 1 | 4 | 0.01 | 0.489 | 0.675 | 0.965 | 6.226 |
| 256 | 64 | 32 | 16 | 16 | 1 | 1 | 0.1 | 0.485 | 0.675 | 0.965 | 6.525 |
| 256 | 128 | 32 | 16 | 16 | 1 | 4 | 0.01 | 0.489 | 0.675 | 0.965 | 6.134 |
| 256 | 128 | 32 | 8 | 8 | 1 | 4 | 0.01 | 0.490 | 0.675 | 0.965 | 6.362 |
| 256 | 128 | 32 | 8 | 16 | 1 | 4 | 0.01 | 0.490 | 0.675 | 0.965 | 6.256 |
| 256 | 256 | 32 | 16 | 16 | 1 | 4 | 0.01 | 0.491 | 0.675 | 0.965 | 6.147 |
| 256 | 64 | 64 | 16 | 16 | 1 | 1 | 0.1 | 0.484 | 0.676 | 0.965 | 6.656 |
| 256 | 256 | 64 | 16 | 16 | 1 | 2 | 1 | 0.486 | 0.676 | 0.964 | 6.236 |
| 256 | 256 | 32 | 16 | 8 | 1 | 4 | 0.01 | 0.491 | 0.676 | 0.965 | 6.243 |
| 256 | 256 | 32 | 8 | 16 | 1 | 4 | 0.01 | 0.492 | 0.676 | 0.965 | 6.248 |
| 256 | 256 | 64 | 16 | 8 | 1 | 2 | 1 | 0.486 | 0.677 | 0.963 | 6.223 |
| 256 | 64 | 64 | 8 | 8 | 1 | 4 | 0.01 | 0.492 | 0.677 | 0.965 | 6.194 |
| 256 | 64 | 64 | 16 | 8 | 1 | 4 | 0.01 | 0.492 | 0.677 | 0.965 | 6.106 |
| 256 | 256 | 64 | 8 | 16 | 1 | 2 | 1 | 0.486 | 0.677 | 0.963 | 6.254 |
| 256 | 64 | 64 | 8 | 16 | 1 | 4 | 0.01 | 0.492 | 0.677 | 0.964 | 6.142 |
| 256 | 256 | 32 | 8 | 8 | 1 | 4 | 0.01 | 0.493 | 0.677 | 0.964 | 6.356 |
| 256 | 64 | 64 | 16 | 16 | 1 | 4 | 0.01 | 0.492 | 0.678 | 0.964 | 6.099 |
| 256 | 256 | 64 | 8 | 8 | 1 | 2 | 1 | 0.487 | 0.678 | 0.963 | 6.243 |
| 256 | 64 | 32 | 16 | 8 | 1 | 4 | 0.01 | 0.491 | 0.678 | 0.964 | 6.222 |
| 256 | 64 | 32 | 8 | 8 | 1 | 4 | 0.01 | 0.492 | 0.678 | 0.964 | 6.372 |
| 256 | 256 | 32 | 16 | 16 | 1 | 2 | 1 | 0.488 | 0.679 | 0.963 | 6.189 |
| 256 | 64 | 32 | 16 | 16 | 1 | 4 | 0.01 | 0.492 | 0.679 | 0.964 | 6.138 |
| 256 | 256 | 32 | 16 | 8 | 1 | 2 | 1 | 0.488 | 0.679 | 0.963 | 6.175 |
| 256 | 64 | 32 | 8 | 16 | 1 | 4 | 0.01 | 0.492 | 0.680 | 0.964 | 6.276 |
| 256 | 256 | 32 | 8 | 16 | 1 | 2 | 1 | 0.489 | 0.680 | 0.963 | 6.204 |
| 256 | 256 | 32 | 8 | 8 | 1 | 2 | 1 | 0.490 | 0.681 | 0.963 | 6.191 |
| 256 | 128 | 64 | 16 | 16 | 1 | 2 | 1 | 0.491 | 0.684 | 0.963 | 6.325 |
| 256 | 128 | 64 | 8 | 8 | 1 | 4 | 0.001 | 0.500 | 0.685 | 0.963 | 6.257 |
| 256 | 128 | 64 | 16 | 8 | 1 | 4 | 0.001 | 0.500 | 0.686 | 0.963 | 6.235 |
| 256 | 128 | 32 | 8 | 8 | 1 | 4 | 0.001 | 0.501 | 0.687 | 0.963 | 6.199 |
| 256 | 128 | 64 | 8 | 16 | 1 | 4 | 0.001 | 0.501 | 0.687 | 0.963 | 6.258 |
| 256 | 128 | 32 | 16 | 8 | 1 | 4 | 0.001 | 0.501 | 0.687 | 0.963 | 6.166 |
| 256 | 128 | 64 | 16 | 16 | 1 | 4 | 0.001 | 0.501 | 0.687 | 0.963 | 6.231 |
| 256 | 64 | 64 | 8 | 8 | 1 | 4 | 0.001 | 0.502 | 0.688 | 0.963 | 6.301 |
| 256 | 256 | 64 | 16 | 8 | 1 | 4 | 0.001 | 0.501 | 0.688 | 0.963 | 6.204 |
| 256 | 64 | 32 | 8 | 8 | 1 | 4 | 0.001 | 0.502 | 0.688 | 0.963 | 6.263 |
| 256 | 256 | 64 | 8 | 8 | 1 | 4 | 0.001 | 0.501 | 0.688 | 0.963 | 6.231 |
| 256 | 256 | 64 | 16 | 16 | 1 | 4 | 0.001 | 0.501 | 0.688 | 0.963 | 6.197 |
| 256 | 256 | 64 | 8 | 16 | 1 | 4 | 0.001 | 0.502 | 0.688 | 0.963 | 6.228 |
| 256 | 64 | 64 | 16 | 8 | 1 | 4 | 0.001 | 0.503 | 0.688 | 0.963 | 6.279 |
| 256 | 64 | 32 | 16 | 8 | 1 | 4 | 0.001 | 0.502 | 0.688 | 0.963 | 6.232 |
| 256 | 128 | 32 | 16 | 16 | 1 | 4 | 0.001 | 0.502 | 0.688 | 0.963 | 6.165 |
| 256 | 128 | 32 | 8 | 16 | 1 | 4 | 0.001 | 0.502 | 0.688 | 0.963 | 6.204 |
| 256 | 64 | 64 | 8 | 16 | 1 | 4 | 0.001 | 0.503 | 0.690 | 0.963 | 6.299 |
| 256 | 64 | 64 | 16 | 16 | 1 | 4 | 0.001 | 0.504 | 0.690 | 0.963 | 6.272 |
| 256 | 256 | 32 | 16 | 8 | 1 | 4 | 0.001 | 0.503 | 0.690 | 0.963 | 6.134 |
| 256 | 64 | 32 | 8 | 16 | 1 | 4 | 0.001 | 0.504 | 0.691 | 0.963 | 6.274 |
| 256 | 256 | 32 | 16 | 16 | 1 | 4 | 0.001 | 0.504 | 0.691 | 0.963 | 6.121 |
| 256 | 64 | 32 | 16 | 16 | 1 | 4 | 0.001 | 0.504 | 0.691 | 0.963 | 6.234 |
| 256 | 256 | 32 | 8 | 8 | 1 | 4 | 0.001 | 0.504 | 0.691 | 0.963 | 6.173 |
| 256 | 256 | 32 | 8 | 16 | 1 | 4 | 0.001 | 0.504 | 0.692 | 0.963 | 6.163 |
| 256 | 256 | 64 | 16 | 16 | 1 | 2 | 0.01 | 0.507 | 0.696 | 0.963 | 6.987 |
| 256 | 128 | 64 | 16 | 16 | 1 | 2 | 0.01 | 0.508 | 0.696 | 0.963 | 6.836 |
| 256 | 128 | 64 | 16 | 8 | 1 | 2 | 0.01 | 0.508 | 0.696 | 0.963 | 6.935 |
| 256 | 256 | 64 | 16 | 8 | 1 | 2 | 0.01 | 0.508 | 0.697 | 0.963 | 7.095 |
| 256 | 128 | 64 | 8 | 16 | 1 | 2 | 0.01 | 0.509 | 0.697 | 0.963 | 6.937 |
| 256 | 128 | 64 | 8 | 8 | 1 | 2 | 0.01 | 0.509 | 0.697 | 0.963 | 7.040 |
| 256 | 256 | 64 | 8 | 16 | 1 | 2 | 0.01 | 0.509 | 0.697 | 0.963 | 7.084 |
| 256 | 256 | 64 | 8 | 8 | 1 | 2 | 0.01 | 0.509 | 0.698 | 0.963 | 7.201 |
| 256 | 128 | 64 | 16 | 8 | 1 | 2 | 0.001 | 0.514 | 0.700 | 0.962 | 6.196 |
| 256 | 256 | 64 | 16 | 16 | 1 | 2 | 0.001 | 0.514 | 0.700 | 0.962 | 6.138 |
| 256 | 128 | 64 | 16 | 16 | 1 | 2 | 0.001 | 0.514 | 0.700 | 0.962 | 6.206 |
| 256 | 128 | 32 | 16 | 8 | 1 | 2 | 0.01 | 0.510 | 0.701 | 0.962 | 7.008 |
| 256 | 64 | 64 | 16 | 8 | 1 | 2 | 0.01 | 0.512 | 0.701 | 0.962 | 6.838 |
| 256 | 256 | 64 | 16 | 8 | 1 | 2 | 0.001 | 0.514 | 0.701 | 0.962 | 6.133 |
| 256 | 128 | 64 | 8 | 8 | 1 | 2 | 0.001 | 0.515 | 0.701 | 0.962 | 6.224 |
| 256 | 128 | 32 | 16 | 16 | 1 | 2 | 0.01 | 0.510 | 0.701 | 0.962 | 6.885 |
| 256 | 64 | 64 | 16 | 16 | 1 | 2 | 0.01 | 0.512 | 0.701 | 0.962 | 6.752 |
| 256 | 64 | 64 | 8 | 8 | 1 | 2 | 0.01 | 0.512 | 0.701 | 0.962 | 6.942 |
| 256 | 128 | 64 | 8 | 16 | 1 | 2 | 0.001 | 0.515 | 0.701 | 0.962 | 6.240 |
| 256 | 256 | 64 | 8 | 16 | 1 | 2 | 0.001 | 0.515 | 0.702 | 0.962 | 6.184 |
| 256 | 128 | 32 | 8 | 8 | 1 | 2 | 0.01 | 0.511 | 0.702 | 0.962 | 7.146 |
| 256 | 64 | 64 | 8 | 16 | 1 | 2 | 0.01 | 0.513 | 0.702 | 0.962 | 6.857 |
| 256 | 256 | 32 | 16 | 16 | 1 | 2 | 0.01 | 0.511 | 0.702 | 0.962 | 7.059 |
| 256 | 256 | 64 | 8 | 8 | 1 | 2 | 0.001 | 0.516 | 0.702 | 0.962 | 6.236 |
| 256 | 128 | 32 | 8 | 16 | 1 | 2 | 0.01 | 0.512 | 0.702 | 0.962 | 7.015 |
| 256 | 256 | 32 | 16 | 8 | 1 | 2 | 0.01 | 0.511 | 0.703 | 0.962 | 7.190 |
| 256 | 256 | 32 | 8 | 16 | 1 | 2 | 0.01 | 0.512 | 0.704 | 0.962 | 7.182 |
| 256 | 64 | 64 | 16 | 8 | 1 | 2 | 0.001 | 0.518 | 0.704 | 0.962 | 6.267 |
| 256 | 64 | 32 | 16 | 8 | 1 | 2 | 0.01 | 0.512 | 0.704 | 0.962 | 6.868 |
| 256 | 256 | 32 | 8 | 8 | 1 | 2 | 0.01 | 0.513 | 0.704 | 0.962 | 7.326 |
| 256 | 64 | 64 | 16 | 16 | 1 | 2 | 0.001 | 0.518 | 0.705 | 0.962 | 6.273 |
| 256 | 64 | 64 | 8 | 8 | 1 | 2 | 0.001 | 0.519 | 0.705 | 0.962 | 6.288 |
| 256 | 128 | 32 | 16 | 8 | 1 | 2 | 0.001 | 0.517 | 0.705 | 0.962 | 6.071 |
| 256 | 64 | 32 | 8 | 8 | 1 | 2 | 0.01 | 0.513 | 0.705 | 0.962 | 7.003 |
| 256 | 64 | 32 | 16 | 16 | 1 | 2 | 0.01 | 0.514 | 0.705 | 0.962 | 6.758 |
| 256 | 128 | 32 | 16 | 16 | 1 | 2 | 0.001 | 0.517 | 0.706 | 0.962 | 6.089 |
| 256 | 64 | 64 | 8 | 16 | 1 | 2 | 0.001 | 0.519 | 0.706 | 0.962 | 6.302 |
| 256 | 256 | 32 | 16 | 16 | 1 | 2 | 0.001 | 0.518 | 0.706 | 0.962 | 6.001 |
| 256 | 64 | 32 | 8 | 16 | 1 | 2 | 0.01 | 0.515 | 0.707 | 0.962 | 6.891 |
| 256 | 128 | 32 | 8 | 8 | 1 | 2 | 0.001 | 0.518 | 0.707 | 0.962 | 6.208 |
| 256 | 256 | 32 | 16 | 8 | 1 | 2 | 0.001 | 0.518 | 0.707 | 0.961 | 6.121 |
| 256 | 128 | 32 | 8 | 16 | 1 | 2 | 0.001 | 0.519 | 0.707 | 0.961 | 6.128 |
| 256 | 64 | 32 | 16 | 8 | 1 | 2 | 0.001 | 0.519 | 0.708 | 0.961 | 6.174 |
| 256 | 256 | 32 | 8 | 16 | 1 | 2 | 0.001 | 0.519 | 0.709 | 0.961 | 6.139 |
| 256 | 256 | 32 | 8 | 8 | 1 | 2 | 0.001 | 0.520 | 0.709 | 0.961 | 6.277 |
| 256 | 64 | 32 | 16 | 16 | 1 | 2 | 0.001 | 0.520 | 0.710 | 0.961 | 6.190 |
| 256 | 64 | 32 | 8 | 8 | 1 | 2 | 0.001 | 0.521 | 0.710 | 0.961 | 6.201 |
| 256 | 64 | 32 | 8 | 16 | 1 | 2 | 0.001 | 0.522 | 0.712 | 0.961 | 6.227 |
| 256 | 256 | 64 | 16 | 16 | 1 | 1 | 1 | 0.533 | 0.735 | 0.956 | 6.378 |
| 256 | 256 | 64 | 16 | 8 | 1 | 1 | 1 | 0.534 | 0.735 | 0.956 | 6.384 |
| 256 | 256 | 64 | 8 | 16 | 1 | 1 | 1 | 0.534 | 0.736 | 0.956 | 6.378 |
| 256 | 256 | 64 | 8 | 8 | 1 | 1 | 1 | 0.535 | 0.737 | 0.956 | 6.387 |
| 256 | 256 | 32 | 16 | 16 | 1 | 1 | 1 | 0.535 | 0.737 | 0.956 | 6.400 |
| 256 | 256 | 32 | 16 | 8 | 1 | 1 | 1 | 0.536 | 0.737 | 0.956 | 6.406 |
| 256 | 256 | 32 | 8 | 16 | 1 | 1 | 1 | 0.536 | 0.738 | 0.956 | 6.399 |
| 256 | 256 | 32 | 8 | 8 | 1 | 1 | 1 | 0.537 | 0.739 | 0.956 | 6.408 |
| 256 | 128 | 64 | 16 | 16 | 1 | 1 | 1 | 0.539 | 0.743 | 0.955 | 6.417 |
| 256 | 256 | 32 | 8 | 8 | 1 | 1 | 0.01 | 0.568 | 0.815 | 0.950 | 9.881 |
| 256 | 256 | 64 | 8 | 8 | 1 | 1 | 0.01 | 0.569 | 0.817 | 0.950 | 9.663 |
| 256 | 128 | 32 | 8 | 8 | 1 | 1 | 0.01 | 0.575 | 0.819 | 0.949 | 9.558 |
| 256 | 64 | 32 | 8 | 8 | 1 | 1 | 0.01 | 0.578 | 0.820 | 0.949 | 9.758 |
| 256 | 128 | 64 | 8 | 8 | 1 | 1 | 0.01 | 0.575 | 0.820 | 0.949 | 9.476 |
| 256 | 64 | 64 | 8 | 8 | 1 | 1 | 0.01 | 0.578 | 0.820 | 0.949 | 9.820 |
| 256 | 256 | 32 | 16 | 8 | 1 | 1 | 0.01 | 0.574 | 0.823 | 0.949 | 10.020 |
| 256 | 256 | 32 | 8 | 16 | 1 | 1 | 0.01 | 0.576 | 0.823 | 0.949 | 9.491 |
| 256 | 256 | 64 | 16 | 8 | 1 | 1 | 0.01 | 0.575 | 0.824 | 0.949 | 9.742 |
| 256 | 256 | 64 | 8 | 16 | 1 | 1 | 0.01 | 0.577 | 0.825 | 0.949 | 9.262 |
| 256 | 64 | 32 | 16 | 8 | 1 | 1 | 0.01 | 0.583 | 0.826 | 0.949 | 10.003 |
| 256 | 128 | 32 | 16 | 8 | 1 | 1 | 0.01 | 0.580 | 0.826 | 0.949 | 9.717 |
| 256 | 64 | 64 | 16 | 8 | 1 | 1 | 0.01 | 0.584 | 0.827 | 0.949 | 10.064 |
| 256 | 128 | 32 | 8 | 16 | 1 | 1 | 0.01 | 0.582 | 0.827 | 0.949 | 9.295 |
| 256 | 128 | 64 | 16 | 8 | 1 | 1 | 0.01 | 0.581 | 0.828 | 0.948 | 9.723 |
| 256 | 64 | 32 | 8 | 16 | 1 | 1 | 0.01 | 0.586 | 0.828 | 0.949 | 9.444 |
| 256 | 64 | 64 | 8 | 16 | 1 | 1 | 0.01 | 0.586 | 0.828 | 0.948 | 9.528 |
| 256 | 128 | 64 | 8 | 16 | 1 | 1 | 0.01 | 0.583 | 0.828 | 0.948 | 9.365 |
| 256 | 256 | 32 | 16 | 16 | 1 | 1 | 0.01 | 0.582 | 0.831 | 0.948 | 9.621 |
| 256 | 256 | 64 | 16 | 16 | 1 | 1 | 0.01 | 0.583 | 0.833 | 0.948 | 9.369 |
| 256 | 64 | 32 | 16 | 16 | 1 | 1 | 0.01 | 0.591 | 0.834 | 0.948 | 9.537 |
| 256 | 128 | 32 | 16 | 16 | 1 | 1 | 0.01 | 0.588 | 0.835 | 0.948 | 9.296 |
| 256 | 64 | 64 | 16 | 16 | 1 | 1 | 0.01 | 0.592 | 0.836 | 0.948 | 9.622 |
| 256 | 128 | 64 | 16 | 16 | 1 | 1 | 0.01 | 0.589 | 0.836 | 0.947 | 9.321 |
| 256 | 64 | 32 | 8 | 8 | 1 | 1 | 0.001 | 0.945 | 1.428 | 0.861 | 25.290 |
| 256 | 64 | 64 | 8 | 8 | 1 | 1 | 0.001 | 0.944 | 1.431 | 0.861 | 25.759 |
| 256 | 128 | 32 | 8 | 8 | 1 | 1 | 0.001 | 0.938 | 1.440 | 0.859 | 30.330 |
| 256 | 128 | 64 | 8 | 8 | 1 | 1 | 0.001 | 0.938 | 1.440 | 0.859 | 30.942 |
| 256 | 256 | 32 | 8 | 8 | 1 | 1 | 0.001 | 0.918 | 1.446 | 0.857 | 35.045 |
| 256 | 256 | 64 | 8 | 8 | 1 | 1 | 0.001 | 0.920 | 1.446 | 0.857 | 35.536 |
| 256 | 64 | 32 | 16 | 8 | 1 | 1 | 0.001 | 0.955 | 1.449 | 0.858 | 27.466 |
| 256 | 64 | 64 | 16 | 8 | 1 | 1 | 0.001 | 0.954 | 1.452 | 0.857 | 27.667 |
| 256 | 64 | 32 | 8 | 16 | 1 | 1 | 0.001 | 0.967 | 1.460 | 0.856 | 25.967 |
| 256 | 128 | 32 | 16 | 8 | 1 | 1 | 0.001 | 0.948 | 1.462 | 0.855 | 32.147 |
| 256 | 128 | 64 | 16 | 8 | 1 | 1 | 0.001 | 0.948 | 1.464 | 0.855 | 32.425 |
| 256 | 64 | 64 | 8 | 16 | 1 | 1 | 0.001 | 0.966 | 1.464 | 0.855 | 25.817 |
| 256 | 256 | 32 | 16 | 8 | 1 | 1 | 0.001 | 0.928 | 1.467 | 0.853 | 36.276 |
| 256 | 256 | 64 | 16 | 8 | 1 | 1 | 0.001 | 0.930 | 1.470 | 0.853 | 36.776 |
| 256 | 128 | 32 | 8 | 16 | 1 | 1 | 0.001 | 0.960 | 1.475 | 0.853 | 31.209 |
| 256 | 128 | 64 | 8 | 16 | 1 | 1 | 0.001 | 0.960 | 1.476 | 0.853 | 31.211 |
| 256 | 64 | 32 | 16 | 16 | 1 | 1 | 0.001 | 0.976 | 1.481 | 0.852 | 28.267 |
| 256 | 256 | 32 | 8 | 16 | 1 | 1 | 0.001 | 0.940 | 1.483 | 0.850 | 36.135 |
| 256 | 256 | 64 | 8 | 16 | 1 | 1 | 0.001 | 0.942 | 1.484 | 0.851 | 36.396 |
| 256 | 64 | 64 | 16 | 16 | 1 | 1 | 0.001 | 0.976 | 1.486 | 0.852 | 28.019 |
| 256 | 128 | 32 | 16 | 16 | 1 | 1 | 0.001 | 0.970 | 1.497 | 0.849 | 33.101 |
| 256 | 128 | 64 | 16 | 16 | 1 | 1 | 0.001 | 0.970 | 1.499 | 0.849 | 32.781 |
| 256 | 256 | 32 | 16 | 16 | 1 | 1 | 0.001 | 0.951 | 1.505 | 0.847 | 37.561 |
| 256 | 256 | 64 | 16 | 16 | 1 | 1 | 0.001 | 0.953 | 1.508 | 0.846 | 37.667 |

**Table S4.** Optimization of kernel weights and GPR parameters ($\sigma_n$ and $\zeta$) for multi-scale kernel prediction of $^1$H chemical shifts. The optimization was carried out on the CSD-2k set, using 3-fold cross validation. For each configuration are reported the corresponding mean absolute error (MAE), root-mean-square error (RMSE), the R-squared ($R^2$) coefficient and the supremum (SUP). In bold is shown the set of parameters that we selected.

| Multi-Scale Kernel Weights | | | | | | $\zeta$ | $\sigma_n$ | MAE (ppm) | RMSE (ppm) | $R^2$ | SUP (ppm) |
|---|---|---|---|---|---|---|---|---|---|---|---|
| $r_c$ =2Å | $r_c$ =3Å | $r_c$ =4Å | $r_c$ =5Å | $r_c$ =6Å | $r_c$ =7Å | | | | | | |
| **256** | **512** | **64** | **8** | **8** | **1** | **2** | **2** | **3.38** | **4.63** | **0.99** | **32.85** |
| 256 | 256 | 32 | 8 | 8 | 1 | 2 | 2 | 3.39 | 4.63 | 0.99 | 32.33 |
| 256 | 512 | 32 | 8 | 8 | 1 | 2 | 2 | 3.38 | 4.64 | 0.99 | 31.37 |
| 256 | 256 | 64 | 8 | 8 | 1 | 2 | 2 | 3.40 | 4.64 | 0.99 | 33.79 |
| 256 | 512 | 64 | 16 | 8 | 1 | 2 | 2 | 3.39 | 4.64 | 0.99 | 33.10 |
| 256 | 512 | 32 | 16 | 8 | 1 | 2 | 2 | 3.39 | 4.64 | 0.99 | 31.76 |
| 256 | 512 | 128 | 8 | 8 | 1 | 2 | 2 | 3.40 | 4.65 | 0.99 | 34.51 |
| 256 | 256 | 32 | 16 | 8 | 1 | 2 | 2 | 3.41 | 4.65 | 0.99 | 32.68 |
| 256 | 512 | 128 | 16 | 8 | 1 | 2 | 2 | 3.41 | 4.65 | 0.99 | 34.63 |
| 256 | 256 | 64 | 16 | 8 | 1 | 2 | 2 | 3.42 | 4.65 | 0.99 | 33.93 |
| 256 | 512 | 64 | 32 | 8 | 1 | 2 | 2 | 3.41 | 4.66 | 0.99 | 33.37 |
| 256 | 128 | 32 | 8 | 8 | 1 | 2 | 2 | 3.42 | 4.66 | 0.99 | 33.56 |
| 256 | 512 | 64 | 8 | 16 | 1 | 2 | 2 | 3.41 | 4.66 | 0.99 | 33.22 |
| 256 | 512 | 32 | 32 | 8 | 1 | 2 | 2 | 3.41 | 4.66 | 0.99 | 32.20 |
| 256 | 512 | 32 | 8 | 16 | 1 | 2 | 2 | 3.40 | 4.66 | 0.99 | 32.00 |
| 256 | 512 | 64 | 16 | 16 | 1 | 2 | 2 | 3.41 | 4.67 | 0.99 | 33.43 |
| 128 | 512 | 64 | 8 | 8 | 1 | 2 | 2 | 3.41 | 4.67 | 0.99 | 33.70 |
| 128 | 512 | 32 | 8 | 8 | 1 | 2 | 2 | 3.40 | 4.67 | 0.99 | 32.38 |
| 256 | 512 | 128 | 32 | 8 | 1 | 2 | 2 | 3.42 | 4.67 | 0.99 | 34.72 |
| 256 | 512 | 128 | 8 | 16 | 1 | 2 | 2 | 3.42 | 4.67 | 0.99 | 34.75 |
| 256 | 128 | 64 | 8 | 8 | 1 | 2 | 2 | 3.44 | 4.67 | 0.99 | 35.02 |
| 256 | 512 | 32 | 16 | 16 | 1 | 2 | 2 | 3.41 | 4.67 | 0.99 | 32.36 |
| 256 | 256 | 32 | 8 | 16 | 1 | 2 | 2 | 3.42 | 4.67 | 0.99 | 33.13 |
| 256 | 256 | 64 | 8 | 16 | 1 | 2 | 2 | 3.43 | 4.67 | 0.99 | 34.22 |
| 128 | 512 | 64 | 16 | 8 | 1 | 2 | 2 | 3.41 | 4.67 | 0.99 | 33.88 |
| 256 | 512 | 128 | 16 | 16 | 1 | 2 | 2 | 3.43 | 4.67 | 0.99 | 34.85 |
| 128 | 512 | 32 | 16 | 8 | 1 | 2 | 2 | 3.41 | 4.68 | 0.99 | 32.70 |
| 256 | 128 | 32 | 16 | 8 | 1 | 2 | 2 | 3.44 | 4.68 | 0.99 | 33.83 |
| 256 | 512 | 64 | 32 | 16 | 1 | 2 | 2 | 3.43 | 4.68 | 0.99 | 33.65 |
| 128 | 512 | 128 | 8 | 8 | 1 | 2 | 2 | 3.43 | 4.68 | 0.99 | 35.03 |
| 256 | 512 | 128 | 8 | 8 | 1 | 2 | 4 | 3.43 | 4.68 | 0.99 | 35.24 |
| 256 | 256 | 64 | 16 | 16 | 1 | 2 | 2 | 3.44 | 4.68 | 0.99 | 34.41 |
| 256 | 256 | 32 | 16 | 16 | 1 | 2 | 2 | 3.44 | 4.68 | 0.99 | 33.43 |
| 256 | 512 | 128 | 16 | 8 | 1 | 2 | 4 | 3.43 | 4.68 | 0.99 | 35.27 |
| 256 | 512 | 64 | 8 | 8 | 1 | 2 | 4 | 3.42 | 4.69 | 0.99 | 34.89 |
| 128 | 512 | 128 | 16 | 8 | 1 | 2 | 2 | 3.44 | 4.69 | 0.99 | 35.10 |
| 256 | 512 | 64 | 16 | 8 | 1 | 2 | 4 | 3.43 | 4.69 | 0.99 | 34.98 |
| 256 | 512 | 128 | 32 | 16 | 1 | 2 | 2 | 3.44 | 4.69 | 0.99 | 34.92 |
| 256 | 512 | 32 | 32 | 16 | 1 | 2 | 2 | 3.43 | 4.69 | 0.99 | 32.78 |
| 256 | 128 | 64 | 16 | 8 | 1 | 2 | 2 | 3.45 | 4.69 | 0.99 | 35.13 |
| 128 | 512 | 64 | 8 | 16 | 1 | 2 | 2 | 3.43 | 4.69 | 0.99 | 34.00 |
| 128 | 512 | 32 | 8 | 16 | 1 | 2 | 2 | 3.43 | 4.69 | 0.99 | 32.79 |
| 256 | 512 | 128 | 32 | 8 | 1 | 2 | 4 | 3.44 | 4.70 | 0.99 | 35.35 |
| 256 | 512 | 64 | 8 | 32 | 1 | 2 | 2 | 3.44 | 4.70 | 0.99 | 33.71 |
| 256 | 256 | 64 | 8 | 8 | 1 | 2 | 4 | 3.44 | 4.70 | 0.99 | 36.44 |
| 256 | 512 | 64 | 32 | 8 | 1 | 2 | 4 | 3.44 | 4.70 | 0.99 | 35.07 |
| 128 | 512 | 64 | 16 | 16 | 1 | 2 | 2 | 3.44 | 4.70 | 0.99 | 34.14 |
| 256 | 512 | 128 | 8 | 16 | 1 | 2 | 4 | 3.44 | 4.70 | 0.99 | 35.64 |
| 256 | 512 | 32 | 8 | 8 | 1 | 2 | 4 | 3.43 | 4.70 | 0.99 | 34.63 |
| 256 | 512 | 128 | 8 | 32 | 1 | 2 | 2 | 3.45 | 4.70 | 0.99 | 35.05 |
| 128 | 512 | 32 | 16 | 16 | 1 | 2 | 2 | 3.44 | 4.70 | 0.99 | 33.05 |
| 256 | 512 | 128 | 16 | 16 | 1 | 2 | 4 | 3.45 | 4.70 | 0.99 | 35.66 |
| 256 | 512 | 64 | 16 | 32 | 1 | 2 | 2 | 3.45 | 4.70 | 0.99 | 33.91 |
| 256 | 512 | 32 | 16 | 8 | 1 | 2 | 4 | 3.44 | 4.70 | 0.99 | 34.76 |
| 128 | 512 | 128 | 8 | 16 | 1 | 2 | 2 | 3.45 | 4.70 | 0.99 | 35.20 |
| 256 | 128 | 32 | 8 | 16 | 1 | 2 | 2 | 3.46 | 4.70 | 0.99 | 34.35 |
| 256 | 512 | 64 | 8 | 16 | 1 | 2 | 4 | 3.44 | 4.70 | 0.99 | 35.36 |
| 256 | 512 | 32 | 8 | 32 | 1 | 2 | 2 | 3.44 | 4.70 | 0.99 | 32.94 |
| 256 | 256 | 64 | 16 | 8 | 1 | 2 | 4 | 3.45 | 4.71 | 0.99 | 36.59 |
| 256 | 512 | 64 | 16 | 16 | 1 | 2 | 4 | 3.45 | 4.71 | 0.99 | 35.43 |
| 256 | 512 | 128 | 16 | 32 | 1 | 2 | 2 | 3.46 | 4.71 | 0.99 | 35.13 |
| 256 | 256 | 32 | 8 | 8 | 1 | 2 | 4 | 3.45 | 4.71 | 0.99 | 36.08 |
| 128 | 512 | 128 | 16 | 16 | 1 | 2 | 2 | 3.46 | 4.71 | 0.99 | 35.25 |
| 256 | 128 | 64 | 8 | 16 | 1 | 2 | 2 | 3.47 | 4.71 | 0.99 | 35.62 |
| 256 | 512 | 32 | 16 | 32 | 1 | 2 | 2 | 3.45 | 4.71 | 0.99 | 33.22 |
| 256 | 512 | 128 | 32 | 16 | 1 | 2 | 4 | 3.46 | 4.71 | 0.99 | 35.68 |
| 256 | 512 | 32 | 32 | 8 | 1 | 2 | 4 | 3.45 | 4.71 | 0.99 | 34.92 |
| 256 | 256 | 32 | 16 | 8 | 1 | 2 | 4 | 3.46 | 4.72 | 0.99 | 36.18 |
| 256 | 512 | 64 | 32 | 16 | 1 | 2 | 4 | 3.46 | 4.72 | 0.99 | 35.51 |
| 256 | 512 | 64 | 32 | 32 | 1 | 2 | 2 | 3.46 | 4.72 | 0.99 | 34.16 |
| 256 | 512 | 128 | 32 | 32 | 1 | 2 | 2 | 3.47 | 4.72 | 0.99 | 35.17 |
| 256 | 512 | 32 | 8 | 16 | 1 | 2 | 4 | 3.45 | 4.72 | 0.99 | 35.15 |
| 256 | 64 | 64 | 8 | 8 | 1 | 2 | 2 | 3.48 | 4.72 | 0.99 | 36.57 |
| 256 | 512 | 32 | 16 | 16 | 1 | 2 | 4 | 3.46 | 4.72 | 0.99 | 35.26 |
| 256 | 128 | 32 | 16 | 16 | 1 | 2 | 2 | 3.48 | 4.72 | 0.99 | 34.58 |
| 256 | 256 | 64 | 8 | 16 | 1 | 2 | 4 | 3.47 | 4.72 | 0.99 | 36.91 |
| 256 | 512 | 32 | 32 | 32 | 1 | 2 | 2 | 3.47 | 4.73 | 0.99 | 33.57 |
| 256 | 128 | 64 | 16 | 16 | 1 | 2 | 2 | 3.49 | 4.73 | 0.99 | 35.72 |
| 256 | 256 | 32 | 8 | 8 | 1 | 2 | 1 | 3.46 | 4.73 | 0.99 | 30.90 |
| 256 | 512 | 128 | 8 | 32 | 1 | 2 | 4 | 3.47 | 4.73 | 0.99 | 36.10 |
| 256 | 256 | 64 | 16 | 16 | 1 | 2 | 4 | 3.47 | 4.73 | 0.99 | 37.00 |
| 256 | 512 | 128 | 16 | 32 | 1 | 2 | 4 | 3.47 | 4.73 | 0.99 | 36.12 |
| 256 | 512 | 32 | 32 | 16 | 1 | 2 | 4 | 3.47 | 4.73 | 0.99 | 35.40 |
| 256 | 512 | 64 | 8 | 32 | 1 | 2 | 4 | 3.47 | 4.74 | 0.99 | 35.93 |
| 256 | 512 | 32 | 8 | 8 | 1 | 2 | 1 | 3.44 | 4.74 | 0.99 | 31.34 |
| 256 | 256 | 32 | 8 | 16 | 1 | 2 | 4 | 3.47 | 4.74 | 0.99 | 36.63 |
| 256 | 512 | 64 | 16 | 32 | 1 | 2 | 4 | 3.47 | 4.74 | 0.99 | 35.99 |
| 256 | 128 | 32 | 8 | 8 | 1 | 2 | 1 | 3.49 | 4.74 | 0.99 | 31.58 |
| 256 | 512 | 128 | 32 | 32 | 1 | 2 | 4 | 3.48 | 4.74 | 0.99 | 36.13 |
| 256 | 128 | 64 | 8 | 8 | 1 | 2 | 4 | 3.49 | 4.74 | 0.99 | 38.88 |
| 256 | 256 | 32 | 16 | 16 | 1 | 2 | 4 | 3.48 | 4.74 | 0.99 | 36.72 |
| 256 | 512 | 64 | 8 | 8 | 1 | 2 | 1 | 3.46 | 4.74 | 0.99 | 32.33 |
| 256 | 512 | 32 | 16 | 8 | 1 | 2 | 1 | 3.46 | 4.74 | 0.99 | 31.08 |
| 256 | 256 | 32 | 16 | 8 | 1 | 2 | 1 | 3.48 | 4.75 | 0.99 | 31.49 |
| 256 | 256 | 64 | 8 | 8 | 1 | 2 | 1 | 3.48 | 4.75 | 0.99 | 33.24 |
| 256 | 64 | 64 | 16 | 8 | 1 | 2 | 2 | 3.51 | 4.75 | 0.99 | 36.55 |
| 256 | 512 | 64 | 32 | 32 | 1 | 2 | 4 | 3.48 | 4.75 | 0.99 | 36.05 |
| 256 | 512 | 64 | 16 | 8 | 1 | 2 | 1 | 3.47 | 4.75 | 0.99 | 32.76 |
| 256 | 128 | 32 | 8 | 8 | 1 | 2 | 4 | 3.49 | 4.75 | 0.99 | 38.27 |
| 256 | 512 | 32 | 8 | 32 | 1 | 2 | 4 | 3.48 | 4.75 | 0.99 | 35.80 |
| 256 | 512 | 32 | 16 | 32 | 1 | 2 | 4 | 3.49 | 4.76 | 0.99 | 35.89 |
| 256 | 128 | 64 | 16 | 8 | 1 | 2 | 4 | 3.50 | 4.76 | 0.99 | 39.04 |
| 256 | 256 | 64 | 16 | 8 | 1 | 2 | 1 | 3.50 | 4.76 | 0.99 | 33.59 |
| 256 | 256 | 32 | 8 | 8 | 1 | 4 | 4 | 3.44 | 4.76 | 0.99 | 30.91 |
| 256 | 128 | 32 | 16 | 8 | 1 | 2 | 4 | 3.50 | 4.76 | 0.99 | 38.53 |
| 256 | 512 | 32 | 32 | 32 | 1 | 2 | 4 | 3.50 | 4.76 | 0.99 | 36.00 |
| 256 | 512 | 32 | 8 | 16 | 1 | 2 | 1 | 3.47 | 4.76 | 0.99 | 31.19 |
| 256 | 512 | 32 | 32 | 8 | 1 | 2 | 1 | 3.49 | 4.77 | 0.99 | 31.68 |
| 256 | 256 | 32 | 8 | 16 | 1 | 2 | 1 | 3.50 | 4.77 | 0.99 | 31.64 |
| 256 | 128 | 64 | 8 | 8 | 1 | 2 | 1 | 3.52 | 4.77 | 0.99 | 34.03 |
| 256 | 128 | 32 | 16 | 8 | 1 | 2 | 1 | 3.52 | 4.77 | 0.99 | 32.02 |
| 256 | 512 | 64 | 32 | 8 | 1 | 2 | 1 | 3.49 | 4.77 | 0.99 | 33.33 |
| 256 | 512 | 64 | 8 | 16 | 1 | 2 | 1 | 3.48 | 4.77 | 0.99 | 32.81 |
| 256 | 512 | 128 | 8 | 8 | 1 | 2 | 1 | 3.49 | 4.77 | 0.99 | 34.68 |
| 256 | 64 | 64 | 8 | 16 | 1 | 2 | 2 | 3.52 | 4.77 | 0.99 | 37.07 |
| 256 | 512 | 32 | 16 | 16 | 1 | 2 | 1 | 3.49 | 4.77 | 0.99 | 31.41 |
| 256 | 512 | 128 | 16 | 8 | 1 | 2 | 1 | 3.50 | 4.77 | 0.99 | 34.91 |
| 256 | 512 | 64 | 16 | 16 | 1 | 2 | 1 | 3.50 | 4.77 | 0.99 | 33.20 |
| 256 | 128 | 64 | 8 | 16 | 1 | 2 | 4 | 3.51 | 4.78 | 0.99 | 39.30 |
| 256 | 128 | 32 | 8 | 8 | 1 | 4 | 4 | 3.46 | 4.78 | 0.99 | 31.96 |
| 256 | 256 | 64 | 8 | 16 | 1 | 2 | 1 | 3.51 | 4.78 | 0.99 | 33.73 |
| 256 | 256 | 32 | 16 | 8 | 1 | 4 | 4 | 3.45 | 4.78 | 0.99 | 31.24 |
| 256 | 256 | 64 | 8 | 8 | 1 | 4 | 4 | 3.46 | 4.78 | 0.99 | 31.59 |
| 256 | 256 | 32 | 16 | 16 | 1 | 2 | 1 | 3.52 | 4.78 | 0.99 | 32.14 |
| 256 | 128 | 32 | 8 | 16 | 1 | 2 | 4 | 3.52 | 4.79 | 0.99 | 38.79 |
| 256 | 128 | 64 | 16 | 8 | 1 | 2 | 1 | 3.54 | 4.79 | 0.99 | 34.24 |
| 256 | 512 | 32 | 8 | 8 | 1 | 4 | 4 | 3.44 | 4.79 | 0.99 | 30.81 |
| 256 | 512 | 128 | 32 | 8 | 1 | 2 | 1 | 3.52 | 4.79 | 0.99 | 35.16 |
| 256 | 128 | 64 | 16 | 16 | 1 | 2 | 4 | 3.53 | 4.79 | 0.99 | 39.45 |
| 256 | 64 | 32 | 16 | 16 | 1 | 2 | 2 | 3.54 | 4.79 | 0.99 | 35.86 |
| 256 | 256 | 64 | 16 | 16 | 1 | 2 | 1 | 3.53 | 4.79 | 0.99 | 34.03 |
| 256 | 128 | 32 | 8 | 16 | 1 | 2 | 1 | 3.53 | 4.79 | 0.99 | 32.22 |
| 256 | 512 | 64 | 32 | 16 | 1 | 2 | 1 | 3.52 | 4.79 | 0.99 | 33.68 |
| 256 | 512 | 128 | 8 | 16 | 1 | 2 | 1 | 3.51 | 4.79 | 0.99 | 34.93 |
| 256 | 512 | 32 | 32 | 16 | 1 | 2 | 1 | 3.51 | 4.79 | 0.99 | 32.14 |
| 256 | 512 | 64 | 8 | 8 | 1 | 4 | 4 | 3.45 | 4.79 | 0.99 | 31.20 |
| 256 | 512 | 128 | 16 | 16 | 1 | 2 | 1 | 3.52 | 4.79 | 0.99 | 35.14 |
| 256 | 64 | 64 | 16 | 16 | 1 | 2 | 2 | 3.55 | 4.79 | 0.99 | 37.06 |
| 256 | 512 | 32 | 16 | 8 | 1 | 4 | 4 | 3.45 | 4.79 | 0.99 | 30.99 |
| 256 | 256 | 64 | 16 | 8 | 1 | 4 | 4 | 3.47 | 4.79 | 0.99 | 31.90 |
| 256 | 512 | 64 | 16 | 8 | 1 | 4 | 4 | 3.46 | 4.80 | 0.99 | 31.51 |
| 256 | 128 | 32 | 16 | 16 | 1 | 2 | 4 | 3.53 | 4.80 | 0.99 | 39.02 |
| 256 | 256 | 32 | 8 | 16 | 1 | 4 | 4 | 3.47 | 4.80 | 0.99 | 31.19 |
| 256 | 128 | 32 | 16 | 8 | 1 | 4 | 4 | 3.48 | 4.80 | 0.99 | 32.00 |
| 256 | 512 | 128 | 32 | 16 | 1 | 2 | 1 | 3.54 | 4.80 | 0.99 | 35.36 |
| 256 | 512 | 64 | 8 | 32 | 1 | 2 | 1 | 3.52 | 4.81 | 0.99 | 33.48 |
| 256 | 512 | 32 | 8 | 32 | 1 | 2 | 1 | 3.52 | 4.81 | 0.99 | 31.79 |
| 256 | 128 | 64 | 8 | 8 | 1 | 4 | 4 | 3.49 | 4.81 | 0.99 | 32.40 |
| 256 | 128 | 64 | 8 | 16 | 1 | 2 | 1 | 3.55 | 4.81 | 0.99 | 34.44 |
| 256 | 64 | 64 | 8 | 8 | 1 | 2 | 1 | 3.56 | 4.81 | 0.99 | 34.75 |
| 256 | 256 | 32 | 8 | 8 | 1 | 4 | 2 | 3.47 | 4.81 | 0.99 | 31.89 |
| 256 | 512 | 64 | 16 | 32 | 1 | 2 | 1 | 3.53 | 4.81 | 0.99 | 33.80 |
| 256 | 512 | 32 | 8 | 16 | 1 | 4 | 4 | 3.46 | 4.81 | 0.99 | 30.77 |
| 256 | 128 | 32 | 16 | 16 | 1 | 2 | 1 | 3.56 | 4.81 | 0.99 | 32.65 |
| 256 | 256 | 32 | 16 | 16 | 1 | 4 | 4 | 3.48 | 4.81 | 0.99 | 31.47 |
| 256 | 256 | 64 | 8 | 16 | 1 | 4 | 4 | 3.49 | 4.81 | 0.99 | 31.76 |
| 256 | 512 | 32 | 16 | 32 | 1 | 2 | 1 | 3.53 | 4.81 | 0.99 | 32.24 |
| 256 | 512 | 64 | 8 | 16 | 1 | 4 | 4 | 3.47 | 4.81 | 0.99 | 31.36 |
| 256 | 512 | 32 | 32 | 8 | 1 | 4 | 4 | 3.47 | 4.81 | 0.99 | 31.51 |
| 256 | 512 | 64 | 32 | 8 | 1 | 4 | 4 | 3.47 | 4.82 | 0.99 | 31.90 |
| 256 | 64 | 64 | 8 | 8 | 1 | 2 | 4 | 3.55 | 4.82 | 0.99 | 41.56 |
| 256 | 128 | 32 | 8 | 8 | 1 | 4 | 2 | 3.50 | 4.82 | 0.99 | 31.11 |
| 256 | 512 | 32 | 16 | 16 | 1 | 4 | 4 | 3.47 | 4.82 | 0.99 | 31.15 |
| 256 | 512 | 64 | 16 | 16 | 1 | 4 | 4 | 3.48 | 4.82 | 0.99 | 31.64 |
| 256 | 512 | 128 | 8 | 8 | 1 | 4 | 4 | 3.48 | 4.82 | 0.99 | 31.91 |
| 256 | 512 | 128 | 8 | 32 | 1 | 2 | 1 | 3.54 | 4.82 | 0.99 | 35.26 |
| 256 | 512 | 128 | 16 | 32 | 1 | 2 | 1 | 3.55 | 4.82 | 0.99 | 35.45 |
| 256 | 256 | 32 | 16 | 8 | 1 | 4 | 2 | 3.49 | 4.82 | 0.99 | 31.74 |
| 256 | 512 | 128 | 16 | 8 | 1 | 4 | 4 | 3.49 | 4.82 | 0.99 | 32.12 |
| 256 | 128 | 64 | 16 | 16 | 1 | 2 | 1 | 3.57 | 4.82 | 0.99 | 34.63 |
| 256 | 512 | 64 | 32 | 32 | 1 | 2 | 1 | 3.55 | 4.82 | 0.99 | 34.18 |
| 256 | 128 | 32 | 8 | 16 | 1 | 4 | 4 | 3.50 | 4.83 | 0.99 | 32.45 |
| 256 | 256 | 64 | 16 | 16 | 1 | 4 | 4 | 3.50 | 4.83 | 0.99 | 32.03 |
| 256 | 128 | 64 | 16 | 8 | 1 | 4 | 4 | 3.51 | 4.83 | 0.99 | 32.55 |
| 256 | 512 | 32 | 32 | 32 | 1 | 2 | 1 | 3.55 | 4.83 | 0.99 | 32.84 |
| 256 | 256 | 64 | 8 | 8 | 1 | 4 | 2 | 3.50 | 4.83 | 0.99 | 31.64 |
| 256 | 512 | 128 | 32 | 32 | 1 | 2 | 1 | 3.56 | 4.83 | 0.99 | 35.63 |
| 256 | 64 | 64 | 16 | 8 | 1 | 2 | 4 | 3.56 | 4.83 | 0.99 | 41.61 |
| 256 | 64 | 64 | 16 | 8 | 1 | 2 | 1 | 3.59 | 4.84 | 0.99 | 34.80 |
| 256 | 512 | 32 | 32 | 16 | 1 | 4 | 4 | 3.49 | 4.84 | 0.99 | 31.62 |
| 256 | 512 | 64 | 32 | 16 | 1 | 4 | 4 | 3.49 | 4.84 | 0.99 | 32.00 |
| 256 | 512 | 128 | 8 | 16 | 1 | 4 | 4 | 3.50 | 4.84 | 0.99 | 32.03 |
| 256 | 128 | 32 | 16 | 8 | 1 | 4 | 2 | 3.52 | 4.84 | 0.99 | 31.55 |
| 256 | 512 | 32 | 8 | 8 | 1 | 4 | 2 | 3.47 | 4.84 | 0.99 | 33.25 |
| 256 | 512 | 128 | 32 | 8 | 1 | 4 | 4 | 3.50 | 4.84 | 0.99 | 32.39 |
| 256 | 256 | 64 | 16 | 8 | 1 | 4 | 2 | 3.51 | 4.84 | 0.99 | 32.06 |
| 256 | 256 | 32 | 8 | 16 | 1 | 4 | 2 | 3.50 | 4.84 | 0.99 | 31.45 |
| 256 | 512 | 128 | 16 | 16 | 1 | 4 | 4 | 3.50 | 4.84 | 0.99 | 32.22 |
| 256 | 512 | 32 | 16 | 8 | 1 | 4 | 2 | 3.48 | 4.85 | 0.99 | 33.30 |
| 256 | 512 | 64 | 8 | 8 | 1 | 4 | 2 | 3.49 | 4.85 | 0.99 | 32.34 |
| 256 | 128 | 32 | 16 | 16 | 1 | 4 | 4 | 3.52 | 4.85 | 0.99 | 32.47 |
| 256 | 512 | 64 | 16 | 8 | 1 | 4 | 2 | 3.50 | 4.85 | 0.99 | 32.44 |
| 256 | 128 | 64 | 8 | 16 | 1 | 4 | 4 | 3.53 | 4.85 | 0.99 | 32.66 |
| 256 | 512 | 32 | 8 | 32 | 1 | 4 | 4 | 3.50 | 4.85 | 0.99 | 31.08 |
| 256 | 512 | 64 | 8 | 32 | 1 | 4 | 4 | 3.50 | 4.85 | 0.99 | 31.58 |
| 256 | 128 | 64 | 8 | 8 | 1 | 4 | 2 | 3.53 | 4.85 | 0.99 | 32.24 |
| 256 | 64 | 64 | 8 | 16 | 1 | 2 | 4 | 3.58 | 4.86 | 0.99 | 41.93 |
| 256 | 256 | 32 | 16 | 16 | 1 | 4 | 2 | 3.52 | 4.86 | 0.99 | 31.55 |
| 256 | 512 | 64 | 16 | 32 | 1 | 4 | 4 | 3.51 | 4.86 | 0.99 | 31.83 |
| 256 | 512 | 32 | 16 | 32 | 1 | 4 | 4 | 3.50 | 4.86 | 0.99 | 31.40 |
| 256 | 64 | 64 | 8 | 16 | 1 | 2 | 1 | 3.60 | 4.86 | 0.99 | 35.13 |
| 256 | 256 | 64 | 8 | 16 | 1 | 4 | 2 | 3.53 | 4.86 | 0.99 | 31.89 |
| 256 | 512 | 128 | 32 | 16 | 1 | 4 | 4 | 3.52 | 4.86 | 0.99 | 32.47 |
| 256 | 512 | 32 | 8 | 16 | 1 | 4 | 2 | 3.50 | 4.86 | 0.99 | 32.92 |
| 256 | 128 | 32 | 8 | 16 | 1 | 4 | 2 | 3.54 | 4.86 | 0.99 | 31.47 |
| 256 | 512 | 32 | 32 | 8 | 1 | 4 | 2 | 3.51 | 4.86 | 0.99 | 33.04 |
| 256 | 512 | 64 | 32 | 8 | 1 | 4 | 2 | 3.51 | 4.87 | 0.99 | 32.35 |
| 256 | 64 | 32 | 16 | 16 | 1 | 2 | 1 | 3.61 | 4.87 | 0.99 | 33.48 |
| 256 | 512 | 64 | 8 | 16 | 1 | 4 | 2 | 3.51 | 4.87 | 0.99 | 32.08 |
| 256 | 64 | 64 | 8 | 8 | 1 | 4 | 4 | 3.54 | 4.87 | 0.99 | 34.01 |
| 256 | 512 | 32 | 16 | 16 | 1 | 4 | 2 | 3.51 | 4.87 | 0.99 | 32.96 |
| 256 | 128 | 64 | 16 | 16 | 1 | 4 | 4 | 3.54 | 4.87 | 0.99 | 32.91 |
| 256 | 512 | 64 | 16 | 16 | 1 | 4 | 2 | 3.51 | 4.87 | 0.99 | 32.18 |
| 256 | 256 | 64 | 16 | 16 | 1 | 4 | 2 | 3.54 | 4.87 | 0.99 | 32.22 |
| 256 | 64 | 64 | 16 | 16 | 1 | 2 | 4 | 3.59 | 4.87 | 0.99 | 41.98 |
| 256 | 128 | 64 | 16 | 8 | 1 | 4 | 2 | 3.55 | 4.87 | 0.99 | 32.51 |
| 256 | 512 | 128 | 8 | 32 | 1 | 4 | 4 | 3.53 | 4.87 | 0.99 | 32.19 |
| 256 | 512 | 64 | 32 | 32 | 1 | 4 | 4 | 3.52 | 4.88 | 0.99 | 32.13 |
| 256 | 512 | 32 | 32 | 32 | 1 | 4 | 4 | 3.52 | 4.88 | 0.99 | 31.79 |
| 256 | 512 | 128 | 8 | 8 | 1 | 4 | 2 | 3.52 | 4.88 | 0.99 | 32.10 |
| 256 | 512 | 128 | 16 | 8 | 1 | 4 | 2 | 3.53 | 4.88 | 0.99 | 32.44 |
| 256 | 512 | 128 | 16 | 32 | 1 | 4 | 4 | 3.53 | 4.88 | 0.99 | 32.36 |
| 256 | 64 | 64 | 16 | 16 | 1 | 2 | 1 | 3.62 | 4.88 | 0.99 | 35.29 |
| 256 | 128 | 32 | 16 | 16 | 1 | 4 | 2 | 3.56 | 4.88 | 0.99 | 31.74 |
| 256 | 512 | 64 | 32 | 16 | 1 | 4 | 2 | 3.53 | 4.88 | 0.99 | 32.46 |
| 256 | 512 | 32 | 32 | 16 | 1 | 4 | 2 | 3.53 | 4.88 | 0.99 | 32.73 |
| 256 | 64 | 32 | 16 | 16 | 1 | 2 | 4 | 3.60 | 4.89 | 0.99 | 41.82 |
| 256 | 256 | 32 | 8 | 8 | 1 | 4 | 8 | 3.53 | 4.89 | 0.99 | 32.72 |
| 256 | 512 | 128 | 32 | 8 | 1 | 4 | 2 | 3.54 | 4.89 | 0.99 | 32.80 |
| 256 | 256 | 64 | 8 | 8 | 1 | 4 | 8 | 3.54 | 4.89 | 0.99 | 33.05 |
| 256 | 128 | 64 | 8 | 16 | 1 | 4 | 2 | 3.56 | 4.89 | 0.99 | 32.33 |
| 256 | 512 | 128 | 8 | 16 | 1 | 4 | 2 | 3.54 | 4.89 | 0.99 | 32.30 |
| 256 | 64 | 64 | 16 | 8 | 1 | 4 | 4 | 3.57 | 4.89 | 0.99 | 34.33 |
| 256 | 512 | 128 | 16 | 16 | 1 | 4 | 2 | 3.54 | 4.89 | 0.99 | 32.59 |
| 256 | 512 | 128 | 32 | 32 | 1 | 4 | 4 | 3.54 | 4.90 | 0.99 | 32.57 |
| 256 | 256 | 32 | 16 | 8 | 1 | 4 | 8 | 3.54 | 4.90 | 0.99 | 32.80 |
| 256 | 512 | 64 | 8 | 32 | 1 | 4 | 2 | 3.54 | 4.90 | 0.99 | 32.00 |
| 256 | 512 | 32 | 8 | 32 | 1 | 4 | 2 | 3.53 | 4.90 | 0.99 | 32.42 |
| 256 | 256 | 64 | 16 | 8 | 1 | 4 | 8 | 3.55 | 4.90 | 0.99 | 33.16 |
| 256 | 512 | 64 | 16 | 32 | 1 | 4 | 2 | 3.54 | 4.90 | 0.99 | 32.29 |
| 256 | 512 | 32 | 16 | 32 | 1 | 4 | 2 | 3.54 | 4.90 | 0.99 | 32.45 |
| 256 | 512 | 128 | 32 | 16 | 1 | 4 | 2 | 3.56 | 4.91 | 0.99 | 32.90 |
| 256 | 128 | 64 | 16 | 16 | 1 | 4 | 2 | 3.58 | 4.91 | 0.99 | 32.56 |
| 256 | 64 | 64 | 8 | 8 | 1 | 4 | 2 | 3.58 | 4.91 | 0.99 | 32.98 |
| 256 | 256 | 32 | 8 | 8 | 1 | 4 | 1 | 3.55 | 4.91 | 0.99 | 33.94 |
| 256 | 128 | 32 | 8 | 8 | 1 | 4 | 1 | 3.57 | 4.91 | 0.99 | 32.79 |
| 256 | 256 | 32 | 16 | 8 | 1 | 4 | 1 | 3.56 | 4.91 | 0.99 | 33.68 |
| 256 | 256 | 32 | 8 | 16 | 1 | 4 | 8 | 3.55 | 4.91 | 0.99 | 32.91 |
| 256 | 128 | 32 | 8 | 8 | 1 | 4 | 8 | 3.56 | 4.92 | 0.99 | 34.72 |
| 256 | 512 | 64 | 32 | 32 | 1 | 4 | 2 | 3.56 | 4.92 | 0.99 | 32.58 |
| 256 | 64 | 32 | 16 | 16 | 1 | 4 | 4 | 3.58 | 4.92 | 0.99 | 34.20 |
| 256 | 64 | 64 | 8 | 16 | 1 | 4 | 4 | 3.58 | 4.92 | 0.99 | 34.44 |
| 256 | 256 | 64 | 8 | 16 | 1 | 4 | 8 | 3.56 | 4.92 | 0.99 | 33.08 |
| 256 | 512 | 32 | 32 | 32 | 1 | 4 | 2 | 3.56 | 4.92 | 0.99 | 32.25 |
| 256 | 512 | 128 | 8 | 32 | 1 | 4 | 2 | 3.56 | 4.92 | 0.99 | 32.53 |
| 256 | 512 | 128 | 16 | 32 | 1 | 4 | 2 | 3.57 | 4.92 | 0.99 | 32.77 |
| 256 | 256 | 64 | 8 | 8 | 1 | 4 | 1 | 3.57 | 4.92 | 0.99 | 32.61 |
| 256 | 256 | 32 | 16 | 16 | 1 | 4 | 8 | 3.56 | 4.92 | 0.99 | 32.98 |
| 256 | 128 | 32 | 16 | 8 | 1 | 4 | 1 | 3.59 | 4.92 | 0.99 | 32.35 |
| 256 | 256 | 64 | 16 | 8 | 1 | 4 | 1 | 3.58 | 4.93 | 0.99 | 32.80 |
| 256 | 128 | 64 | 8 | 8 | 1 | 4 | 8 | 3.58 | 4.93 | 0.99 | 34.92 |
| 256 | 256 | 64 | 16 | 16 | 1 | 4 | 8 | 3.57 | 4.93 | 0.99 | 33.22 |
| 256 | 256 | 32 | 8 | 16 | 1 | 4 | 1 | 3.57 | 4.93 | 0.99 | 33.25 |
| 256 | 128 | 32 | 16 | 8 | 1 | 4 | 8 | 3.58 | 4.93 | 0.99 | 34.95 |
| 256 | 64 | 64 | 16 | 8 | 1 | 4 | 2 | 3.61 | 4.93 | 0.99 | 33.32 |
| 256 | 512 | 128 | 32 | 32 | 1 | 4 | 2 | 3.58 | 4.94 | 0.99 | 33.01 |
| 256 | 256 | 32 | 16 | 16 | 1 | 4 | 1 | 3.58 | 4.94 | 0.99 | 33.07 |
| 256 | 128 | 64 | 8 | 8 | 1 | 4 | 1 | 3.60 | 4.94 | 0.99 | 32.92 |
| 256 | 256 | 64 | 8 | 16 | 1 | 4 | 1 | 3.59 | 4.94 | 0.99 | 32.58 |
| 256 | 64 | 64 | 16 | 16 | 1 | 4 | 4 | 3.60 | 4.94 | 0.99 | 34.73 |
| 256 | 512 | 64 | 16 | 8 | 1 | 4 | 1 | 3.57 | 4.94 | 0.99 | 34.20 |
| 256 | 256 | 64 | 16 | 16 | 1 | 4 | 1 | 3.59 | 4.94 | 0.99 | 32.85 |
| 256 | 128 | 64 | 16 | 8 | 1 | 4 | 8 | 3.59 | 4.94 | 0.99 | 35.08 |
| 256 | 128 | 32 | 8 | 16 | 1 | 4 | 1 | 3.60 | 4.94 | 0.99 | 32.01 |
| 256 | 512 | 32 | 16 | 8 | 1 | 4 | 1 | 3.56 | 4.95 | 0.99 | 35.30 |
| 256 | 512 | 64 | 32 | 8 | 1 | 4 | 1 | 3.58 | 4.95 | 0.99 | 33.90 |
| 256 | 64 | 32 | 16 | 16 | 1 | 4 | 2 | 3.62 | 4.95 | 0.99 | 32.66 |
| 256 | 512 | 64 | 8 | 8 | 1 | 4 | 1 | 3.56 | 4.95 | 0.99 | 34.18 |
| 256 | 128 | 64 | 16 | 8 | 1 | 4 | 1 | 3.61 | 4.95 | 0.99 | 33.15 |
| 256 | 512 | 32 | 8 | 8 | 1 | 4 | 1 | 3.56 | 4.95 | 0.99 | 35.41 |
| 256 | 512 | 32 | 32 | 8 | 1 | 4 | 1 | 3.58 | 4.95 | 0.99 | 34.79 |
| 256 | 128 | 32 | 8 | 16 | 1 | 4 | 8 | 3.59 | 4.95 | 0.99 | 35.10 |
| 256 | 512 | 64 | 16 | 16 | 1 | 4 | 1 | 3.58 | 4.95 | 0.99 | 33.77 |
| 256 | 64 | 64 | 8 | 16 | 1 | 4 | 2 | 3.62 | 4.96 | 0.99 | 33.39 |
| 256 | 128 | 32 | 16 | 16 | 1 | 4 | 1 | 3.62 | 4.96 | 0.99 | 32.18 |
| 256 | 512 | 64 | 8 | 16 | 1 | 4 | 1 | 3.58 | 4.96 | 0.99 | 33.74 |
| 256 | 512 | 64 | 32 | 16 | 1 | 4 | 1 | 3.59 | 4.96 | 0.99 | 33.53 |
| 256 | 512 | 32 | 16 | 16 | 1 | 4 | 1 | 3.58 | 4.96 | 0.99 | 34.77 |
| 256 | 512 | 32 | 8 | 16 | 1 | 4 | 1 | 3.57 | 4.96 | 0.99 | 34.86 |
| 256 | 512 | 128 | 16 | 8 | 1 | 4 | 1 | 3.59 | 4.96 | 0.99 | 33.29 |
| 256 | 128 | 64 | 8 | 16 | 1 | 4 | 8 | 3.60 | 4.96 | 0.99 | 35.24 |
| 256 | 512 | 128 | 32 | 8 | 1 | 4 | 1 | 3.60 | 4.97 | 0.99 | 33.47 |
| 256 | 512 | 32 | 32 | 16 | 1 | 4 | 1 | 3.59 | 4.97 | 0.99 | 34.34 |
| 256 | 128 | 64 | 8 | 16 | 1 | 4 | 1 | 3.63 | 4.97 | 0.99 | 32.90 |
| 256 | 128 | 32 | 16 | 16 | 1 | 4 | 8 | 3.60 | 4.97 | 0.99 | 35.34 |
| 256 | 512 | 128 | 8 | 8 | 1 | 4 | 1 | 3.59 | 4.97 | 0.99 | 33.17 |
| 256 | 512 | 128 | 16 | 16 | 1 | 4 | 1 | 3.60 | 4.97 | 0.99 | 33.42 |
| 256 | 512 | 128 | 32 | 16 | 1 | 4 | 1 | 3.61 | 4.97 | 0.99 | 33.54 |
| 256 | 512 | 64 | 16 | 32 | 1 | 4 | 1 | 3.60 | 4.98 | 0.99 | 33.19 |
| 256 | 128 | 64 | 16 | 16 | 1 | 4 | 1 | 3.64 | 4.98 | 0.99 | 33.10 |
| 256 | 512 | 128 | 8 | 16 | 1 | 4 | 1 | 3.60 | 4.98 | 0.99 | 33.37 |
| 256 | 64 | 64 | 16 | 16 | 1 | 4 | 2 | 3.64 | 4.98 | 0.99 | 33.76 |
| 256 | 512 | 64 | 8 | 32 | 1 | 4 | 1 | 3.60 | 4.98 | 0.99 | 33.23 |
| 256 | 128 | 64 | 16 | 16 | 1 | 4 | 8 | 3.62 | 4.98 | 0.99 | 35.41 |
| 256 | 512 | 64 | 32 | 32 | 1 | 4 | 1 | 3.61 | 4.98 | 0.99 | 33.23 |
| 256 | 512 | 32 | 16 | 32 | 1 | 4 | 1 | 3.60 | 4.98 | 0.99 | 34.06 |
| 256 | 512 | 32 | 8 | 32 | 1 | 4 | 1 | 3.60 | 4.99 | 0.99 | 34.11 |
| 256 | 512 | 128 | 16 | 32 | 1 | 4 | 1 | 3.62 | 4.99 | 0.99 | 33.52 |
| 256 | 512 | 32 | 32 | 32 | 1 | 4 | 1 | 3.62 | 4.99 | 0.99 | 33.71 |
| 256 | 512 | 128 | 8 | 32 | 1 | 4 | 1 | 3.62 | 4.99 | 0.99 | 33.54 |
| 256 | 512 | 128 | 32 | 32 | 1 | 4 | 1 | 3.63 | 4.99 | 0.99 | 33.60 |
| 256 | 64 | 64 | 8 | 8 | 1 | 4 | 1 | 3.65 | 4.99 | 0.99 | 33.23 |
| 256 | 64 | 64 | 8 | 8 | 1 | 4 | 8 | 3.64 | 5.00 | 0.99 | 37.06 |
| 256 | 256 | 64 | 8 | 8 | 1 | 2 | 8 | 3.67 | 5.00 | 0.99 | 42.45 |
| 256 | 256 | 64 | 16 | 8 | 1 | 2 | 8 | 3.68 | 5.01 | 0.99 | 42.57 |
| 256 | 64 | 64 | 16 | 8 | 1 | 4 | 1 | 3.67 | 5.01 | 0.99 | 33.38 |
| 256 | 64 | 32 | 16 | 16 | 1 | 4 | 1 | 3.68 | 5.02 | 0.99 | 32.48 |
| 256 | 64 | 64 | 16 | 8 | 1 | 4 | 8 | 3.65 | 5.02 | 0.99 | 37.13 |
| 256 | 256 | 64 | 8 | 16 | 1 | 2 | 8 | 3.69 | 5.02 | 0.99 | 42.79 |
| 256 | 256 | 64 | 16 | 16 | 1 | 2 | 8 | 3.69 | 5.02 | 0.99 | 42.90 |
| 256 | 256 | 32 | 16 | 8 | 1 | 2 | 8 | 3.69 | 5.03 | 0.99 | 42.70 |
| 256 | 256 | 32 | 8 | 8 | 1 | 2 | 8 | 3.69 | 5.03 | 0.99 | 42.53 |
| 256 | 64 | 64 | 8 | 16 | 1 | 4 | 1 | 3.68 | 5.03 | 0.99 | 33.24 |
| 256 | 64 | 64 | 16 | 16 | 1 | 4 | 1 | 3.70 | 5.04 | 0.99 | 33.59 |
| 256 | 256 | 32 | 8 | 16 | 1 | 2 | 8 | 3.70 | 5.04 | 0.99 | 42.92 |
| 256 | 64 | 64 | 8 | 16 | 1 | 4 | 8 | 3.67 | 5.04 | 0.99 | 37.32 |
| 256 | 256 | 32 | 16 | 16 | 1 | 2 | 8 | 3.71 | 5.04 | 0.99 | 43.06 |
| 256 | 64 | 32 | 16 | 16 | 1 | 4 | 8 | 3.68 | 5.06 | 0.99 | 37.66 |
| 256 | 64 | 64 | 16 | 16 | 1 | 4 | 8 | 3.68 | 5.06 | 0.99 | 37.42 |
| 256 | 128 | 64 | 8 | 8 | 1 | 2 | 8 | 3.73 | 5.07 | 0.99 | 45.46 |
| 256 | 128 | 64 | 16 | 8 | 1 | 2 | 8 | 3.73 | 5.08 | 0.99 | 45.53 |
| 256 | 128 | 64 | 8 | 16 | 1 | 2 | 8 | 3.75 | 5.09 | 0.99 | 45.82 |
| 256 | 128 | 32 | 8 | 8 | 1 | 2 | 8 | 3.74 | 5.09 | 0.99 | 45.79 |
| 256 | 128 | 32 | 16 | 8 | 1 | 2 | 8 | 3.75 | 5.10 | 0.99 | 45.91 |
| 256 | 128 | 64 | 16 | 16 | 1 | 2 | 8 | 3.75 | 5.10 | 0.99 | 45.86 |
| 256 | 128 | 32 | 8 | 16 | 1 | 2 | 8 | 3.76 | 5.12 | 0.99 | 46.20 |
| 256 | 128 | 32 | 16 | 16 | 1 | 2 | 8 | 3.77 | 5.12 | 0.99 | 46.28 |
| 256 | 64 | 64 | 8 | 8 | 1 | 2 | 8 | 3.80 | 5.17 | 0.99 | 48.43 |
| 256 | 64 | 64 | 16 | 8 | 1 | 2 | 8 | 3.81 | 5.18 | 0.99 | 48.40 |
| 256 | 64 | 64 | 8 | 16 | 1 | 2 | 8 | 3.83 | 5.20 | 0.99 | 48.76 |
| 256 | 64 | 64 | 16 | 16 | 1 | 2 | 8 | 3.83 | 5.20 | 0.99 | 48.70 |
| 256 | 64 | 32 | 16 | 16 | 1 | 2 | 8 | 3.86 | 5.24 | 0.99 | 49.55 |
| 256 | 512 | 32 | 8 | 8 | 1 | 1 | 1 | 3.94 | 5.28 | 0.99 | 35.88 |
| 256 | 512 | 64 | 8 | 8 | 1 | 1 | 1 | 3.94 | 5.28 | 0.99 | 35.98 |
| 256 | 512 | 128 | 8 | 8 | 1 | 1 | 1 | 3.95 | 5.29 | 0.99 | 36.06 |
| 256 | 512 | 32 | 16 | 8 | 1 | 1 | 1 | 3.96 | 5.31 | 0.99 | 35.93 |
| 256 | 512 | 64 | 16 | 8 | 1 | 1 | 1 | 3.97 | 5.31 | 0.99 | 36.00 |
| 256 | 256 | 32 | 8 | 8 | 1 | 1 | 1 | 3.97 | 5.31 | 0.99 | 36.11 |
| 256 | 256 | 64 | 8 | 8 | 1 | 1 | 1 | 3.98 | 5.31 | 0.99 | 36.14 |
| 256 | 512 | 128 | 16 | 8 | 1 | 1 | 1 | 3.97 | 5.31 | 0.99 | 36.05 |
| 256 | 512 | 32 | 8 | 16 | 1 | 1 | 1 | 3.97 | 5.32 | 0.99 | 36.07 |
| 256 | 512 | 64 | 8 | 16 | 1 | 1 | 1 | 3.98 | 5.33 | 0.99 | 36.15 |
| 256 | 512 | 128 | 8 | 16 | 1 | 1 | 1 | 3.98 | 5.33 | 0.99 | 36.21 |
| 256 | 512 | 128 | 8 | 8 | 1 | 1 | 2 | 3.99 | 5.33 | 0.99 | 40.19 |
| 256 | 512 | 64 | 8 | 8 | 1 | 1 | 2 | 3.98 | 5.33 | 0.99 | 39.53 |
| 256 | 512 | 32 | 32 | 8 | 1 | 1 | 1 | 3.99 | 5.34 | 0.99 | 35.91 |
| 256 | 256 | 32 | 16 | 8 | 1 | 1 | 1 | 4.00 | 5.34 | 0.99 | 36.07 |
| 256 | 128 | 32 | 8 | 8 | 1 | 1 | 1 | 4.01 | 5.34 | 0.99 | 36.07 |
| 256 | 128 | 64 | 8 | 8 | 1 | 1 | 1 | 4.01 | 5.34 | 0.99 | 36.08 |
| 256 | 512 | 64 | 32 | 8 | 1 | 1 | 1 | 4.00 | 5.34 | 0.99 | 35.95 |
| 256 | 512 | 32 | 8 | 8 | 1 | 1 | 2 | 3.99 | 5.34 | 0.99 | 38.98 |
| 256 | 256 | 64 | 16 | 8 | 1 | 1 | 1 | 4.00 | 5.34 | 0.99 | 36.08 |
| 256 | 512 | 128 | 32 | 8 | 1 | 1 | 1 | 4.00 | 5.35 | 0.99 | 35.98 |
| 256 | 512 | 32 | 16 | 16 | 1 | 1 | 1 | 4.00 | 5.35 | 0.99 | 36.11 |
| 256 | 512 | 128 | 16 | 8 | 1 | 1 | 2 | 4.00 | 5.35 | 0.99 | 40.56 |
| 256 | 512 | 64 | 16 | 8 | 1 | 1 | 2 | 4.00 | 5.35 | 0.99 | 39.91 |
| 256 | 512 | 64 | 16 | 16 | 1 | 1 | 1 | 4.00 | 5.35 | 0.99 | 36.16 |
| 256 | 256 | 64 | 8 | 8 | 1 | 1 | 2 | 4.00 | 5.35 | 0.99 | 41.54 |
| 256 | 512 | 128 | 16 | 16 | 1 | 1 | 1 | 4.00 | 5.35 | 0.99 | 36.20 |
| 256 | 512 | 32 | 16 | 8 | 1 | 1 | 2 | 4.00 | 5.35 | 0.99 | 39.37 |
| 256 | 256 | 32 | 8 | 8 | 1 | 1 | 2 | 4.00 | 5.36 | 0.99 | 40.91 |
| 256 | 256 | 32 | 8 | 16 | 1 | 1 | 1 | 4.01 | 5.36 | 0.99 | 36.26 |
| 256 | 256 | 64 | 8 | 16 | 1 | 1 | 1 | 4.01 | 5.36 | 0.99 | 36.28 |
| 256 | 512 | 128 | 8 | 16 | 1 | 1 | 2 | 4.01 | 5.36 | 0.99 | 40.78 |
| 256 | 512 | 64 | 8 | 16 | 1 | 1 | 2 | 4.01 | 5.36 | 0.99 | 40.11 |
| 256 | 512 | 128 | 32 | 8 | 1 | 1 | 2 | 4.02 | 5.37 | 0.99 | 40.96 |
| 256 | 256 | 64 | 16 | 8 | 1 | 1 | 2 | 4.02 | 5.37 | 0.99 | 41.93 |
| 256 | 64 | 64 | 8 | 8 | 1 | 1 | 1 | 4.04 | 5.37 | 0.99 | 35.79 |
| 256 | 512 | 64 | 32 | 8 | 1 | 1 | 2 | 4.02 | 5.37 | 0.99 | 40.32 |
| 256 | 512 | 32 | 8 | 16 | 1 | 1 | 2 | 4.01 | 5.37 | 0.99 | 39.54 |
| 256 | 128 | 32 | 16 | 8 | 1 | 1 | 1 | 4.03 | 5.37 | 0.99 | 35.95 |
| 256 | 128 | 64 | 16 | 8 | 1 | 1 | 1 | 4.03 | 5.37 | 0.99 | 35.95 |
| 256 | 256 | 32 | 16 | 8 | 1 | 1 | 2 | 4.02 | 5.37 | 0.99 | 41.32 |
| 256 | 512 | 128 | 16 | 16 | 1 | 1 | 2 | 4.02 | 5.37 | 0.99 | 41.10 |
| 256 | 512 | 32 | 32 | 8 | 1 | 1 | 2 | 4.02 | 5.37 | 0.99 | 39.81 |
| 256 | 512 | 64 | 16 | 16 | 1 | 1 | 2 | 4.02 | 5.37 | 0.99 | 40.45 |
| 256 | 512 | 32 | 8 | 32 | 1 | 1 | 1 | 4.02 | 5.37 | 0.99 | 36.24 |
| 256 | 512 | 32 | 32 | 16 | 1 | 1 | 1 | 4.02 | 5.38 | 0.99 | 36.07 |
| 256 | 512 | 64 | 8 | 32 | 1 | 1 | 1 | 4.02 | 5.38 | 0.99 | 36.30 |
| 256 | 512 | 64 | 32 | 16 | 1 | 1 | 1 | 4.03 | 5.38 | 0.99 | 36.11 |
| 256 | 512 | 128 | 8 | 32 | 1 | 1 | 1 | 4.02 | 5.38 | 0.99 | 36.34 |
| 256 | 512 | 32 | 16 | 16 | 1 | 1 | 2 | 4.02 | 5.38 | 0.99 | 39.90 |
| 256 | 128 | 64 | 8 | 8 | 1 | 1 | 2 | 4.03 | 5.38 | 0.99 | 43.89 |
| 256 | 256 | 32 | 16 | 16 | 1 | 1 | 1 | 4.03 | 5.38 | 0.99 | 36.22 |
| 256 | 256 | 64 | 8 | 16 | 1 | 1 | 2 | 4.03 | 5.38 | 0.99 | 42.15 |
| 256 | 512 | 128 | 32 | 16 | 1 | 1 | 1 | 4.03 | 5.38 | 0.99 | 36.13 |
| 256 | 256 | 64 | 16 | 16 | 1 | 1 | 1 | 4.04 | 5.39 | 0.99 | 36.22 |
| 256 | 128 | 32 | 8 | 8 | 1 | 1 | 2 | 4.03 | 5.39 | 0.99 | 43.16 |
| 256 | 256 | 32 | 8 | 16 | 1 | 1 | 2 | 4.03 | 5.39 | 0.99 | 41.50 |
| 256 | 128 | 32 | 8 | 16 | 1 | 1 | 1 | 4.05 | 5.39 | 0.99 | 36.20 |
| 256 | 128 | 64 | 8 | 16 | 1 | 1 | 1 | 4.05 | 5.39 | 0.99 | 36.20 |
| 256 | 512 | 128 | 32 | 16 | 1 | 1 | 2 | 4.04 | 5.39 | 0.99 | 41.47 |
| 256 | 512 | 64 | 32 | 16 | 1 | 1 | 2 | 4.04 | 5.39 | 0.99 | 40.84 |
| 256 | 512 | 32 | 16 | 32 | 1 | 1 | 1 | 4.04 | 5.40 | 0.99 | 36.26 |
| 256 | 512 | 128 | 8 | 32 | 1 | 1 | 2 | 4.04 | 5.40 | 0.99 | 41.41 |
| 256 | 512 | 64 | 16 | 32 | 1 | 1 | 1 | 4.04 | 5.40 | 0.99 | 36.32 |
| 256 | 512 | 64 | 8 | 32 | 1 | 1 | 2 | 4.04 | 5.40 | 0.99 | 40.73 |
| 256 | 256 | 64 | 16 | 16 | 1 | 1 | 2 | 4.04 | 5.40 | 0.99 | 42.51 |
| 256 | 512 | 32 | 32 | 16 | 1 | 1 | 2 | 4.04 | 5.40 | 0.99 | 40.33 |
| 256 | 64 | 64 | 16 | 8 | 1 | 1 | 1 | 4.06 | 5.40 | 0.99 | 35.62 |
| 256 | 64 | 32 | 16 | 8 | 1 | 1 | 1 | 4.06 | 5.40 | 0.99 | 35.58 |
| 256 | 512 | 128 | 16 | 32 | 1 | 1 | 1 | 4.04 | 5.40 | 0.99 | 36.34 |
| 256 | 128 | 64 | 16 | 8 | 1 | 1 | 2 | 4.05 | 5.40 | 0.99 | 44.29 |
| 256 | 256 | 32 | 16 | 16 | 1 | 1 | 2 | 4.05 | 5.40 | 0.99 | 41.89 |
| 256 | 128 | 32 | 16 | 8 | 1 | 1 | 2 | 4.05 | 5.41 | 0.99 | 43.58 |
| 256 | 512 | 32 | 8 | 32 | 1 | 1 | 2 | 4.04 | 5.41 | 0.99 | 40.14 |
| 256 | 512 | 128 | 16 | 32 | 1 | 1 | 2 | 4.05 | 5.41 | 0.99 | 41.69 |
| 256 | 512 | 64 | 16 | 32 | 1 | 1 | 2 | 4.05 | 5.41 | 0.99 | 41.03 |
| 256 | 128 | 32 | 16 | 16 | 1 | 1 | 1 | 4.07 | 5.42 | 0.99 | 36.09 |
| 256 | 512 | 32 | 16 | 32 | 1 | 1 | 2 | 4.05 | 5.42 | 0.99 | 40.47 |
| 256 | 128 | 64 | 16 | 16 | 1 | 1 | 1 | 4.07 | 5.42 | 0.99 | 36.09 |
| 256 | 128 | 64 | 8 | 16 | 1 | 1 | 2 | 4.06 | 5.42 | 0.99 | 44.55 |
| 256 | 64 | 64 | 8 | 16 | 1 | 1 | 1 | 4.08 | 5.42 | 0.99 | 35.93 |
| 256 | 512 | 32 | 32 | 32 | 1 | 1 | 1 | 4.06 | 5.42 | 0.99 | 36.22 |
| 256 | 128 | 32 | 8 | 16 | 1 | 1 | 2 | 4.06 | 5.43 | 0.99 | 43.80 |
| 256 | 512 | 128 | 32 | 32 | 1 | 1 | 2 | 4.07 | 5.43 | 0.99 | 42.04 |
| 256 | 512 | 64 | 32 | 32 | 1 | 1 | 1 | 4.07 | 5.43 | 0.99 | 36.27 |
| 256 | 512 | 64 | 32 | 32 | 1 | 1 | 2 | 4.06 | 5.43 | 0.99 | 41.41 |
| 256 | 64 | 64 | 8 | 8 | 1 | 1 | 2 | 4.08 | 5.43 | 0.99 | 46.56 |
| 256 | 512 | 128 | 32 | 32 | 1 | 1 | 1 | 4.07 | 5.43 | 0.99 | 36.28 |
| 256 | 512 | 32 | 32 | 32 | 1 | 1 | 2 | 4.07 | 5.44 | 0.99 | 40.89 |
| 256 | 128 | 64 | 16 | 16 | 1 | 1 | 2 | 4.08 | 5.44 | 0.99 | 44.92 |
| 256 | 128 | 32 | 16 | 16 | 1 | 1 | 2 | 4.08 | 5.44 | 0.99 | 44.20 |
| 256 | 64 | 64 | 16 | 16 | 1 | 1 | 1 | 4.10 | 5.45 | 0.99 | 35.77 |
| 256 | 64 | 32 | 16 | 16 | 1 | 1 | 1 | 4.10 | 5.45 | 0.99 | 35.73 |
| 256 | 64 | 64 | 16 | 8 | 1 | 1 | 2 | 4.10 | 5.46 | 0.99 | 46.96 |
| 256 | 64 | 64 | 8 | 16 | 1 | 1 | 2 | 4.11 | 5.48 | 0.99 | 47.26 |
| 256 | 64 | 64 | 16 | 16 | 1 | 1 | 2 | 4.13 | 5.50 | 0.99 | 47.62 |
| 256 | 64 | 32 | 16 | 16 | 1 | 1 | 2 | 4.13 | 5.50 | 0.99 | 46.86 |
| 256 | 512 | 128 | 8 | 8 | 1 | 1 | 4 | 4.14 | 5.56 | 0.99 | 49.47 |
| 256 | 512 | 64 | 8 | 8 | 1 | 1 | 4 | 4.14 | 5.56 | 0.99 | 49.22 |
| 256 | 512 | 128 | 16 | 8 | 1 | 1 | 4 | 4.15 | 5.56 | 0.99 | 49.88 |
| 256 | 512 | 64 | 16 | 8 | 1 | 1 | 4 | 4.14 | 5.56 | 0.99 | 49.56 |
| 256 | 512 | 32 | 8 | 8 | 1 | 1 | 4 | 4.15 | 5.57 | 0.99 | 49.16 |
| 256 | 512 | 128 | 32 | 8 | 1 | 1 | 4 | 4.15 | 5.57 | 0.99 | 50.38 |
| 256 | 512 | 64 | 32 | 8 | 1 | 1 | 4 | 4.15 | 5.57 | 0.99 | 50.11 |
| 256 | 512 | 32 | 16 | 8 | 1 | 1 | 4 | 4.15 | 5.57 | 0.99 | 49.52 |
| 256 | 512 | 128 | 8 | 16 | 1 | 1 | 4 | 4.16 | 5.57 | 0.99 | 49.91 |
| 256 | 512 | 64 | 8 | 16 | 1 | 1 | 4 | 4.15 | 5.57 | 0.99 | 49.60 |
| 256 | 256 | 64 | 8 | 8 | 1 | 1 | 4 | 4.16 | 5.57 | 0.99 | 51.94 |
| 256 | 512 | 128 | 16 | 16 | 1 | 1 | 4 | 4.16 | 5.57 | 0.99 | 50.26 |
| 256 | 512 | 64 | 16 | 16 | 1 | 1 | 4 | 4.16 | 5.57 | 0.99 | 49.94 |
| 256 | 512 | 32 | 32 | 8 | 1 | 1 | 4 | 4.16 | 5.58 | 0.99 | 49.98 |
| 256 | 256 | 32 | 8 | 8 | 1 | 1 | 4 | 4.16 | 5.58 | 0.99 | 51.58 |
| 256 | 256 | 64 | 16 | 8 | 1 | 1 | 4 | 4.16 | 5.58 | 0.99 | 52.42 |
| 256 | 512 | 128 | 32 | 16 | 1 | 1 | 4 | 4.17 | 5.58 | 0.99 | 50.71 |
| 256 | 512 | 32 | 8 | 16 | 1 | 1 | 4 | 4.16 | 5.58 | 0.99 | 49.53 |
| 256 | 512 | 64 | 32 | 16 | 1 | 1 | 4 | 4.17 | 5.58 | 0.99 | 50.45 |
| 256 | 512 | 32 | 16 | 16 | 1 | 1 | 4 | 4.16 | 5.58 | 0.99 | 49.82 |
| 256 | 256 | 32 | 16 | 8 | 1 | 1 | 4 | 4.17 | 5.58 | 0.99 | 52.13 |
| 256 | 512 | 32 | 32 | 16 | 1 | 1 | 4 | 4.17 | 5.59 | 0.99 | 50.27 |
| 256 | 256 | 64 | 8 | 16 | 1 | 1 | 4 | 4.17 | 5.59 | 0.99 | 52.43 |
| 256 | 512 | 128 | 8 | 32 | 1 | 1 | 4 | 4.18 | 5.59 | 0.99 | 50.43 |
| 256 | 512 | 64 | 8 | 32 | 1 | 1 | 4 | 4.18 | 5.60 | 0.99 | 50.09 |
| 256 | 512 | 128 | 16 | 32 | 1 | 1 | 4 | 4.18 | 5.60 | 0.99 | 50.70 |
| 256 | 256 | 64 | 16 | 16 | 1 | 1 | 4 | 4.18 | 5.60 | 0.99 | 52.84 |
| 256 | 512 | 64 | 16 | 32 | 1 | 1 | 4 | 4.18 | 5.60 | 0.99 | 50.40 |
| 256 | 256 | 32 | 8 | 16 | 1 | 1 | 4 | 4.18 | 5.60 | 0.99 | 52.10 |
| 256 | 512 | 128 | 32 | 32 | 1 | 1 | 4 | 4.19 | 5.61 | 0.99 | 51.09 |
| 256 | 256 | 32 | 16 | 16 | 1 | 1 | 4 | 4.18 | 5.61 | 0.99 | 52.56 |
| 256 | 512 | 32 | 8 | 32 | 1 | 1 | 4 | 4.18 | 5.61 | 0.99 | 49.99 |
| 256 | 512 | 64 | 32 | 32 | 1 | 1 | 4 | 4.19 | 5.61 | 0.99 | 50.85 |
| 256 | 512 | 32 | 16 | 32 | 1 | 1 | 4 | 4.19 | 5.61 | 0.99 | 50.19 |
| 256 | 512 | 32 | 32 | 32 | 1 | 1 | 4 | 4.20 | 5.62 | 0.99 | 50.68 |
| 256 | 128 | 64 | 8 | 8 | 1 | 1 | 4 | 4.19 | 5.62 | 0.99 | 55.65 |
| 256 | 128 | 32 | 8 | 8 | 1 | 1 | 4 | 4.19 | 5.62 | 0.99 | 55.33 |
| 256 | 128 | 64 | 16 | 8 | 1 | 1 | 4 | 4.20 | 5.63 | 0.99 | 56.10 |
| 256 | 128 | 32 | 16 | 8 | 1 | 1 | 4 | 4.20 | 5.63 | 0.99 | 55.86 |
| 256 | 128 | 64 | 8 | 16 | 1 | 1 | 4 | 4.22 | 5.65 | 0.99 | 56.16 |
| 256 | 128 | 32 | 8 | 16 | 1 | 1 | 4 | 4.22 | 5.65 | 0.99 | 55.87 |
| 256 | 128 | 64 | 16 | 16 | 1 | 1 | 4 | 4.22 | 5.66 | 0.99 | 56.54 |
| 256 | 128 | 32 | 16 | 16 | 1 | 1 | 4 | 4.23 | 5.66 | 0.99 | 56.32 |
| 256 | 64 | 64 | 8 | 8 | 1 | 1 | 4 | 4.25 | 5.70 | 0.99 | 59.81 |
| 256 | 64 | 64 | 16 | 8 | 1 | 1 | 4 | 4.26 | 5.72 | 0.99 | 60.19 |
| 256 | 64 | 64 | 8 | 16 | 1 | 1 | 4 | 4.28 | 5.73 | 0.99 | 60.31 |
| 256 | 64 | 64 | 16 | 16 | 1 | 1 | 4 | 4.29 | 5.75 | 0.99 | 60.62 |
| 256 | 64 | 32 | 16 | 16 | 1 | 1 | 4 | 4.29 | 5.75 | 0.99 | 60.63 |
| 256 | 256 | 64 | 8 | 8 | 1 | 1 | 8 | 4.47 | 6.04 | 0.99 | 61.56 |
| 256 | 256 | 64 | 16 | 8 | 1 | 1 | 8 | 4.47 | 6.04 | 0.99 | 61.82 |
| 256 | 256 | 32 | 8 | 8 | 1 | 1 | 8 | 4.47 | 6.04 | 0.99 | 61.74 |
| 256 | 256 | 32 | 16 | 8 | 1 | 1 | 8 | 4.47 | 6.04 | 0.99 | 61.95 |
| 256 | 256 | 64 | 8 | 16 | 1 | 1 | 8 | 4.48 | 6.05 | 0.99 | 61.88 |
| 256 | 256 | 64 | 16 | 16 | 1 | 1 | 8 | 4.48 | 6.05 | 0.99 | 62.12 |
| 256 | 256 | 32 | 8 | 16 | 1 | 1 | 8 | 4.48 | 6.05 | 0.99 | 62.00 |
| 256 | 256 | 32 | 16 | 16 | 1 | 1 | 8 | 4.49 | 6.05 | 0.99 | 62.17 |
| 256 | 128 | 64 | 8 | 8 | 1 | 1 | 8 | 4.52 | 6.10 | 0.99 | 65.03 |
| 256 | 128 | 32 | 8 | 8 | 1 | 1 | 8 | 4.52 | 6.11 | 0.99 | 65.01 |
| 256 | 128 | 64 | 16 | 8 | 1 | 1 | 8 | 4.52 | 6.11 | 0.99 | 65.29 |
| 256 | 128 | 32 | 16 | 8 | 1 | 1 | 8 | 4.52 | 6.11 | 0.99 | 65.32 |
| 256 | 128 | 64 | 8 | 16 | 1 | 1 | 8 | 4.53 | 6.12 | 0.99 | 65.38 |
| 256 | 128 | 32 | 8 | 16 | 1 | 1 | 8 | 4.53 | 6.12 | 0.99 | 65.39 |
| 256 | 128 | 64 | 16 | 16 | 1 | 1 | 8 | 4.53 | 6.13 | 0.99 | 65.60 |
| 256 | 128 | 32 | 16 | 16 | 1 | 1 | 8 | 4.54 | 6.13 | 0.99 | 65.65 |
| 256 | 64 | 64 | 8 | 8 | 1 | 1 | 8 | 4.60 | 6.21 | 0.99 | 68.36 |
| 256 | 64 | 64 | 16 | 8 | 1 | 1 | 8 | 4.60 | 6.22 | 0.99 | 68.55 |
| 256 | 64 | 64 | 8 | 16 | 1 | 1 | 8 | 4.61 | 6.23 | 0.99 | 68.68 |
| 256 | 64 | 64 | 16 | 16 | 1 | 1 | 8 | 4.62 | 6.24 | 0.99 | 68.83 |
| 256 | 64 | 32 | 16 | 16 | 1 | 1 | 8 | 4.62 | 6.25 | 0.99 | 69.12 |

**Table S5.** Optimization of kernel weights and GPR parameters ($\sigma_n$ and $\zeta$) for multi-scale kernel prediction of $^{13}C$ chemical shifts. The optimization was carried out on the CSD-2k set, using 3-fold cross validation. For each configuration are reported the corresponding mean absolute error (MAE), root-mean-square error (RMSE), the R-squared ($R^2$) coefficient and the supremum (SUP). In bold is shown the set of parameters that we selected.

| Multi-Scale Kernel Weights | | | | | | $\zeta$ | $\sigma_n$ | MAE (ppm) | RMSE (ppm) | $R^2$ | SUP (ppm) |
|---|---|---|---|---|---|---|---|---|---|---|---|
| $r_c$ =2Å | $r_c$ =3Å | $r_c$ =4Å | $r_c$ =5Å | $r_c$ =6Å | $r_c$ =7Å | | | | | | |
| **256** | **128** | **32** | **8** | **8** | **1** | **2** | **0.1** | **9.96** | **14.25** | **0.99** | **132.00** |
| 256 | 128 | 32 | 8 | 8 | 1 | 2 | 1 | 9.95 | 14.28 | 0.99 | 133.20 |
| 256 | 128 | 32 | 8 | 8 | 1 | 4 | 1 | 9.67 | 14.35 | 0.98 | 137.10 |
| 256 | 128 | 32 | 8 | 8 | 1 | 4 | 0.1 | 9.74 | 14.36 | 0.98 | 135.69 |
| 256 | 128 | 32 | 8 | 8 | 1 | 2 | 0.01 | 10.36 | 14.72 | 0.98 | 133.68 |
| 256 | 128 | 32 | 8 | 8 | 1 | 4 | 0.01 | 10.34 | 15.12 | 0.98 | 141.99 |
| 256 | 128 | 32 | 8 | 8 | 1 | 1 | 1 | 11.68 | 15.94 | 0.98 | 115.98 |
| 256 | 128 | 32 | 8 | 8 | 1 | 1 | 0.1 | 14.59 | 19.36 | 0.97 | 114.82 |
| 256 | 128 | 32 | 8 | 8 | 1 | 1 | 0.01 | 14.64 | 19.43 | 0.97 | 118.61 |

**Table S6.** Optimization of kernel weights and GPR parameters ($\sigma_n$ and $\zeta$) for multi-scale kernel prediction of $^{15}N$ chemical shifts. The optimization was carried out on the CSD-2k set, using 3-fold cross validation. For each configuration are reported the corresponding mean absolute error (MAE), root-mean-square error (RMSE), the R-squared ($R^2$) coefficient and the supremum (SUP). In bold is shown the set of parameters that we selected.

| Multi-Scale Kernel Weights | | | | | | $\zeta$ | $\sigma_n$ | MAE (ppm) | RMSE (ppm) | $R^2$ | SUP (ppm) |
|---|---|---|---|---|---|---|---|---|---|---|---|
| $r_c$ =2Å | $r_c$ =3Å | $r_c$ =4Å | $r_c$ =5Å | $r_c$ =6Å | $r_c$ =7Å | | | | | | |
| **256** | **128** | **32** | **8** | **8** | **1** | **2** | **5** | **14.80** | **20.09** | **0.99** | **117.92** |
| 256 | 128 | 32 | 8 | 8 | 1 | 2 | 10 | 14.62 | 20.27 | 0.99 | 154.74 |
| 256 | 128 | 32 | 8 | 8 | 1 | 4 | 10 | 14.67 | 20.35 | 0.99 | 145.67 |
| 256 | 128 | 32 | 8 | 8 | 1 | 4 | 5 | 14.84 | 20.39 | 0.99 | 132.32 |
| 256 | 128 | 32 | 8 | 8 | 1 | 4 | 2 | 14.97 | 20.49 | 0.99 | 129.65 |
| 256 | 128 | 32 | 8 | 8 | 1 | 2 | 2 | 15.26 | 20.52 | 0.99 | 110.01 |
| 256 | 128 | 32 | 8 | 8 | 1 | 4 | 1 | 15.04 | 20.54 | 0.99 | 127.63 |
| 256 | 128 | 32 | 8 | 8 | 1 | 2 | 1 | 15.43 | 20.68 | 0.99 | 108.26 |
| 256 | 128 | 32 | 8 | 8 | 1 | 1 | 5 | 16.22 | 22.20 | 0.99 | 149.10 |
| 256 | 128 | 32 | 8 | 8 | 1 | 1 | 10 | 16.27 | 23.00 | 0.99 | 201.11 |
| 256 | 128 | 32 | 8 | 8 | 1 | 1 | 2 | 17.55 | 23.52 | 0.99 | 141.85 |
| 256 | 128 | 32 | 8 | 8 | 1 | 1 | 1 | 19.43 | 25.78 | 0.98 | 147.58 |

**Table S7.** Optimization of kernel weights and GPR parameters ($\sigma_n$ and $\zeta$) for multi-scale kernel prediction of $^{17}O$ chemical shifts. The optimization was carried out on the CSD-2k set, using 3-fold cross validation. For each configuration are reported the corresponding mean absolute error (MAE), root-mean-square error (RMSE), the R-squared ($R^2$) coefficient and the supremum (SUP). In bold is shown the set of parameters that we selected.

## V. Comparison to Experiments

Comparison between $^1$H experimental chemical shifts and $^1$H chemical shifts calculated with ShiftML were carried out analysing 68 chemical shifts obtained from 6 crystal structures. The names, IUPAC IDs, CSD refererence codes (when available) and references to the experimental NMR data of the analysed crystal structures are the following:

(i) Naproxen, (2S)-2-(6-Methoxy-2-naphthyl)propanoic acid, COYRUD11, Ref. 22
(ii) Uracil, Pyrimidine-2,4(1H,3H)-dione, URACIL, Ref. 23
(iii) Co-crystal of 3,5-dimethylimidazole and 4,5-dimethylimidazole, Ref. 24
(iv) Theophylline, 1,3-Dimethyl-3,7-dihydro-1H-purine-2,6-dione, BAPLOT0, Ref. 3
(v) Cocaine, methyl (1R,2R,3S,5S)-3- (benzoyloxy)-8-methyl-8-azabicyclo[3.2.1] octane-2-carboxylate, COCAIN10, Ref. 3
(vi) AZD8329, 4-[4-(2-adamantylcarbamoyl)-5-tert-butylpyrazol-1-yl]benzoic acid, Ref. 4

The crystal structures (i-iv) were obtained from Ref. 25, where the experimentally determined crystal structures were subjected to all-atom geometry optimization with fixed lattice parameters, as described in the reference. Crystal structures (v) and (vi) were obtained from Refs. [3] and [4] respectively.

We used assigned chemical shift values and we account for rotational dynamics of the methyl groups by averaging the chemical shift values of the three $^1$H positions to a single value for each methyl group. The calculated chemical shieldings $\sigma$ are converted to the corresponding chemical shifts $\delta$ through the relationship $\delta = \sigma_{\text{ref}} - \beta\sigma$. For each structure, we calculated the value of $\sigma_{\text{ref}}$ and $\beta$ by a linear regression between calculated and experimental shifts. The calculations were carried out in MATLAB using a home-written script. The chemical structures, together with the assigned experimental chemical shifts and the parameters for conversion between shieldings and shifts are shown in Figure S9 and Table S8.

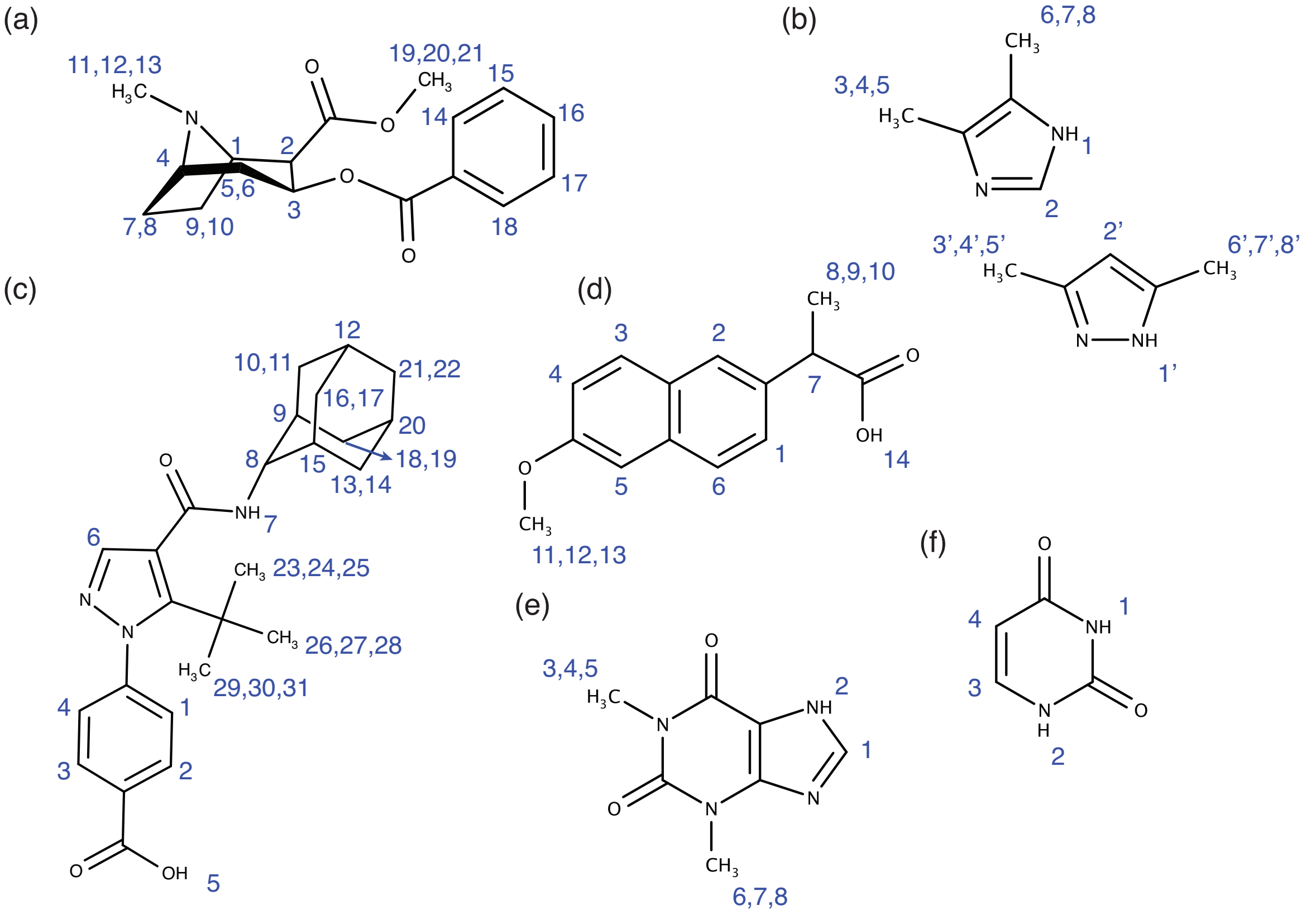

**Figure S9**. Chemical structure of cocaine (a), 3,5-dimethylimidazole and 4,5-dimethylimidazole (b), AZD8329 (c), naproxen (d), theophylline (e) and uracil (f), and the labelling scheme used here.

**Naproxen**

| Atom Label | Experimental $^1H$ δ (ppm) | ShiftML $^1H$ δ (ppm) |
|---|---|---|
| 1 | 7 | 6.58 |
| 2 | 6.1 | 5.78 |
| 3 | 3.8 | 3.79 |
| 4 | 4.5 | 4.52 |
| 5 | 4.1 | 4.69 |
| 6 | 5.9 | 5.29 |
| 7 | 3.2 | 3.20 |
| 8,9,10 | 1.8 | 1.83 |
| 11,12,13 | 2.3 | 2.69 |
| 14 | 11.5 | 11.83 |
| $\sigma_{ref}$ 24.77 | | β 0.79 |

**Uracil**

| Atom Label | Experimental $^1H$ δ (ppm) | ShiftML $^1H$ δ (ppm) |
|---|---|---|
| 1 | 11.2 | 11.15 |
| 2 | 10.8 | 10.82 |
| 3 | 7.5 | 7.60 |
| 4 | 6 | 5.93 |
| $\sigma_{ref}$ 24.09 | | β 0.75 |

**3,5-dimethylimidazole & 4,5-dimethylimidazole**

| Atom Label | Experimental $^1H$ δ (ppm) | ShiftML $^1H$ δ (ppm) |
|---|---|---|
| 2 | 4.8 | 4.81 |
| 3,4,5 | 1.4 | 0.85 |
| 6,7,8 | 0.7 | 1.01 |
| 2 | 13 | 12.76 |
| 2’ | 5.2 | 6.05 |
| 3’,4’,5’ | 1.5 | 1.44 |
| 6’,7’,8’ | 1.4 | 1.19 |
| 1’ | 15 | 14.89 |
| $\sigma_{ref}$ 29.07 | | β 0.95 |

**Theophylline**

| Atom Label | Experimental $^1H$ δ (ppm) | ShiftML $^1H$ δ (ppm) |
|---|---|---|
| 1 | 7.7 | 7.18 |
| 2 | 14.6 | 14.77 |
| 3,4,5 | 3.4 | 3.56 |
| 6,7,8 | 3.4 | 3.59 |
| $\sigma_{ref}$ 26.53 | | β 0.84 |

**Cocaine**

| Atom Label | Experimental $^1H$ δ (ppm) | ShiftML $^1H$ δ (ppm) |
|---|---|---|
| 1 | 3.76 | 3.95 |
| 2 | 3.78 | 3.21 |
| 3 | 5.63 | 6.11 |
| 4 | 3.32 | 2.98 |
| 5 | 3.49 | 3.73 |
| 6 | 3.06 | 2.55 |
| 7 | 2.91 | 2.69 |
| 8 | 3.38 | 3.19 |
| 9 | 2.56 | 2.44 |
| 10 | 2.12 | 2.36 |
| 11,12,13 | 1.04 | 1.80 |

**AZD8329**

| Atom Label | Experimental $^1H$ δ (ppm) | ShiftML $^1H$ δ (ppm) |
|---|---|---|
| 1 | 6.92 | 6.53 |
| 2 | 8.69 | 7.85 |
| 3 | 9.01 | 9.35 |
| 4 | 8.47 | 7.91 |
| 5 | 15.37 | 15.95 |
| 6 | 7.73 | 7.60 |
| 7 | 9.64 | 9.37 |
| 8 | 2.90 | 2.79 |
| 9 | 1.78 | 1.98 |
| 10 | 1.88 | 1.79 |
| 11 | 1.88 | 2.61 |

| | | | | | | | |
|---|---|---|---|---|---|---|---|
| 14 | | 8.01 | 8.40 | 12 | | 1.8 | 1.68 |
| 15 | | 8.01 | 7.39 | 13 | | 1.6 | 1.28 |
| 15 | | 8.01 | 7.66 | 14 | | 0.44 | 0.87 |
| 17 | | 8.01 | 8.09 | 15 | | 1.54 | 1.94 |
| 18 | | 8.01 | 8.04 | 16 | | 1.88 | 2.76 |
| 19,20,21 | | 3.78 | 4.28 | 17 | | 1.88 | 1.69 |
| | | | | 18 | | 0.8 | 1.21 |
| | | | | 19 | | 0.8 | 0.43 |
| | | | | 20 | | 1 | 1.42 |
| | | | | 21 | | 1.74 | 1.47 |
| | | | | 22 | | 1.74 | 1.21 |
| | | | | 23,24,25 | | 0.73 | 0.84 |
| | | | | 26,27,28 | | 0.73 | 1.02 |
| | | | | 29,30,31 | | 0.73 | 0.14 |
| $\sigma_{ref}$ | 30.08 | $\beta$ | 0.96 | $\sigma_{ref}$ | 28.39 | $\beta$ | 0.91 |

**Table S8**. Experimental and calculated chemical shifts of naproxen, uracil, the co-crystal of 3,5-dimethylimidazole and 4,5-dimethylimidazole, theophylline, cocaine and AZD8329. Labelling scheme is given in Figure S3. When more than one atom corresponds to a single chemical shift value, their values were averaged

## VI. Structures and Chemical Shifts of the CSD-6 Set

For all of the structures in CSD-6 we removed atoms with partial occupations, leaving only one conformation in the structure file. Missing Hydrogen atoms were added with the program IQmol. Prior to the chemical shift calculations all the coordinates of the structures were DFT optimized using the same parameters as for the CSD-2k set. Chemical shieldings predicted for each structures are given as separate .xyz files, named according to the corresponding CSD Refcodes. For each atom we report: atom type, Cartesian coordinates and predicted isotropic chemical shielding.

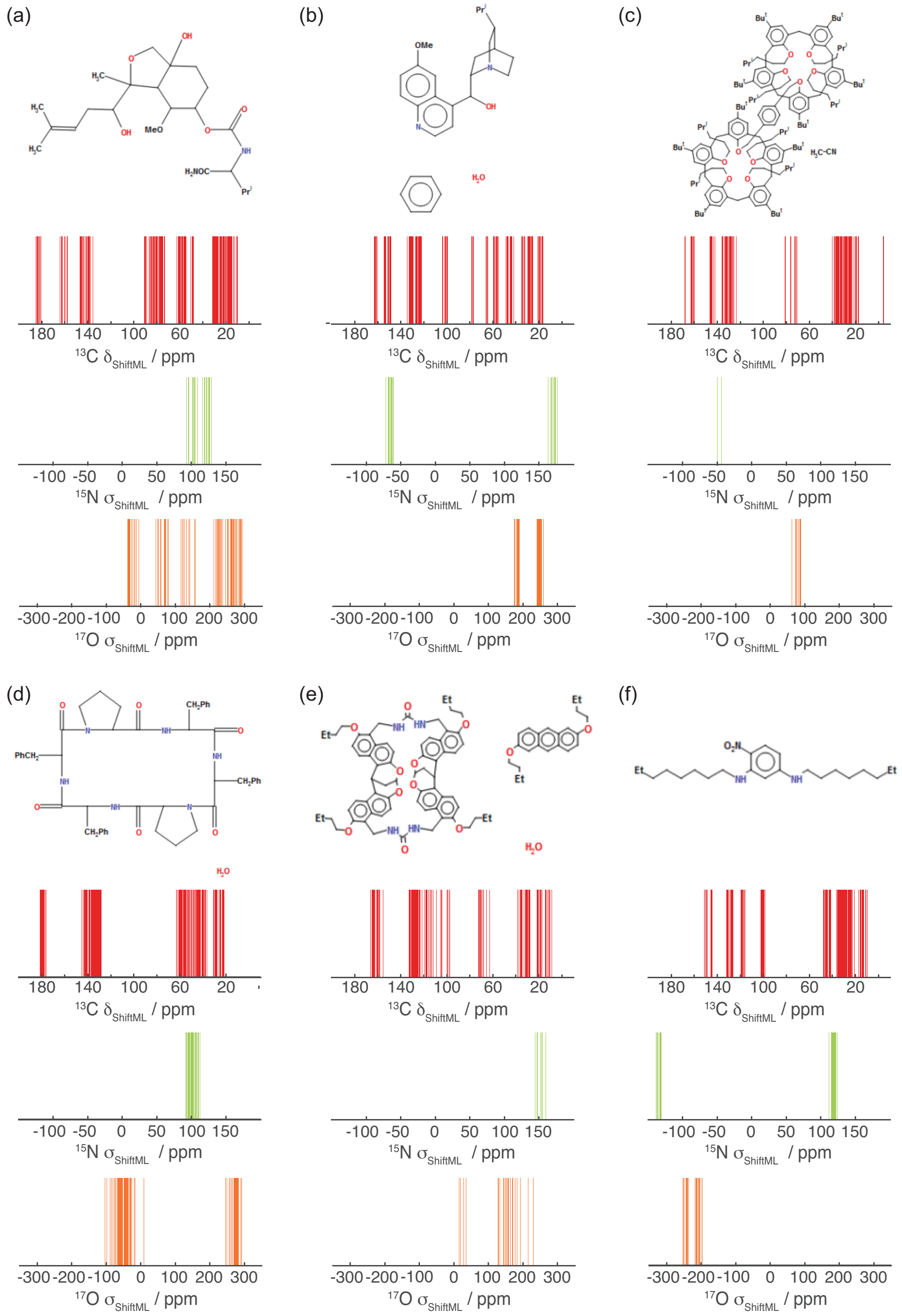

**Figure S10.** Chemical formula and corresponding $^{13}$C, $^{15}$N and $^{17}$O NMR spectra predicted using ShiftML of the six large molecular crystals with CSD Refcodes: (a) CAJVUH,[26] $N_{atoms}$ = 828, (b) RUKTOI,[27] $N_{atoms}$ = 768, (c) EMEMUE,[28] $N_{atoms}$ = 860, (d) GOKXOV,[29] $N_{atoms}$ = 945, (e) HEJBUW,[30] $N_{atoms}$ = 816, (f) RAYFEF,[31] $N_{atoms}$ = 1,584.

## VII. CSD-2k Set

The CSD-2k set contains the structures with the following CSD Refcodes (in order of FPS selection):
XISJIT, COVSEM, NOETNA02, KESVOT, AMFORM, GUMMUW04, ETDIAM17, LAKRAR, ZUTKEG, CAFMIF, DAFGAS, CUPZEU, NOZKES, RIZFAI, VINYLC, NTROMA01, TILJOP, RIWTIB01, TETZOL, DOJKUJ, FOKSUT, PARBAC03, UREAOH12, IMAZOL33, DHNAPH17, DAYHAN, EJIQEU02, JOJWOU03, BCBANN06, GLYCIN35, LEPFES, HIXHIF02, NUQPOF, CILBII, OTEBEV, CUZJEO, MOKPUX, MOSGEI, CALFOM, HUQRAM, AMACAM01, CTMTNA06, AJAPIL03, NAYPAF, KUTHAI, DOVCAT, KEMZIL01, AVAKAK, VIWYEH, TAPZAN, KICCOO11, ZIKMOV01, DIOXAZ10, OGIQEB, CITVIL, QAHVUQ, MOMFAV, TAYHOR, GANHUY, GACLEE, CAMGAA, YEKHUT, SIBVUU, VANXIS, QIQKIK, FEDCOI, WEVVOJ01, MEVDAT, XEXVEB, EMUFAU, BECJEY, QOPKUB, FUSVAQ01, LUKCIF, FUNSOW, WIDJOI, FIZZOD, KULMUA, IMAZOL15, ARCLAM03, LUVPAU, UMERAF, NAFZEC, CBOHAZ02, MALIAC02, MUYHIY, COXYAR01, INIJEW, QOXPEZ, MOZSEZ, PODHOH, QQQFGS10, CERMOB, BZDMAZ05, PHGLOL01, ZACSOO, DMDFHZ, LUVTOL, SOJMOT, DUFBOV10, QUDNOU, DAJVUG01, MOQDED, ZDGLPN, GOKNOK, UJEREG01, ZAXLIV, GEFBAV, INEZUY, TEDROL, HYDRZN11, COWHUT, CITSED10, NANQUO02, FAYBUC, CSURCD10, AHEYAO, CALVAM, TIGBIU, FESNEW15, WOSGIU, ZTNONX, UDAYUT01, ZUNTOR, SEDTUQ09, NADHOP01, SAFKAL, CIFROY, ABEWAG, BOQQUT08, PYRTHA10, MELAMI01, KECYBU02, NEPGCL02, FEKVOG, UREAXX27, NACXEU01, TICJAS, EJEHEG, QUVFIX, FOBSEU, MOYHAJ, COTPAC, GEZBAO, MUGNUX, NIKVUZ, PUBMUU23, HEHVAR, IMIWOQ, ACETAC09, EKUJUR, JAPBIL, CUPYUJ, JABCIB, SEQBUN, LETGIA10, MEOXAL, ADALAU, METACM02, CIWMAW01, KEMDOW, OXHEPA, GACGAU, CILWUP11, JEDSUG, TALSOP, EBUKUH, YARZUN03, VOQXAA, FUJTUZ, BONGIW, BICMAD, ACXMPR, ACRLAC03, TETZIB, MEPCHX, IVEZEO, JEXZUH, FEMCEF, AJAQAE, IKIGUE, UREANT03, OFUKOR, XESHOT, DUMQAF, RIZZEH, GOFPUN02, LEGZII, WOCREL, VOBDOF, WUFTUO, MUQHOW, ECADER, BUWCAX, OXAZDO, VIPKIO03, FEXLEZ, YAHCIV, MNPYDO29, RIKPEG, XUSGOI, PORPIN01, LAKDAD, GUHROQ01, JACPOS01, EFUMEY, CIKWOJ, ESIWUZ, YIFXIV, AGUSEB, ATOXAL03, BUKJOG10, ADEZUF, RAVFEA, GLYCNI, APENIU, ZEFPUX01, PIDHIU, GAQWAX, ATZCBX, YOKJOZ, SALOXM09, NUMXAV, WETGAE, NOFBIT, FORRUB, KAXWIQ, DNBOIM, PARGIZ, HEMLAM, QODBOC, ALUQIH, GLYCIN81, ESOYUI, KAYWIR01, LONJAA, QORROF, BIBVAK, OSEXEQ, MENFXN, PECYON, FOHMUK, QIGVAD, TAFKUG, FEHVUK, QONRAO, PELJOG, FORAMO, QIHHIY, YABFEO, EFOBOR, MEUREA, ZERZIH, PYRDHN01, YECVUX, AMTETZ, TURPUR01, DIHIXL10, FEPGEM, RIWJEM, WIHJAZ, SAGLAQ, WEZCAG, HONSIO, KOTWOH, GEJSET, PRMDIN20, HAMCEF, OMIWIQ, AMOXAM, FEPNAP, HIFKUC01, NTRACD, OKEBAJ, OBUPUY, URPRBN10, RIHJIB, AXOSOW, NOHTAZ, PYRDNA06, GIPPAW, IKEGAH, GEXGIZ01, NIPYAZ, JUNQIS, AFZPYM, LETGIA03, TRAZOL01, DZBASK, DUHJIB, AMTETA, IWERUX, PYRAZI06, URICAC, FAFKEC, CYAMPD01, YOVVIP, XOKVIE, IMEGIR09, ZESSOF, KUKHOP, XOCZUL, BUXWIA, GOVZEW01, UDOZUJ, LEVXOB, LUHSIR, CISPOJ, OMIWEM, KUPCOP, BIJJIP, EKUKEC, BATWOK, UNUSED, KUQTIB, NTRGUA01, QAGDAG, ITIPEG, GEDXOD, DUROQN10, YANREM01, BIDJIH, ULEZES, WIKQOX, KABBEW, KENMOF, ZAVTEV, XZBTZO, MAZGAX, HIHJEN, NEPBEQ, PINXAL, ALXANM01, ANIMUH, HOZPET01, XAYYUS, REWHOR, VAZCEG, QAKKOE, IXAQAB, THZTAZ, GELWOL, NBFURX, JEMSUP, NEHJER01, LALNIN28, CASYIG, NUJLUB, ASUWOB01, NOPPOY, BEMYEY, MAMHOX, PUFFOL01, OKUXUP, LABJON01, XIQCEF, ZUNTAD, LAGHOQ, WULYIN, NIBZAM, CREATH04, NEZNAH, EDAWIP, IFUVOW, UTAKUX, LIVZIA, GUMMUW10, IBURAM, BINROF, MTZOPZ10, XIHLIL, VITBAC, YARYAS, JUVMUI, XIBTUY, MEWSUE, VOFDEZ, PYZTAZ, WETFOR, YOTSUV, FONMEB, ONAVEF, GIVFIC, NEBPUF, YIQXAY, PIMELA08, NAXTIQ, AMMALA, EWAMIZ, CPRDCA, CYHEXO, JOHBUD, WABNEV, DEBSOS, HEHQOC, LAYSOV, MOPPUC01, UFORUE, QARWAI, DLGYAH, DUTTAN10, SAPDAQ, JOPNAD, IBOLUG, EXIGAU, DUNVEN, VOFVAN26, EBUKOB, GUKVOX, DOXXAP, HAWYIN, PYRZIN14, ARADEN10, TESDOL01, XADBOW, VEHXAH, JABLUW01, LUSNOC, SIGFOE01, XYLTOL02, VEXREV, HOYMAJ, TUQDUF, NOCNEA, HINXAF, SALOXM04, DOZXEW, PEKSAZ, KUQTEX, MSORAM10, YUZTET, DAZDOZ, QOQGOT, SOJCUP, ITIXUE01, FIBYEW, MUTVEE, TUBYIY, QAJQAV, FUMRAM01, NAPHET, BZAMID06, GUNTEP, OHAHAI, ECARBM10, TEDFEP, SUFHUY, GUHOXM02, BOXBEV, POFLAX01, OXAZIL10, GADZER, HEXDAQ, PUFJIJ, LEGLEQ, DUYREU, GIRRUV, MUDDEU, QIMFAT, SECMOE, ODEJEN, COTTAI, JORBUN, YUWMUY, QUGGAA, HUSTIY, FIBGEE, KAMFAH, SUSYAI, BUFMOG, COWJAB, DUHTUW, UCOTEM, DNITPY, AMACET, HDNBPP, KUHDIA, QEXJIN, CEVGIT, DNPMTA, BZDMAZ16, LACFON, DNPIMZ, YODHIJ01, LESLAX, LENBAH, HIVMOP, AQUHAX, JUVMOC, WOWFUL, THYMIN02, ZIPSIC, OJOLAC, DAVARO, RIVKAJ, JOWWOH, XAVZUP, METHOL04, BZCBUO, JUBSOQ, ZOVHID, BARBAD07, KOVGIL, QOYJOD06, KUQCAB, ACAMOX, ZZZJII01, KIXXEW, HXIBAC, CUMBUH, NEZMUA, KOBYOP, BISMEV10, NOACTD, VUBQAM01, CUZJOY, VODWIU, VONNET, VOFVAN23, OGOKEC, FEHDAX01, ACARBM, FIXPUX, AYAHUE, EMUPAE, GEXMED, TIJKEC, KECPIR, YUSWAM, ACEMID07, FOLVIL, OGAJOV, RATMEF01, ZEXKEU, QARVUB, ZIVTOP, JESNAX, FIMWII, WIGXAN, PUPBAD02, VUCBAW, KEPKIZ, RIGJOG, NURQEW, WEPPUE, LOXBAD, XUDRUL, SAGRUN, LUNYID, LUQPOE01, FOLXIN, POQQIW, DALGON03, OROMIT, TECNUL, NUPTIC, GAFRAJ, ZZZTXI11, GATCAG, PIKYEN, BISKET, GLYCIN18, WOQPEY, TAPHEY, WOLGIN, OREHOK, KEHYAZ, BARBAC01, TOCLUU, PEJBAJ, QAGDEK, KATKIA, HABFUL, TATNBZ, ACETAC01, ZZZRJG01, CDECOL11, CINSOH, TRIRED, PMXCDO10, ZACSII, TROPOL10, IJEPOE, DALQOW, TACQUJ, NMALAM01, ACIDOH, ZZZRTW01,

KIJNUM, POWBEI, CPCOHA, PULGOU, IYINAF, MOLMUV, SAWJUW, TIQROC, NIGNOH, BELBUP02, JEMFUD, HUKROU, HEWQOQ, NOCNOK, YUYWEU, METAMI, MEQPAZ, JEZZUJ, UGEBOZ, KUCPED, OXAYAO, DACETA11, BECYAL, UVIZEG, EVIJUP, NPHDZB, UPONAQ, SOQNOC, GESNUN, QEYCOM, ZENMUA, ITEGIX01, KEPFUI, JIXGUS, XENQUD, FAQROE, SERVOA, OJUMIR, WOFVAQ, XELNAE, VAZMEP, VANCIX, XUDSOG, TEZPDZ10, XACSOL, DOMQUS, SESCAV, CANXIA, KOMHAV, BUVBUR, BAHSUY, TIMCHX, HEWKAW, JIPFAQ, XACCOW, DEBMOM, BUYRUI02, ZOTDAP, KEHWUR, TRURET20, DIPMUK, FENCIK, FOACAM, JUVKAM, HUMNEI, KOWTEV, DIPICA10, XIMKUA, PIZGUA, JOWGAF01, DODCHD, YICMIG, QEKSOP, LSERMH15, KOHMOJ, MUHFOK, MLEICA, CIZSEL, YAGQOM, YIHHON13, PUBLON, YITBOS, JIHVEB, XEPNUD, MUKLIO, DUROQN11, TRMTRZ, DOPLOL05, POWBAE, COGDEH, NIPLED, NILQAC, QIGCIS, FUTFIJ, VOHCAY, JOHKIC, BIZBET, WAJQAC, MFZCHZ, WIKBAT, FEDCUO, IPRPOL03, YALLOO, VIOLME05, GAMGIL, ROBYAI, HABTEJ, CAKZAP, KOVHIM, UPANIJ, MALNAC02, QIWJAI, TAFKAM, BAKFUO, AJEREM, FIBWUI, IQUJEK, HAMCIJ, VASDUO, QONVIZ, EREREY, COYZEX, FIBXUL, QONRUI, EMIMUI, QOBHIA, KIKVAC01, COHLUI, UCONIJ, LAJZAZ, HOQHAW, CAHOAM, EYOVIX, MTETAC01, CTMTNA12, AXUDED03, BUSFUQ, XESRIW, JINYIP, MISGAX, GEJCAB, TURRAZ, ECIBIB, DOGQUL, KUZWOS, KANQAR01, ISOMUS, JUXLEV, YEKGUS, VUWXAM, EFARIO, QOXMUM, WECCAK, DUSBUQ, LESZIT, VABLIT, TEOXDE01, APITIE, DOYTUG, NRURAM11, VIKVOA, SAGMOE, JUDZIR, YEXNEU01, CILZII, SAWNAH, FEDCIC, NIVBUP, FOZLIQ, DOCMEO, FIHNIV, DETYLE, UKIHON, BPHENO14, FOHFOY, FULPIM, BIGUAN, MIHJIW, MAPCIQ, XEBTEF, RAJDEM, CUFFUG, WOKPAO, KEPSED, YEKJEF, JAXVEK, LIKYOV, NIMFOE02, EDAJIE, NUZKAU, OMIZOB, HUYYAB01, XAJSAE, KOXBOO, ADMNTB04, MIYPEP, LEMVUU, PUQJOC, PUHCOL, AFEXUF, INEZIM, OGIQOL, SAZGOP, UKIHUS, NUFDOJ, QEFBUY01, GOYLOV, SUCCIN, ZERLUD, QEGSOL, CORYIR10, QAFFOS, HIWJIG, BABBUB, TAXTER, FUZYUU, EXUJAI, CAZPUQ, EJEQUF, DMOCDO10, CEZSEF, LEFYUS, VOPTOL, AHUZUZ, RATMIJ, QUDNAG, VIZWEI, QAHCUA, DAYHIV, COKJOB, XEGPII01, MINGEX01, QONROC, NTPYRO12, GOHVUU, XIVBEK, UQAFAV, NENZOX, KIWJIK, YEKJAB, NIFBOT02, OLOBIA, TALJEV, JIWQIQ, BEVXEF01, XANAZH01, TABBEE, UQALUU, IHANUB, CINMER04, TEDNIB, LIJFUG, UHUMEP01, HMHOCN02, DUFDOX10, NEHQAU, KIKTEE, BOCZIE, MAQDAJ, ZOYGUP, TUGSUL, QEXBAX, RIDKIZ, JEMGIS, FOPDUL, NADQOZ, VOTMEY, LAQSIG, ONOQUE, BAXVEB, NAYNUX01, ULAVEK, PAKMAR, QARVOV, WAZDEG, CIKRET, HATYEG02, WIGKIH, IFUWIR, CAPKIP, KUXSEC, EXIHAV, COHWIF01, RIQWIZ, EXIGOI, ELIGOX, LIKLAU, WUHYIH, GIZYIY01, VOGWAQ, VEHCUG, APENOA, HECHOO, TMZBCO10, DIYHOJ, FIPZIN01, GAVKIX, KIZCUT, AZALIX, AZASER11, CECWAK, GOTSOX, TEMDOG, IJIVOM, HIZTEQ, PEFLUI, BIGFEC, TRZPUR, YUJDOW, LIWCUP, BALNIN04, UMOBAB, AFASOQ, XURCIW, HAZQUV, YOJQUL01, LUZFUJ, MOQLUA, ZZZDUI02, BEPNIT, GLICAC01, OCPNDN, LIPWEM, NIJHUJ, DUSHAB, ZZZKAY02, SIWDAD, ZBCNON01, HARJIV01, UREWAQ, QURYUZ01, OKINAY, WUMGAO, LONMEH, RESKIK, TIPHIL, TIPVIZ, YENDEC, GOGYUX, CIRXEH, TACGIN, DATFIO, HXANTH10, YINDUU, SEYJAH, QAKNIB, GAUTAM15, CUYXAV, URNFRO, HACMUT, XODBAU, AHIMOU, EFUMAU03, WAHBOX, SETFUS, AWIHOE02, NUPVUR, KEPNOI, XEFSIK, QODQEF, UTUVAI, HOJLIC, COXXUJ, UKIKEG, NBONAN01, ACOWOE, HOHTAZ02, GIRKOI, JAXSOR, OSOLEO, PAZRUF, SIBFOA, PHENOL11, MINPUR, TUNMUL, RONHEH, AJAPOR, EKUJIF, XOZHEZ, QEDFOW, IPIPAY, IYAWAG, MOZWII, VALIDL03, TABFAG, FOJMOG01, HTMTZC, FEYLUQ01, LESPUW, HEWFEW, VALWOW, LUHPAF, UFIYUD, XUHPIB, EMIJIT, FOYGIL, BUFKIX, ATOMAY, DAFLUO, REQQUA, AMTRAZ, NOKXEQ, QEDROG, QUICNA02, RATJOM, LEPMID, KONWOB, FUQYIB, YAMFOI, ULANUS, QELLOI, FEBKAZ, MUXGIX, SAWNIP, CAKREM, VOPJEP, JUMWUL, BCBANN02, PIKDAO, QEKQOM, CROTAC, SEYDIJ02, VOBJEB, BUTGAA, QONRAN, VIGLAA, HATWII, XARYIY, SITQET, MENPDL, KAJRIY, QOSRUN, MFCBXA, LETDUJ, PEPYIS, VUWXIU, IKALAJ01, MAFXAT, HAMHUZ, CXHIMZ20, KISJEC, WOQXOR, IMUWIX, PRUVAC, CEGXES, NOKYER, NAMVEE, EJIQEU, AMPYRZ, TAZPYR, TOLDAM01, LEZJAB, VAZZEB, QAFFIM, QOTWUR, GALDEC01, IWUGIQ01, TECKAO, ZOQJAQ, UNEWUG, ODEVUQ, VOHCEC, SUCANH16, IJARES, METRIM, LASCAC12, AROMAV, HOCWAX, ZEVJAL, TIPNEN, JONPUZ, KABWUF, DUFBOV11, YOBBEW, NAKVUT, CEDMUV, ACICAR, TIVSUM, DAZFLU11, FUTLEN, IHEPIU, HOYSOF, LSERIN12, YITKAP, YUFKUF, PUHJIM, NICGEL, FABFAQ, DAYQEZ, LUVNAS, REVMIP, QAGRIC, CADBOB, AMPHOM03, CABHUL, YUHTIE, GAGSIR01, IBUWAD, RIYZEF, YEKHIH, WOLNIW, MUYBAK, SAZFIJ, BODSAO, GLYCIN52, JIXHUU, GOCHIQ, YOVJUP01, XADBAI, BOVLEF, ELULOM, JOWWIB, LURYOM, FUGLEA, FUVDOP, MABZNA01, FETNUO, LODWOR, JEVPIJ, WIQDIJ, SOBBOA, SINMEI, FATZEH, VEFMOI, MILYIP, HMTOFA07, GIQRAZ, DMEBQU01, ULOBAA, USIZOM, HIVGAW, BACRAA, XIMQUG, POHHUQ, CAZDOY, NOGHIA, WASRAJ02, COFHOW, GULHEA, KEMHOB, SUCROS27, DILSEY, JOPNOR, BIDJON, GOMJUO, CEDXUE01, MCPRAC01, LIKKUN, RAMVUW, IMAZOL19, BEJTAL, GUDSUV, DMANAP02, SAFMAQ01, VEXGOU, MOGYOX, LIJJIY, IMICUE, SEGBUB, FOVYAQ01, ZAQJEI, KULGOO, TIHHOJ, FULYEQ10, IFIJOW, NTRTAC01, WIKGEC, MEPYRZ, VAZBIJ, SEZJEN, VANCOD, NIMVAG, XOHCEE, WIGBOE, KUDZAL, COWYET, QEYFUW, SARCEU, XIDQIL, TENZOD, GUZMUK, JICLAI, GIVXUE, WURLAY, PIFZAG, LIWJOQ, DUPDAU, HEPHUF, VITNIW, IVAGUH, PAGYUS, ANOLUL, RUGJUZ, LORJAD, TUGJUB, TARCOE, PHGLYA, KEPDIU, WOCVUF, JECVUI, OJUMOX, FEFNOT08, NNAPAN10, SOLHAD, GIZSOX, IHEHIO, IZARAC, NASQED01, IMEZIL, BIDRUB11, FEMFUY01, MODYOU, KEHYED, JERJOF, HUFFEU, JUXHUF,

GENZEF, TBPMTZ, KOHPUT, AZOXCH, KEPFAO, LIJLEW, KOFKIZ, AMPYRE01, GEJCEF, LOBHIV,
OKESAY, COYMUY, PNOIZA, HADFEY, ACACOX, PACFIL, THYMMH, EKATOB, CAXNIY, JEVPOQ,
NABMIM, CAZBAI, ZAQHAC, HFULCA01, MOWBUX, SUPVOO, TOYLUO, ENINIA, HMBQOX10,
GUSVOH, TEKQAB01, DOTTUB, HISYEO, WIFKEB01, PEDLER, DMAHOX01, SOQNIW, WUDHOU,
HXOXAM10, MEGLAN01, YOGJUB, DAFPUV, DIXRIL, SOJGED01, AKEZIB05, EDOREW, KOWRUJ,
TUHHUZ, JUKWIV, LIVROY, YEHPIK, NOKLOP, KOBFUD, UNUWOQ, ASOTIL, TIGKUQ, PYMDON,
MOHCUI, CEBGOF06, LEMSIF, ZOBYAQ, ALOPUR, QOLRIT, EVIXEM, CODPAN, LILLEY, ROFQIO,
MALEHY11, YASSIV, JOHJIB01, MAZFOK, SASPAE, MCBPCX, ZEXJAP, MUKGAA, KEPFOC, APMCOX,
XOZBUK01, ZELQOW, DHPROX, IMZMAL11, VIZWAE, OCEBAA, MOVNAN, FAJFII, GEZXAK, EROGOI,
LADLOT, PICANO04, POXYAC, GUNSIS, CACGUI, YITJES, REYGUZ, FUQSAN, DEDNEF01, SOWSEC,
BAKQUA, UHAMEV, SOJHEE, HEVLUP, YIRDAG, GOVKEH, EMOHAQ, CAZCAJ, YILNOY, HEYQUY,
FIMQAU, TUPWEI, DEZBOB, IFIDUW, GAQKOA, YIHHON09, AHXGLP, YEKSOX, ENUNOR, FADCUI,
REHCAK, IBPRAC14, WEJCEU, JABCAT, XIXQUR, AYUKIP, INOSND10, TIJLIH, JENHEP, ITIZUG,
BIRGEO, LOHQAA, SEKPED, PITMAH, TAHHOC, IFOSON, FUSJOT, NAYKOO, JAYRAC, PYRCOX01,
ZEQVIC, VEBTIG, QIJDES, LONRAI, APENUG, QUDPOW, FADTIQ, BENZAC18, HOQHUQ, PDABZA01,
KEDBAX, FOXJIN, INEZOS, AGEQIO, AHPHAL01, ZILQOA01, OBAYOF, MOVLUG, EHIXOI, IHURUA,
KEQQED, DOXKOR, GEJCIJ, CEBMUR, LIZCEC, XOSJUK, YIZXOV, QAPYUC, IKASUK, XUDTIB,
APITEA, POVKAO, HEXTIN02, CEDHUQ, TADQAR, GEMZAZ, TOAZCD, IYAQOP01, LUYJIZ, DATGEL,
XAGWIM02, RAWBAS06, FECVEP, TAZPYD10, DOZNOV, ACOGAA, UMEREJ, ACRLAC04, ZEGDIA,
ODEZEE, DIHRUH10, HODMOD, QAMSII, EFAXIS, KAYSUY, ZILNOX01, HIXLEG, VAWMAJ, WADPEX,
LSERIN21, NADQUF, TANPEE, TANCUI, BEXNIC, KEVXUG, BIZTOV, DACLOL, CEVTUU, KIXWIZ,
RIZBOS, RATGEZ, OHOYUH, JUTGEK, HRFPZO10, ALATIN02, NUMBII, YAXHUC, ECADIV, QQQAUJ07,
COWSOY, VUHDEJ, AXOSIQ, ERETAW, LISGUR, NIWCAY, DUVXIB, MNHCHA, ZIZRAD, NIPQUA,
DOJNAS, XACZIN, ISEDOB, IPABUX, QIJDAO01, BAZYAC01, ACEQIK, QEJJUK, BENZIE01, HYXBUR10,
MOAZCD, TEVJIP01, MEMALA, TORQIC, MUQXIH01, XADPEA, KADPOU02, BUWJEK, GOHHER,
LEZQUE, VAZBUV, DIQCEN, NERJIE, GEDYAQ, QIWJEM, COJMUJ, HMNCXC01, PUSZOU, TALHET,
RAGGOX, PAPHOD, WEHCAO, XEXYII, YANVUG, XAVPOZ01, PAPJUM, WUJQEY, ANGULF, KOJZUE,
BOZGIH, NIRJEE04, BODFEF, DOJHOA, FEYTOU, FACQAE, KUSJIR, NOLDOJ01, XOMWAX,
HEYJOK01, ALIWID, UPAWEP, UPOMIX, XABFOY, MIOZPO01, XOWNON, MMXPDZ10, OXOGEO,
MEQPON, XULDUD01, JIXCOI01, UDAZIJ, PYRZAL10, KIKNEY, KAFLEI, FEZHAV, NINHYD02,
DIMEDO03, YOLQOF, POVKIW, DIKXEB, AZETAC, DHANQU02, HAXTAD, AIPMGH, DOVDIC, HEKTEW,
DOKQEA, BAGQIJ, GILCIP, HISTPU, MUDVUE, APITUQ, GUBZOS, XOYWEO01, LICROE, HUZLUJ,
FURDCB01, KETVEK03, SUMKIU, MORPHN, LAQDUE, SIGFEU, XOKXIE, RAMGUH01, WIKRAL,
AMOXAC, COHQEV, NEYKIN, SOWREB, QUSKOG, DZHPDO, PUVPED, ESAREV, HUNWUI,
HOPKUT01, BOMHOA, NANQUO, XUDQUJ, MEXZOE, NADNIP, DASTAS, DIDFAY, DUBTEA, QIZBIL,
NAPDCX, DMADEN10, GOGXUV, GADCIX, DIFPEO, MOJMEE, AMXBPM10, PYRIDO04, RAQVUB,
NIYMAJ, KICCOO, SODBIW, VUXYEU, OCASIV, XAZDUY, GLUTAM08, XDHURC, MUKVUJ, IFAYIZ,
KAMRAS, HMBFUR, DOZSUG, XOJQUK, EFUMAU02, YIKQOY, XEBHET, HOPLAA, TCMPCQ01,
WIKCAU, IJUMEG, MEUREA01, WAKSUX, GAVLUK, LOKFIA, XOVWEK, ZERMOY, QEMHIA,
ZZZNWQ01, ZAWDIM, MUVDOX, ZECWOV, HOLQUU, VEMFEY01, FUFMAU, XAJHEX, QANRUT01,
GAVMUM, KAHHIM02, NABKIN, TAFKIU, PELDER, FIFBAY, YIFLOP, QIRRUG, BAWPOE, AMHPYR,
UBEGUF, IHINEU, XAYROG, AZOBEN04, LOZJUF, DEXBEN, KUZKOG, CIGBAV, HEQWOR, RAYLAE,
DOGQOF, UREJAD, GOLDIV, TEFMEW, LAYYOB, YUWGOO, ISIBER, ZEFPAD, SIXYEE, ULAVIO,
WERHUX, GAFWIV, PEDERL, DIPGIS11, KECZAU, FEMMOA, CONXIN, CIRXAD, NUMRAP, DAHMII,
TAHNEV, FOMNAX, VUBZUN, DEKRUG, RAYPOY, HOFGUF, QOWFUF, EVIQAB, QUCGEC,
CBOHAZ04, TEDQAU, WAWVUN, NESFIA, SEGCOW, WIRYEB02, PHALIM, RUHGEI, ROFHAV, ZIYGUL,
IKIDAH, XAJTIM, NSMACM, VACTIF, YENGOP, WAHGUI, NENFES, DOXDAN, CINWEC, GILZUW,
MUXGOD, PORXUQ, SUCPYR11, UHAMOF, VEDLOG, UPOCEJ, WEHXOX, ZIWXEK, VIKNIM, DELXEX,
KEKHEO01, CEYJUN, FINWOO, SAQZAK, ZACLEX, ZZZNYY01, QIFDOZ, NUVXIM, ANPRAL, ZEGBIY,
EXETEI, PYRZOL05, QAMSUU, PUSYIM, YODSIT, VEBFIR, BEHZIX, PTZCOC, QEHBEL, DIKQUJ,
SEDTUQ06, DEZBIV, MAYXEP, PORXEA, LIWCOJ, QAKFAK, ZERJIR, SOQNUI, WAKNUS, NOLCOG,
BABXUA, KAQPUN, WAJPUV, SOVGIT, HIPXAF, DAFNON, GEDXIX, ILOFIA, VIGTUA, KECXIZ, FEHVIY,
EMOGUJ, XUHREX, AMIPYZ, FIJKEP, YAMXUF, TABRET, ICRFRA10, WIQCUU, NEQWEN, KIFJAK,
VUGMIV, ETANOL, IJEFAG, LEYXIW, JUMXEW, PEVKUX, QINFUO, SAYBID01, ULIZAS, OPISUC,
BEVYOS, BZTZAD, BUYZAW, UGECOA, SABMAL, YEJYIV, PUYTEI, UDURUH, JASLEV, ZOQWOT,
QALQIE, VIZZIP, NINDOD, BOGGAH, PEHXIK, ROFFEY, VALWIQ, HYZMAC, GLYALB02, DOXHPX,
LEMGEQ, DUPNOT, OMEJOG, LAGVAT, ELUFOH, ORAVUZ, OQAHEV, VETQUI, AZIGAR, VUNKEU,
EWUKIQ01, LEJYIJ, PUHHOQ, JOTKUY01, ANENUD01, DOHDEJ, CUNGAT, FERDUD, CASCIJ01,
DIBGEA01, BANRUE, TIMBAU, KUYPAV02, DUDXIL, CEMGIK, EGAFAU, LAJGUA, FASFAG,
KOVHAE01, BUKNEA, XISNET01, ZUHKIW, WAHNUO, MAQCEN, KERPII, MDXTCU, TIHQOR, YOWLUS,
LIWGUU, BAZGOY05, OTOGEL, WAPPUY, NBZOAO05, DOJLAQ, DASRIY, WEJBIW, JOSXUK,
ACYGLY12, SEWHUY, SUYQEK, ZUTKAC, CIDROW, VATJIK, CAZCIR, SORBPY20, DAZVEF, IGAMEJ,
FIZJON01, FIKXAY, WIKCIE, MIRTOW, CATNUI, DOLTED, XARQAK, EYIJAY, HAXMAW01, LIWGOO,

ABIVUD, UGUXAW, TAQXAL, JOQTAM, XOHXUP, FIGYID01, ACSALA10, GAQSEY, QAPZAJ, JUPROB, MIPHOL, BOXGEA, UNURUS, XISPOF, DOTRAG, FENTOH01, NICGAH, HOKSOR, SILDOH, MACLEI, VARGAX, AJASEK, EPAHOT, DAJXIV, HOFLAQ01, BUGDAI, PHTHAO, ROKBOI, URAWER, HAMJEL, MNIANL11, DISBUE, LANYAA, RAQRUX, SEGCAI01, CUTWIX, ELEYOJ, NITPOL02, YODHUV, UXAZEZ, INEJOA, FOVVUI, LENYOS, SUYWOA, TODSAG, AMCYTS, PUNLOB, KUBPOM, AJODOU, PAGVAU, KENMIA, NUYJEW, OHEMAR, REDUCA, HOTNAF, EBULAO, VAYJIO, CAXKOB11, ROGDIB, ABUMAL, AJUKIZ01, OTAKEB, SIKGIC, QUHSUJ, EROWEO, MUZNEC, FUNXER, DNBENZ15, DUHYIP, GEYSEI, DNCOOC, WALDES, QOPYOK, TOXCTD, KOYLIT, FOXJOT, BUHGOA, EHUPAZ, ETBARB, PODFIZ, CMPDZB, EQIXEI, QQQFCY01, WECRED, VADQAU, YUYSIU, AJASAG, DMXBZQ01, HAVKOE, SEPNUX, YUKFUG, FUFSIK, LUZZEN, CIQSIE, JODPIC, BUDXII, WEMQUC, POLYUM, WUFVAW, GUDKAS, CUXXAW, XIVCUC, FEHCUQ, MAXYOZ, PEHXAC, MAZMEH, HIJFIP01, ROKTAO, TIMFUS, PPYRED, NOLDUP, PHTHIB, RODLOM, UNEVOZ01, AJAKOL, CBUDCX03, EACLID, CEKVIX, SOZHUM, QQQGEM01, XATZAT, BEXGAM, MIKTOQ, COVXAN01, YIHJOP, KIBDOO, PUTTUU, DIDJUX01, CROTAM, BUBNAN, CUBRUN, BAKTOX, VEZNEV, GORXOA, EDEBUK, YIFVAL, DORKEA, AMFURZ, ERIVAC, PAZDID, LOYXAA, FIGMAJ, ZEXYOQ, LUVPEY, HAJXIB, RAZXEV, WAWWAU, MAYCEW, ECUXUW, TAZWIB, FULTEM, XUBZID, PRONAC02, GOCJAK, ROFLAZ01, NOGUNA, CONNID, EQUXOD, LAFGIL, UKIJUV, TIWHAK, JAKKIP, VUXYAO, DUMPUW, UNEPOT, LOCVEE, XUMDUE, KUSLIU, GOMXIR, ADOJEK, PUMVOJ, LAPSIG, YITKIX, CYTOSM12, ACEMID03, SALMID05, CEDSAH, EVUMOX, SIGBOZ, PICNOE, YANSEN, NURWOM03, HMALAC01.

## VIII. CSD-500 set

The CSD-500 set contains the structures with the following CSD Refcodes:
COBHUW, QUGWUK, MELYUY, RAVFOK, UMUKUJ, TAVJAD, DOMNEY, QEXKUA, KOGWUZ, HAMTIZ, MAHPUJ, POLJEF, DXCYTD, BUFNEV01, COWPUZ, VIMKOT, ASPARM10, JIPCUG10, AZOBEN12, GUJGEX, XOWJUP, NUYWUB, RECYIH, XOHMAI, UJUKIT, MEYBIB, POQSEU, ENIMET, QOMVUK, AHOWOL, YIMPOB, LEVSIO, PANLEZ10, WIQZOL, CORTPY, FOSLEG, CINCHO10, NOPHKN, UWEZED, RUSGOD, MENDAL01, NEQPEG, MATQOO, HODKEQ, MEHLER, EDAXOW, FADHOJ, ROHJED, KOFKAR, YAZDEI, KABKIJ, XUJKUK, OXUJUN, QUWFIZ, VAFPAV, ITINEG, LILDEP, VONNOB, AROKUN, HUVWOL, NEZFON, ONBZAM, VOGDIE, IDURIJ, WEPTIV01, CBMZPN21, AMEXOH, IVEZAK, EFIBAX, MOSLAI, AXADAF, MOBNUM, RIQSER, DASNIV, ROGRIQ, YODXAS, MEJDOU, COCYAW, SOMNIT, KUZJIA, BZAMID08, THYDIN05, XINHIL, TALVAD, SEGROL, FUPWES, VIDMAX02, YEHWUD, NANJIW, MEHNAP, DAFTAF, ITIREI, FACZIU, TOPRIB, AZIDES, ROJXOD, QIYLAM, ZEYLAS, UQOLIW, VANFEV, ZOFNUD, HUDHEU, PIHBOZ, VOCHUR, LIXQEO, SAWVET, MUBBAN, XAZYIG, ODOROO, IBOPIA, YOCWUK, KEMFIS, BUMNOM, EWOBIB, WEGPEF, WECZEJ, RULHOX, FELDOR, BUZJIR, VUHZEE, DILDUZ, AXOSOW03, ANOSAY, PACWAU, YODPAJ, QECNAP, PUMSEV, ZZZFFY01, DAJZEU, ZEMHOO, EKAHOP, PMPZOL, FIHLEO, AKUBIT, NUZPUT01, ARONOM, BEGDIB01, UNURIF, IMEQUO, IQIZEO, SIMYOE, FESNEW05, RIFZAI, COSPEG, PELXAG10, UTICIK, RUVPIJ, SUKNIW02, WIFQEI, SEFNOG, HUDYUA, ANAHII, AJIXUM, LUXSAY, ZIYYUD, UBUXOG, RAKTOO, ACRDIN07, EMODUG, IDUJEW, KOTMUB, DUTKOU, QAKDAJ, NORFUW, QQQFDJ20, AFIQUC, SAZFOO, DITZOX, MUJGEE, BIKNUE, TIWZUV, KUYWEH, EABZBU, IQIKOI, OXAROV, MAJJUD, LAFHEH, FALHIJ, XAQTUF, COLBAG, MELAMI05, ISIJIE, EMISUQ, ESESEA, HIMSUS, ZICKIF, GIZRUE, IYASUW, WEVVEZ, KUZQIG, YAWWOK, SIHZAM, CAZCOX, LOPLUZ, EBAXOW, GIDHUW, RIHFIY, EXEWEJ, MODXUZ, BEHWER, VIDDAO, OJICUF, HUVZUV, WUCVIB, SENKUR, EZISUC, PEFGIS, GAWFEQ, NCUBEB10, ORADUH, ZEBXOV, JOHKEY, VEZCUY, IBEHII, REJVAE, XASHUW, NBZOAC11, OFEVOL, JEBQOW, WAFBIQ, WULTUT, BOMSIH, VASLOR, OGIMIC, VINZUP, SOGCUN, DOTFOI, OPIZAQ, KAHJEK, GADSIO, GAQJUF, XEZFUF, ZIKQIT, BAJCIY03, SUWKEC, EVINII, BERSOG, SOYMEZ, OCIPAR, GIZFEB, UNAMOL, TICLIC, ALOSEZ, PIWBUS, AYUNEO, ECODUV, PUNFAH, BOLGOZ, UBASAT, RIMHEC, WAWQUH, MOTNUF, QEPRIO, SAJCAJ, XABFUE, XIMGAB, SEYCUU, DOVWAM, RICTIG, MISDAT, SATPUZ, UQAMIK, KIVZIZ, BASNOZ, CUTCUQ, SORFIQ, TESDOL, EMIPUM, UBUVIY, EVIHUM02, DUZLUF, HOMKIF, IPINIE, RAZVUJ, TEMKAZ, XUVBAT, VUGWAX, HUYYOP, NEVDOH, YERTIZ01, KAKHEL, SAVREN, AHOXOL, IVABEO, JOQTUE, PIJREF, ZAYPOE, WIZWOS, TOPXUT, XINLAJ, COYBOJ, REKMEZ, WOKJOV, BIXQEF, GASXON, HECNOS, RACGEK, EHAHAY, BAYPAT, ZULCEQ, UNUVEF, JAYFOF, BEDLEB01, MAQWIM23, HOMZUG, EPHEDR01, ZIWMOJ, GUFYOX, MBPHOL02, SATPEI02, QIQCOI, CIKSAQ, SUYYIV, RUCNOU, NIQTAJ, HISNII, IJEZUS, NASZAJ, APODUG, DILKIT, GIXKOP, INAVIC, BAPPUF, NURZOP, NUPPOD, RIZBAF, RACGEJ, PEPLAX, SIHCES, SIGSAD, ZOYMOP07, PILFIB, XOFFEF, GOVQOX, XULNOI, YOFTOE, SEZXIG, FEZLUT, AXAWIG, IQUBZA, VEQMUA, FEMGAF, YOXRIO, FEPTID, EXUVUP, PUPBAD01, QQQAMS02, AHATEK, ZZZBPY10, CXMTUN, WOYTAH, AQEYAW, AMUQOQ, ZOGTIY, VORMEV, JOTKIM01, ZOXYOA, QUBQOT, PUMQEV, QUYZIV, NAPTPR, XIYTIJ, GERPOJ01, PEXPEN, HIZHOP, KETYUF, PRMDIN05, KOJTOT, SUHFEH, FAJDEC, BAQNEM, DIZWEQ, EKAWAQ, VOXNOL, DOHPEV, WOBRIP, YIPPOC, QUFJUY, LUDZIT, DORKOK, BZTROP11, VILPUB, FEMXOK, CEDVIS, HUVPAR, MESQOR, IZAKOK, YIXPUR, OMSTER01, IROZIY, OPUTEA, FAHXUH, VEMBAS, TAJSOM, SULROG, NUQLES, LAVSIL, GADVAJ, EVIQEF, XANJUS, HIWYIV, YOWYOY02, ZAVXUP, OCAWOF, GUCJUK, HABNED, LUQDOS,

PYAZAC, ONOTIW, TICKAT, POKKAD10, BUGQUQ, ZATDOP, MEJQEY, RULDAF, KEZNUZ, MUTWON, DOLBIR10, HIGCIK01, NOQHAE, NAJLUF, WAGTII, ZOSVEI, IQULUC, EBOVEX, ARUZUK, MALSOH, OTAKAW, QUFCEZ, HAXREE, ZIGBAS, JULGOO, JESHIZ, OHEWOP, AQAGII, ECACER, DENXUP02, LIZHEJ, EKOGAO, AWAVEZ, YUNYUC, BAWRAT, NUKSAO, XESYEA, QAMKEW, KUTKAL, HESTOO, FAHLAB, KUJZIY, WIHBEW, HONKEC, UWOCAM, PEWNIQ, APUPIK, PUWNIG, PHBZAC01, EDIZUM, RUKTAU, YEGGIA, OMABEK, DIWWEN, XACTEC, XAVQOB, VUDDUV, XOTFAN, GUTZOM, SOLGIL, SAFQAU, AFIKAC, POHCAS, EMEFOT, XIZVAD, WIGWOA, RELJAU, ROJHOP, HUMNEK, HODLOC, PUPGUD, ALEXEW, ZEMNAG, YUXCUP, EXEYUD, VETJIO, OWIWUN, EYASAZ, UCANIV, XICCAO, BOPJAS, SEBVAW, XUHZOR, UKUROJ, PEDHAJ, YIDTIQ, EVILEB, ELAWIX, AMHTAR02, OCATOC, PETRAH, BEDFUM, LADNEL.

## IX. Environments Eliminated with the Unusual Environment Detection Procedure

### $^{1}$H

Of the 76,214 $^{1}$H environments of the CSD-2k, the following 211 environments were detected as unusual (the numbering follows the FPS order listed in section V):
1536, 1537, 1538, 1539, 1792, 3295, 3296, 6807, 6808, 6809, 6810, 6831, 6832, 6833, 6834, 6843, 6844, 6845, 6846, 13876, 13877, 13884, 13885, 13896, 13897, 16544, 16545, 16546, 16547, 18285, 18286, 18765, 18766, 18767, 18768, 22192, 22193, 22194, 22195, 24448, 24449, 24450, 24451, 24464, 24465, 24466, 24467, 25466, 25467, 25468, 25469, 30674, 30675, 30676, 30677, 31670, 31671, 31672, 31673, 31674, 31675, 31676, 31677, 31678, 31679, 31680, 31681, 31682, 31683, 31684, 31685, 31686, 31687, 31688, 31689, 31690, 31691, 31692, 31693, 31694, 31695, 31696, 31697, 31698, 31699, 31700, 31701, 31702, 31703, 31790, 31791, 32742, 32743, 34296, 34297, 35394, 35395, 35810, 35811,35812, 35813, 35814, 35815, 35816, 35817, 36460, 36461, 36462, 36463, 36464, 36465, 36466, 36467, 36468, 36469, 36470, 36471, 36472, 36473, 36474, 36475, 37880, 37881, 37882, 37883, 37884, 37885, 37886, 37887, 38488, 38489, 38490, 38491, 38492, 38493, 38494, 38495, 42945, 42946, 43449, 43450, 43451, 43452, 44037, 44038, 44039, 44040, 44041, 44042, 44043, 44044, 44045, 44046, 44047, 44048, 44049, 44050, 44051, 44052, 49022, 49023, 49024, 49025, 49026, 49027, 49028, 49029, 52146, 52147, 52148, 52149, 54656, 54657, 54658, 54659, 58410, 58411, 62970, 62971, 62972, 62973, 63714, 63715, 66522, 66523, 66524, 66525, 68668, 68669, 68670, 68671, 68672, 68673, 68674, 68675, 68688, 68689, 68690, 68691, 70782, 70783, 70828, 70829, 72408, 72409, 72410, 72411, 74442, 74443, 74444, 74445.
These environments belong to the structures VIWYEH, QAHVUQ, ZACSOO, PORPIN01, VOFVAN26, CEVGIT, TIJKEC, PUPBAD02, TIMCHX, UPANIJ, MTETAC01, WUHYIH, HAZQUV, YOJQUL01, MOQLUA, CIRXEH, EKUJIF, QELLOI, SEYDIJ02, MAFXAT, NAKVUT, AMPHOM03, IMEZIL, AMPYRE01, JEVPOQ, OBAYOF, BAZYAC01, KAFLEI, QEMHIA, ZEGBIY, NOLCOG, CASCIJ01, ABIVUD, KENMIA, SEPNUX, FIGMAJ.

### $^{13}$C

Of the 58,148 $^{13}$C environments of the CSD-2k, the following 1,419 environments were detected as unusual (the numbering follows the FPS order listed in section V):
96, 97, 470, 471, 472, 473, 676, 677, 764, 765, 770, 771, 918, 1027, 1028, 1029, 1030, 1311, 1312, 1367, 1468, 1469, 1470, 1471, 1484, 1485, 1486, 1487, 1500, 1501, 1502, 1503, 1516, 1517, 1518, 1519, 1766, 1796, 1882, 1883, 1884, 1885, 1886, 1887, 1888, 1889, 1894, 1895, 1896, 1897, 1898, 1899, 1900, 1901, 1902, 1903, 1904, 1905, 1906, 1907, 1908, 1909, 1914, 1915, 1916, 1917, 1918, 1919, 1920, 1921, 2148, 2149, 2150, 2151, 2156, 2157, 2158, 2159, 2164, 2165, 2166, 2167, 2172, 2173, 2174, 2175, 2180, 2181, 2182, 2183, 2264, 2265, 2270, 2271, 2352, 2353, 2354, 2355, 2754, 2755, 2756, 2757, 2790, 2791, 2792, 2793, 2832, 2833, 3038, 3039, 3040, 3041, 3514, 3515, 3516, 3517, 3780, 3781, 3942, 3943, 3944, 3945, 3946, 3947, 3948, 3949, 3990, 3991, 3992, 3993, 3994, 3995, 3996, 3997, 4006, 4007, 4008, 4009, 4010, 4011, 4012, 4013, 4108, 4109, 4110, 4111, 4112, 4113, 4114, 4115, 4132, 4133, 4134, 4135, 4136, 4137, 4138, 4139, 4298, 4299, 4300, 4301, 4314, 4315, 4316, 4317, 4394, 4395, 4396, 4397, 5034, 5035, 5036, 5037, 5618, 5619, 5620, 5621, 5785, 5786, 5805, 5806, 5809, 5810, 5821, 5822, 5860, 5861, 5862, 5863, 6606, 6607, 6656, 6657, 6658, 6659, 6672, 6673, 6674, 6675, 6684, 6685, 6686, 6687, 6688, 6689, 6690, 6691, 6692, 6693, 6694, 6695, 6700, 6701, 6702, 6703, 6716, 6717, 6718, 6719, 6728, 6729, 6730, 6731, 6732, 6733, 6734, 6735, 6792, 6793, 6794, 6795, 6828, 6829, 6830, 6831, 6852, 6853, 6854, 6855, 6856, 6857, 6858, 6859, 7086, 7087, 7088, 7089, 7090, 7091, 7092, 7093, 7376, 7377, 7378, 7379, 7388, 7389, 7390, 7391, 7456, 7457, 7458, 7459, 7460, 7461, 7462, 7463, 7567, 7568, 7569, 7570, 7571, 7572, 7591, 7592, 7593, 7594, 7595, 7596, 7597, 7598, 7599, 7600, 7601, 7602, 7639, 7640, 7641, 7642, 7643, 7644, 7645, 7646, 7981, 7982, 7983, 7984, 7985, 7986, 7987, 7988, 8175, 8176, 8179, 8180, 8187, 8188, 8341, 8342, 8343, 8344, 8345, 8346, 8347, 8348, 8465, 8466, 8467, 8468, 8469, 8470, 8471, 8472, 8591, 8592, 8593, 8594, 8599, 8600, 8601, 8602, 8847, 8848, 8849, 8850, 9099, 9100, 9101, 9102, 9103, 9104, 9105, 9106, 9107, 9108, 9109, 9110, 9171, 9172, 9173, 9174, 9187, 9188, 9189, 9190, 9191, 9192, 9193, 9194,

9307, 9308, 9309, 9310, 9315, 9316, 9317, 9318, 9463, 9464, 9465, 9466, 9591, 9592, 9927, 9928, 10143, 10144, 10147, 10148, 10353, 10354, 10367, 10368, 10371, 10372, 10849, 10850, 10851, 10852, 11457, 11458, 11571, 11572, 11573, 11574, 11961, 11962, 12407, 12408, 12721, 12722, 12723, 12724, 12725, 12726, 12727, 12728, 12845, 12846, 12849, 12850, 12883, 12884, 12885, 12886, 12969, 12970, 12971, 12972, 13319, 13320, 13329, 13330, 13331, 13332, 13349, 13350, 13351, 13352, 13485, 13486, 13487, 13488, 13727, 13728, 13729, 13730, 13907, 13908, 13909, 13910, 13911, 13912, 13913, 13914, 13919, 13920, 13921, 13922, 14093, 14094, 14247, 14248, 14249, 14250, 14829, 14830, 14839, 14840, 14979, 14980, 14981, 14982, 15525, 15526, 15527, 15528, 15537, 15538, 15539, 15540, 15767, 15768, 15769, 15770, 15771, 15772, 15773, 15774, 16275, 16276, 16277, 16278, 16355, 16356, 16357, 16358, 16359, 16360, 16361, 16362, 16371, 16372, 16373, 16374, 16411, 16412, 16413, 16414, 16497, 16498, 16499, 16500, 16501, 16502, 16503, 16504, 16565, 16566, 16567, 16568, 16569, 16570, 16571, 16572, 16573, 16574, 16575, 16576, 16577, 16578, 16579, 16580, 16581, 16582, 16583, 16584, 16585, 16586, 16587, 16588, 16945, 16946, 16947, 16948, 16949, 16950, 16951, 16952, 16953, 16954, 16955, 16956, 16957, 16958, 16959, 16960, 17167, 17168, 17181, 17182, 17461, 17462, 18805, 18806, 18807, 18808, 19363, 19364, 19365, 19366, 19539, 19540, 19541, 19542, 20143, 20144, 20145, 20146, 20159, 20160, 20161, 20162, 20857, 20858, 20859, 20860, 20929, 20930, 21279, 21280, 21281, 21282, 21671, 21672, 21673, 21674, 21675, 21676, 22085, 22086, 22087, 22088, 22119, 22120, 22121, 22122, 22123, 22124, 22125, 22126, 22277, 22278, 22279, 22280, 22305, 22306, 22307, 22308, 22317, 22318, 22319, 22320, 22597, 22598, 22599, 22600, 22701, 22702, 22703, 22704, 22765, 22766, 22767, 22768, 22815, 22816, 22817, 22818, 22819, 22820, 22821, 22822, 22927, 22928, 22929, 22930, 22947, 22948, 22957, 22958, 23005, 23006, 23325, 23326, 23327, 23328, 23329, 23330, 23331, 23332, 23333, 23334, 23335, 23336, 23337, 23338, 23339, 23340, 23341, 23342, 23343, 23344, 23345, 23346, 23347, 23348, 23427, 23428, 23877, 23878, 23879, 23880, 24281, 24282, 24283, 24284, 24301, 24302, 24303, 24304, 24305, 24306, 24307, 24308, 24309, 24310, 24311, 24312, 24329, 24330, 24331, 24332, 24333, 24334, 24335, 24336, 24523, 24524, 24787, 24788, 25371, 25372, 25601, 25602, 25603, 25604, 25605, 25606, 25759, 25760, 25761, 25762, 25763, 25764, 25765, 25766, 26093, 26094, 26095, 26096, 26295, 26296, 26299, 26300, 26303, 26304, 26309, 26310, 26313, 26314, 26317, 26318, 26347, 26348, 26349, 26350, 26449, 26450, 26451, 26452, 26561, 26562, 26563, 26564, 26715, 26716, 26803, 26804, 26805, 26806, 27015, 27016, 27017, 27018, 27019, 27020, 27021, 27022, 27039, 27040, 27041, 27042, 27043, 27044, 27045, 27046, 27425, 27426, 27427, 27428, 27935, 27936, 27965, 27966, 27967, 27968, 27997, 27998, 27999, 28000, 28001, 28002, 28003, 28004, 28123, 28124, 28129, 28130, 28209, 28210, 28211, 28212, 28229, 28230, 28231, 28232, 28429, 28430, 28455, 28456, 28508, 28553, 28554, 28555, 28556, 28843, 28844, 28845, 28846, 29313, 29314, 29417, 29418, 29419, 29420, 29425, 29426, 29427, 29428, 29433, 29434, 29435, 29436, 29553, 29554, 29555, 29556, 29573, 29574, 29575, 29576, 29627, 29628, 29629, 29630, 29763, 29764, 29765, 29766, 29775, 29776, 29777, 29778, 29857, 29858, 30257, 30258, 30259, 30260, 30357, 30358, 30359, 30360, 30989, 30990, 30991, 30992, 31249, 31250, 31251, 31252, 31273, 31274, 31275, 31276, 31445, 31446, 31447, 31448, 31489, 31490, 31963, 31964, 31965, 31966, 32083, 32084, 32085, 32086, 32179, 32180, 32181, 32182, 32183, 32184, 32189, 32190, 32381, 32382, 32383, 32384, 32449, 32450, 32451, 32452, 32457, 32458, 32459, 32460, 32725, 32726, 32727, 32728, 32965, 32966, 32967, 32968, 33001, 33002, 33003, 33004, 33009, 33010, 33011, 33012, 33021, 33022, 33023, 33024, 33033, 33034, 33035, 33036, 33093, 33094, 33095, 33096, 33171, 33279, 33280, 33345, 33346, 33843, 33844, 33845, 33846, 33871, 33872, 33873, 33874, 33875, 33876, 33877, 33878, 33887, 33888, 33889, 33890, 33891, 33892, 33893, 33894, 34357, 34358, 34359, 34360, 34761, 34762, 34763, 34764, 35193, 35194, 35195, 35196, 35201, 35202, 35203, 35204, 35809, 35810, 35811, 35812, 35899, 35900, 35901, 35902, 36051, 36052, 36593, 36594, 36595, 36596, 36609, 36610, 36611, 36612, 37081, 37082, 37083, 37084, 37275, 37276, 37277, 37278, 37279, 37280, 37281, 37282, 37747, 37748, 37749, 37750, 37751, 37752, 37753, 37754, 37763, 37764, 37765, 37766, 37767, 37768, 37769, 37770, 37799, 37800, 37801, 37802, 37955, 37956, 37957, 37958, 38049, 38050, 38927, 38928, 38929, 38930, 39241, 39242, 39243, 39244, 39861, 39862, 39863, 39864, 40495, 40496, 40497, 40498, 40827, 40828, 40829, 40830, 41715, 41716, 41717, 41718, 41859, 41860, 41861, 41862, 42308, 42309, 42310, 42311, 42312, 42313, 42314, 42315, 42780, 42781, 42782, 42783, 42808, 42809, 42810, 42811, 42816, 42817, 42818, 42819, 43132, 43133, 43134, 43135, 43808, 43809, 43810, 43811, 43820, 43821, 43822, 43823, 44436, 44437, 44438, 44439, 44784, 44785, 44786, 44787, 44792, 44793, 44794, 44795, 44916, 44917, 44918, 44919, 44928, 44929, 44930, 44931, 45330, 45331, 45436, 45437, 45438, 45439, 45622, 45623, 45628, 45629, 45896, 45897, 45898, 45899, 45904, 45905, 45906, 45907, 46102, 46103, 46104, 46105, 46106, 46107, 46108, 46109, 46650, 46651, 46652, 46653, 46666, 46667, 46668, 46669, 46690, 46691, 46692, 46693, 46698, 46699, 46700, 46701, 46758, 46759, 46760, 46761, 47110, 47111, 47112, 47113, 47234, 47235, 47236, 47237, 47312, 47313, 47314, 47315, 47380, 47381, 47396, 47397, 47618, 47619, 47622, 47623, 47624, 47625, 47628, 47629, 47786, 47787, 47788, 47789, 47802, 47803, 47804, 47805, 48014, 48015, 48210, 48211, 48212, 48213, 48374, 48375, 48376, 48377, 48414, 48415, 48416, 48417, 48484, 48485, 48492, 48493, 48652, 48653, 48654, 48655, 48656, 48657, 48658, 48659, 48758, 48759, 48760, 48761, 49168, 49169, 49170, 49171, 49172, 49173, 49174, 49175, 49622, 49623, 49624, 49625, 49662, 49663, 49664, 49665, 50186, 50187, 50188, 50189, 50190, 50191, 50192, 50193, 50386, 50387, 50388, 50389, 50514, 50515, 50516,

50517, 50554, 50555, 50556, 50557, 50874, 50875, 50876, 50877, 50878, 50879, 50896, 50897, 50904, 50905, 51188, 51189, 51190, 51191, 51220, 51221, 51222, 51223, 51314, 51315, 51316, 51317, 51342, 51343, 51344, 51345, 51438, 51439, 51496, 51497, 51574, 51575, 51576, 51577, 51586, 51587, 51588, 51589, 51590, 51591, 51592, 51593, 51606, 51607, 51608, 51609, 51618, 51619, 51620, 51621, 51634, 51635, 52016, 52017, 52212, 52213, 52278, 52279, 52280, 52281, 52286, 52287, 52288, 52289, 52472, 52473, 52474, 52475, 52788, 52789, 52790, 52791, 52792, 52793, 52794, 52795, 52808, 52809, 52810, 52811, 52828, 52829, 52830, 52831, 52844, 52845, 52846, 52847, 53180, 53181, 53182, 53183, 53200, 53204, 53205, 53206, 53207, 53696, 53697, 53700, 53701, 53702, 53703, 54376, 54377, 54378, 54379, 54388, 54389, 54390, 54391, 54684, 54685, 54686, 54687, 54848, 54849, 55330, 55331, 55368, 55369, 55406, 55407, 55408, 55409, 55410, 55411, 55412, 55413, 55750, 55751, 55832, 55833, 55834, 55835, 55876, 55877, 56024, 56025, 56026, 56027, 56028, 56029, 56030, 56031, 56420, 56421, 56422, 56423, 56824, 56825, 56826, 56827, 57080, 57081, 57082, 57083, 57380, 57381, 57382, 57383, 57440, 57441, 57442, 57443, 57540, 57541, 57542, 57543.
These environments belong to the structures AMFORM, FOKSUT, JOJWOU03, LEPFES, CUZJEO, CALFOM, DIOXAZ10, QAHVUQ, GANHUY, MEVDAT, EMUFAU, UMERAF, MUYHIY, QOXPEZ, GEFBAV, CITSED10, FESNEW15, EJEHEG, JAPBIL, OXHEPA, GACGAU, VOQXAA, TETZIB, IVEZEO, RIKPEG, GAQWAX, DNBOIM, QODBOC, SAGLAQ, HONSIO, GEJSET, PRMDIN20, AXOSOW, DZBASK, AMTETA, FAFKEC, CYAMPD01, OMIWEM, ITIPEG, KABBEW, XZBTZO, NEPBEQ, NBFURX, PUFFOL01, ZUNTAD, NIBZAM, LIVZIA, XIHLIL, NEBPUF, HEHQOC, VOFVAN26, DOZXEW, POFLAX01, LEGLEQ, KAMFAH, LESLAX, BZCBUO, ZOVHID, KOVGIL, KIXXEW, OGOKEC, FEHDAX01, GEXMED, FOLVIL, PUPBAD02, LUNYID, DALGON03, HABFUL, PMXCDO10, HUKROU, BECYAL, WOFVAQ, VANCIX, XUDSOG, XACSOL, BUVBUR, TIMCHX, JUVKAM, DODCHD, YITBOS, MTETAC01, QOXMUM, DOYTUG, WOKPAO, CORYIR10, TAXTER, QAHCUA, TABBEE, NADQOZ, BAXVEB, CIKRET, WUHYIH, VEHCUG, HECHOO, AZALIX, AZASER11, TEMDOG, HAZQUV, MOQLUA, WUMGAO, CUYXAV, WAHBOX, QODQEF, QEDFOW, FOJMOG01, HEWFEW, QUICNA02, CAKREM, JUMWUL, QEKQOM, VOBJEB, XARYIY, MENPDL, MAFXAT, LEZJAB, LASCAC12, AROMAV, HOCWAX, KABWUF, YOBBEW, DAZFLU11, IHEPIU, YITKAP, CABHUL, ELULOM, FUVDOP, LODWOR, JEVPIJ, WIQDIJ, MILYIP, DMEBQU01, DILSEY, CEDXUE01, VANCOD, GUZMUK, PIFZAG, DUPDAU, SOLHAD, IZARAC, FEMFUY01, GENZEF, KOHPUT, PNOIZA, CAXNIY, JEVPOQ, ZAQHAC, MOWBUX, HMBQOX10, DOTTUB, JUKWIV, LIVROY, ZOBYAQ, MUKGAA, LADLOT, YEKSOX, FADCUI, AYUKIP, LONRAI, IHURUA, QAPYUC, UMEREJ, ACRLAC04, EFAXIS, HIXLEG, VUHDEJ, BAZYAC01, DIQCEN, NIRJEE04, HEYJOK01, BAGQIJ, GUBZOS, SOWREB, DASTAS, MOJMEE, TCMPCQ01, WIKCAU, PELDER, LOZJUF, CIGBAV, ULAVIO, GAFWIV, CONXIN, RAYPOY, TEDQAU, DOXDAN, GILZUW, DELXEX, CEYJUN, ZACLEX, QIFDOZ, PUSYIM, PTZCOC, LIWCOJ, SOVGIT, VIGTUA, FEHVIY, YAMXUF, WIQCUU, OPISUC, NINDOD, BOGGAH, PUHHOQ, TIMBAU, CEMGIK, EGAFAU, WAHNUO, WAPPUY, WEJBIW, SEWHUY, CIDROW, DAZVEF, IGAMEJ, HAXMAW01, XOHXUP, GAQSEY, DOTRAG, DAJXIV, RAQRUX, KENMIA, DNCOOC, PODFIZ, WECRED, POLYUM, GUDKAS, HIJFIP01, PPYRED, PHTHIB, CBUDCX03, BUBNAN, LOYXAA, TAZWIB, TIWHAK, VUXYAO, UNEPOT.

## $^{15}N$

Of the 27,814 $^{13}C$ environments of the CSD-2k, the following 514 environments were detected as unusual (the numbering follows the FPS order listed in section V):
20, 21, 26, 27, 32, 33, 1255, 1256, 1257, 1258, 1259, 1260, 1261, 1262, 1263, 1264, 1265, 1266, 1267, 1268, 1269, 1270, 1551, 1552, 1553, 1554, 1555, 1556, 1557, 1558, 1655, 1656, 1673, 1674, 1745, 1746, 1823, 1824, 1833, 1834, 1959, 1960, 1961, 1962, 2535, 2536, 2537, 2538, 2969, 2970, 2971, 2972, 2973, 2974, 2975, 2976, 3013, 3014, 3145, 3146, 3147, 3148, 3189, 3190, 3191, 3192, 3305, 3306, 3307, 3308, 3317, 3318, 3319, 3320, 3354, 3355, 3871, 3872, 3873, 3874, 4137, 4138, 4139, 4140, 4141, 4142, 4143, 4144, 4188, 4189, 4190, 4191, 4192, 4193, 4200, 4201, 4202, 4203, 4204, 4205, 4206, 4207, 4208, 4209, 4210, 4211, 4212, 4213, 4214, 4215, 4252, 4253, 4254, 4255, 4260, 4261, 4262, 4263, 4388, 4389, 4390, 4391, 4468, 4469, 4470, 4471, 4472, 4473, 4474, 4475, 4936, 4937, 4938, 4939, 4940, 4941, 4942, 4943, 5090, 5091, 5100, 5101, 5282, 5283, 5284, 5285, 5286, 5287, 5288, 5289, 5958, 5959, 5960, 5961, 5962, 5963, 5964, 5965, 5998, 5999, 6000, 6001, 6050, 6051, 6052, 6053, 6054, 6055, 6056, 6057, 6400, 6401, 6402, 6403, 6412, 6413, 6414, 6415, 6618, 6619, 6620, 6621, 7252, 7253, 7254, 7255, 7256, 7257, 8003, 8004, 8005, 8006, 8171, 8172, 8173, 8174, 8564, 8565, 8566, 8567, 8968, 8969, 8970, 8971, 9186, 9187, 9188, 9189, 9190, 9191, 9192, 9193, 9478, 9479, 9480, 9481, 9482, 9483, 9484, 9485, 9526, 9527, 9528, 9529, 9530, 9531, 9532, 9533, 9694, 9695, 9696, 9697, 9974, 9975, 10446, 10447, 10448, 10449, 10450, 10451, 10452, 10453, 11132, 11133, 11134, 11135, 11140, 11141, 11142, 11143, 12286, 12287, 12288, 12289, 12468, 12469, 12470, 12471, 12472, 12473, 12474, 12475, 13188, 13189, 13190, 13191, 13290, 13291, 13462, 13463, 13464, 13465, 13466, 13467, 13468, 13469, 13496, 13497, 13498, 13499, 13802, 13803, 13804, 13805, 13806, 13807, 13808, 13809, 14520, 14521, 14522, 14523, 14524, 14525, 14526, 14527, 14752, 14753, 14754, 14755, 14756, 14757, 15002, 15003, 15060, 15061, 15062, 15063, 15216, 15217, 15218, 15219, 15512, 15513, 15514, 15515, 15663, 15667, 15671, 15675, 15708, 15709, 15710, 15711, 16179, 16180, 16181, 16182, 16183, 16184, 16185, 16186, 16239, 16240, 16593, 16594, 16595,

16596, 16905, 16906, 16907, 16908, 16915, 16916, 17391, 17392, 17393, 17394, 17399, 17400, 17401, 17402, 17403, 17404, 17405, 17406, 18916, 18917, 19246, 19247, 19248, 19249, 19254, 19255, 19256, 19257, 19342, 19343, 19592, 19593, 19616, 19617, 19832, 19833, 19834, 19835, 20320, 20321, 20322, 20323, 20324, 20325, 20326, 20327, 20328, 20329, 20330, 20331, 20332, 20333, 20334, 20335, 20436, 20437, 20438, 20439, 20440, 20441, 20442, 20443, 20622, 20623, 20624, 20625, 20900, 20901, 20902, 20903, 21580, 21581, 21896, 21897, 21898, 21899, 22094, 22095, 22096, 22097, 22194, 22195, 22412, 22413, 22414, 22415, 23012, 23013, 23014, 23015, 23448, 23449, 23450, 23451, 23464, 23465, 23466, 23467, 23654, 23655, 23656, 23657, 23874, 23875, 23876, 23877, 23878, 23879, 23880, 23881, 24086, 24087, 24088, 24089, 24090, 24091, 24092, 24093, 24170, 24171, 24172, 24173, 24174, 24175, 24176, 24177, 24246, 24247, 24504, 24505, 24506, 24507, 24524, 24525, 24526, 24527, 24858, 24859, 24860, 24861, 25484, 25485, 25486, 25487, 25504, 25505, 25506, 25507, 25508, 25509, 25510, 25511, 25512, 25513, 25514, 25515, 25516, 25517, 25518, 25519, 25596, 25597, 25598, 25599, 25970, 25971, 26396, 26397, 26398, 26399, 27302, 27303, 27400, 27401, 27402, 27403.
These environments belong to the structures COVSEM, UMERAF, DAJVUG01, UJEREG01, ZAXLIV, CITSED10, TIGBIU, SAFKAL, TETZIB, JACPOS01, ESIWUZ, APENIU, FORRUB, DNBOIM, PRMDIN20, DZBASK, FAFKEC, CYAMPD01, YOVVIP, XOCZUL, OMIWEM, XZBTZO, IXAQAB, NOPPOY, HEHQOC, MOPPUC01, SAPDAQ, LUSNOC, PEKSAZ, BUFMOG, AYAHUE, FIMWII, TOCLUU, JEMFUD, OXAYAO, XENQUD, TIMCHX, XIMKUA, BIZBET, BUSFUQ, QUDNAG, KIWJIK, APENOA, AZASER11, HAZQUV, LUZFUJ, CUYXAV, XUHPIB, CAKREM, VUWXIU, HAMHUZ, NAMVEE, AROMAV, IHEPIU, YITKAP, FUVDOP, LODWOR, BIDJON, VANCOD, NIMVAG, IHEHIO, HEVLUP, REHCAK, AYUKIP, APENUG, DOXKOR, UMEREJ, TANPEE, YAXHUC, BAZYAC01, UDAZIJ, GUBZOS, ESAREV, QIZBIL, IFAYIZ, LAYYOB, NSMACM, VACTIF, DELXEX, PTZCOC, WAKNUS, HIPXAF, FEHVIY, BUYZAW, PUHHOQ, HAXMAW01, ABIVUD, GAQSEY, PUNLOB, PODFIZ, XUBZID, NOGUNA.

### $^{17}$O

Of the 25,924 $^{13}$C environments of the CSD-2k, the following 441 environments were detected as unusual (the numbering follows the FPS order listed in section V):
232, 233, 234, 235, 308, 309, 310, 311, 312, 313, 314, 315, 378, 379, 380, 381, 1390, 1391, 1392, 1393, 1442, 1443, 1550, 1551, 1552, 1553, 1910, 1911, 1912, 1913, 1914, 1915, 1916, 1917, 2058, 2059, 2060, 2061, 2066, 2067, 2068, 2069, 2320, 2321, 2322, 2323, 2540, 2541, 2542, 2543, 2760, 2761, 2762, 2763, 3048, 3049, 3300, 3301, 3440, 3441, 3442, 3443, 3908, 3909, 3910, 3911, 3912, 3913, 3914, 3915, 3968, 3969, 3970, 3971, 4036, 4037, 4042, 4043, 4264, 4268, 4270, 4272, 4276, 4278, 4592, 4593, 4594, 4595, 4596, 4597, 4598, 4599, 4600, 4601, 4602, 4603, 4798, 4799, 4800, 4801, 4802, 4803, 4804, 4805, 5244, 5245, 5246, 5247, 5544, 5545, 5546, 5547, 5856, 5857, 5858, 5859, 6886, 6887, 6888, 6889, 7206, 7207, 7208, 7209, 7350, 7351, 7352, 7353, 7478, 7479, 7480, 7481, 7660, 7661, 7662, 7663, 7762, 7763, 7764, 7765, 8028, 8348, 8349, 8350, 8351, 8352, 8353, 8354, 8355, 8426, 8427, 8748, 8749, 8750, 8751, 8752, 8753, 8754, 8755, 9636, 9637, 9660, 9661, 9662, 9663, 10006, 10007, 10158, 10159, 10160, 10161, 10162, 10163, 10164, 10165, 10288, 10289, 10290, 10291, 10292, 10293, 10294, 10295, 10324, 10325, 10326, 10327, 10712, 10713, 11158, 11159, 11160, 11161, 11558, 11559, 11560, 11561, 11562, 11563, 11564, 11565, 11590, 11591, 11592, 11593, 11594, 11595, 11596, 11597, 11598, 11599, 11600, 11601, 12656, 12657, 12658, 12659, 12802, 12803, 12804, 12805, 12806, 12807, 12808, 12809, 12810, 12811, 12812, 12813, 12814, 12815, 12816, 12817, 12946, 12947, 12948, 12949, 12960, 12961, 12962, 12963, 12964, 12965, 12966, 12967, 13108, 13109, 13110, 13111, 13140, 13141, 13142, 13143, 13160, 13161, 13162, 13163, 13164, 13165, 13166, 13167, 13168, 13169, 13170, 13171, 13438, 13439, 13444, 13445, 13450, 13451, 13452, 13453, 13458, 13459, 13460, 13461, 13502, 13503, 13504, 13505, 13628, 13629, 13630, 13631, 13709, 13712, 13715, 13718, 13765, 13766, 13767, 13768, 14839, 14840, 14841, 14842, 14991, 14992, 14993, 14994, 15325, 15326, 15327, 15328, 15553, 15554, 15555, 15556, 15609, 15610, 15611, 15612, 15891, 15892, 15893, 15894, 15915, 15916, 15917, 15918, 15919, 15920, 15921, 15922, 15935, 15936, 16011, 16012, 16013, 16014, 16015, 16016, 16017, 16018, 16685, 16686, 17157, 17158, 17159, 17160, 17219, 17220, 17411, 17412, 17413, 17414, 17437, 17438, 18213, 18214, 18215, 18216, 18587, 18588, 18589, 18590, 18615, 18616, 18617, 18618, 18619, 18620, 18621, 18622, 18725, 18726, 18727, 18728, 19111, 19112, 19113, 19114, 19115, 19116, 19117, 19118, 20180, 20181, 20182, 20183, 20494, 20495, 20496, 20497, 20708, 20709, 21810, 21811, 21812, 21813, 23088, 23089, 23090, 23091, 23108, 23109, 23110, 23111, 23412, 23413, 23414, 23415, 23532, 23533, 23534, 23535, 23688, 23689, 23690, 23691, 23694, 23695, 24258, 24259, 24260, 24261, 24552, 24553, 24554, 24555, 24556, 24557, 24558, 24559, 24792, 24793, 24794, 24795, 24796, 24797, 24798, 24799, 25144, 25145, 25430, 25431, 25556, 25557, 25558, 25559.
These environments belong to the structures PARBAC03, BCBANN06, NUQPOF, TEDROL, CITSED10, FESNEW15, NIKVUZ, MEOXAL, BONGIW, OFUKOR, RIKPEG, PARGIZ, ZERZIH, HONSIO, UDOZUJ, OMIWEM, EKUKEC, ZAVTEV, PUFFOL01, UTAKUX, JOPNAD, PEKSAZ, GADZER, ZEXKEU, NUPTIC, PEJBAJ, TRIRED, BELBUP02, OXAYAO, CANXIA, JUVKAM, XIMKUA, NILQAC, DUSBUQ, DOYTUG, YEKJEF, PUQJOC, CORYIR10, QAFFOS, TABBEE, IFUWIR, LIWCUP, HAZQUV, HEWFEW, QUICNA02, CAKREM, BCBANN02, KAJRIY, MFCBXA, LETDUJ, PEPYIS, TECKAO, ZOQJAQ, IJARES, KABWUF,

IHEPIU, YITKAP, VANCOD, LIWJOQ, IHEHIO, KOHPUT, GEJCEF, HMBQOX10, TEKQAB01, DOTTUB, HXOXAM10, POXYAC, FADCUI, AYUKIP, LONRAI, QUDPOW, TANPEE, DOJNAS, ISEDOB, XADPEA, NIRJEE04, IFAYIZ, HOLQUU, FIFBAY, FINWOO, TIMBAU, JOSXUK, FIZJON01, HAXMAW01, KUBPOM, DUHYIP, DMXBZQ01, CEKVIX, RAZXEV, NOGUNA.

## X. References

1 Groom, C. R., Bruno, I. J., Lightfoot, M. P. & Ward, S. C. The Cambridge structural database. *Acta Crystallogr., Sect. B: Struct. Sci., Cryst. Eng. Mater.* **72**, 171-179, doi:10.1107/S2052520616003954 (2016).

2 Day, G. M., Motherwell, W. S. & Jones, W. A strategy for predicting the crystal structures of flexible molecules: the polymorphism of phenobarbital. *Phys. Chem. Chem. Phys.* **9**, 1693-1704, doi:10.1039/b612190j (2007).

3 Baias, M. *et al.* Powder crystallography of pharmaceutical materials by combined crystal structure prediction and solid-state 1H NMR spectroscopy. *Phys. Chem. Chem. Phys.* **15**, 8069-8069, doi:10.1039/c3cp41095a (2013).

4 Baias, M. *et al.* De novo determination of the crystal structure of a large drug molecule by crystal structure prediction-based powder NMR crystallography. *J. Am. Chem. Soc.* **135**, 17501-17507, doi:10.1021/ja4088874 (2013).

5 Pickard, C. J. & Mauri, F. All-electron magnetic response with pseudopotentials: NMR chemical shifts. *Phys. Rev. B* **63**, doi:ARTN 245101
DOI 10.1103/PhysRevB.63.245101 (2001).

6 Yates, J. R., Pickard, C. J. & Mauri, F. Calculation of NMR chemical shifts for extended systems using ultrasoft pseudopotentials. *Phys. Rev. B* **76**, 024401, doi:ARTN 024401 DOI 10.1103/PhysRevB.76.024401 (2007).

7 Clark, S. J. *et al.* First principles methods using CASTEP. *Zeitschrift Fur Kristallographie* **220**, 567-570, doi:DOI 10.1524/zkri.220.5.567.65075 (2005).

8 Giannozzi, P. *et al.* QUANTUM ESPRESSO: a modular and open-source software project for quantum simulations of materials. *J. Phys.: Condens. Matter* **21**, 395502, doi:10.1088/0953-8984/21/39/395502 (2009).

9 Giannozzi, P. *et al.* Advanced capabilities for materials modelling with Quantum ESPRESSO. *J. Phys.: Condens. Matter* **29**, 465901, doi:10.1088/1361-648X/aa8f79 (2017).

10 Lejaeghere, K. *et al.* Reproducibility in density functional theory calculations of solids. *Science* **351**, aad3000 (2016).

11 Perdew, J. P., Burke, K. & Ernzerhof, M. Generalized gradient approximation made simple. *Phys. Rev. Lett.* **77**, 3865, doi:DOI 10.1103/PhysRevLett.77.3865 (1996).

12 Grimme, S. Semiempirical GGA-type density functional constructed with a long-range dispersion correction. *J. Comput. Chem.* **27**, 1787-1799, doi:10.1002/jcc.20495 (2006).

13 Monkhorst, H. J. & Pack, J. D. Special points for Brillouin-zone integrations. *Phys. Rev. B* **13**, 5188, doi:DOI 10.1103/PhysRevB.13.5188 (1976).

14 Rasmussen, C. E. & Williams, C. K. *Gaussian processes for machine learning*. Vol. 1 (MIT press Cambridge, 2006).

15 Bartók, A. P., Kondor, R. & Csányi, G. On representing chemical environments. *Phys. Rev. B: Condens. Matter Mater. Phys.* **87**, 1-16, doi:10.1103/PhysRevB.87.184115 (2013).

16 De, S., Bartók, A. P., Csányi, G. & Ceriotti, M. Comparing molecules and solids across structural and alchemical space. *Phys. Chem. Chem. Phys.* **18**, 13754-13769, doi:10.1039/C6CP00415F (2016).

17 Bartok, A. P. *et al.* Machine learning unifies the modeling of materials and molecules. *Sci. Adv.* **3**, e1701816, doi:ARTN e1701816 DOI 10.1126/sciadv.1701816 (2017).

18 Ceriotti, M., Tribello, G. A. & Parrinello, M. Demonstrating the transferability and the descriptive power of sketch-map. *J. Chem. Theory Comput.* **9**, 1521-1532, doi:10.1021/ct3010563 (2013).

19 Campello, R. J., Moulavi, D., Zimek, A. & Sander, J. Hierarchical density estimates for data clustering, visualization, and outlier detection. *ACM Trans. Knowl. Discov. Data* **10**, 5, doi:Artn 5 DOI 10.1145/2733381 (2015).

20 Salager, E. *et al.* Powder Crystallography by Combined Crystal Structure Prediction and High-Resolution H-1 Solid-State NMR Spectroscopy. *J. Am. Chem. Soc.* **132**, 2564-+, doi:10.1021/ja909449k (2010).

21 Hofstetter, A. & Emsley, L. Positional Variance in NMR Crystallography. *J. Am. Chem. Soc* **139**, 2573-2576, doi:10.1021/jacs.6b12705 (2017).

22 Carignani, E., Borsacchi, S., Bradley, J. P., Brown, S. P. & Geppi, M. Strong intermolecular ring current influence on 1H chemical shifts in two crystalline forms of naproxen: a combined solid-state NMR and DFT study. *The Journal of Physical Chemistry C* **117**, 17731-17740 (2013).

23 Uldry, A.-C. *et al.* Quantifying weak hydrogen bonding in uracil and 4-Cyano-4 '-ethynylbiphenyl: a combined computational and experimental investigation of NMR chemical shifts in the solid state. *J. Am. Chem. Soc.* **130**, 945-954 (2008).

24 Sardo, M. *et al.* Diazole-based powdered cocrystal featuring a helical hydrogen-bonded network: Structure determination from PXRD, solid-state NMR and computer modeling. *Solid State Nucl Mag* **65**, 49-63 (2015).

25 Hartman, J. D., Kudla, R. A., Day, G. M., Mueller, L. J. & Beran, G. J. O. Benchmark fragment-based H-1, C-13, N-15 and O-17 chemical shift predictions in molecular crystals. *Phys. Chem. Chem. Phys.* **18**, 21686-21709, doi:10.1039/c6cp01831a (2016).

26 Arico-Muendel, C. C. *et al.* Orally Active Fumagillin Analogues: Transformations of a Reactive Warhead in the Gastric Environment. *ACS Med. Chem. Lett.* **4**, 381-386, doi:10.1021/ml3003633 (2013).

27 Dao, H. T., Li, C., Michaudel, Q., Maxwell, B. D. & Baran, P. S. Hydromethylation of Unactivated Olefins. *J. Am. Chem. Soc.* **137**, 8046-8049, doi:10.1021/jacs.5b05144 (2015).

28 Garozzo, D. *et al.* Inclusion networks of a calix[5]arene-based exoditopic receptor and long-chain alkyldiammonium ions. *Org. Lett.* **5**, 4025-4028, doi:10.1021/ol035310b (2003).

29 Bats, J. W. *CSD Communication* (2010).

30 Huang, G. B. *et al.* Selective recognition of aromatic hydrocarbons by endo-functionalized molecular tubes via C/N-H center dot center dot center dot pi interactions. *Chin. Chem. Lett.* **29**, 91-94, doi:10.1016/j.cclet.2017.07.005 (2018).

31 Plater, M. J., Harrison, W. T. A., de los Toyos, L. M. M. & Hendry, L. The consistent hexameric paddle-wheel crystallisation motif of a family of 2,4-bis(n-alkylamino)nitrobenzenes: alkyl = pentyl, hexyl, heptyl and octyl. *J. Chem. Res.*, 235-238, doi:10.3184/174751917x14902201357356 (2017).